\documentclass[longauth]{aa}

\usepackage{graphicx}
\usepackage{natbib}
\usepackage{scalerel}
\usepackage{comment}
\usepackage{caption}

\usepackage[table]{xcolor}

\usepackage{txfonts}
\usepackage[pdfencoding=auto,psdextra]{hyperref}
\hypersetup{
    colorlinks=true,
    linkcolor=blue,
    filecolor=magenta,      
    urlcolor=blue,
    citecolor=blue
}
\makeatletter
\renewcommand*\aa@pageof{, page \thepage{} of \pageref*{LastPage}}
\makeatother

\usepackage[utf8]{inputenc}

\usepackage[switch, modulo]{linenoaa}

\usepackage{euclid}

\renewcommand{\linenumbers}[0]{} 

\usepackage{siunitx}  

\usepackage{float}
\usepackage[dvipsnames,table]{xcolor}
\usepackage{enumitem}
\usepackage{mathtools}  
\usepackage{booktabs}  
\usepackage{colortbl}  
\usepackage{tabularx}  
\usepackage{csvsimple}  
\usepackage[normalem]{ulem}
\usepackage{cancel}

\usepackage{needspace} 

\usepackage[section]{placeins} 
\usepackage{afterpage} 

\usepackage[acronym]{glossaries}
\makeglossaries
\glsdisablehyper   

\newcommand{\mrm}[1]{\mathrm{#1}}
\newcommand{\mbf}[1]{\mathbf{#1}}
\newcommand{\mcal}[1]{\mathcal{#1}}

\newcommand{\mbs}[1]{\boldsymbol{#1}}

\usepackage{cleveref}
\crefname{section}{Sect.}{Sects.}
\Crefname{section}{Section}{Sections}
\crefname{subsection}{Sect.}{Sects.}
\Crefname{subsection}{Section}{Sections}
\crefname{subsubsection}{Sect.}{Sects.}
\Crefname{subsubsection}{Section}{Sections}
\crefname{figure}{Fig.}{Figs.}
\Crefname{figure}{Figure}{Figures}
\crefname{table}{Table}{Tables}
\Crefname{table}{Table}{Tables}
\crefname{appendix}{Appendix}{Appendices}
\Crefname{appendix}{Appendix}{Appendices}

\crefformat{equation}{Eq.~(#2#1#3)}
\crefmultiformat{equation}
  {Eq.~(#2#1#3)}
  { and~(#2#1#3)}
  {, (#2#1#3)}
  { and~(#2#1#3)}
\crefrangeformat{equation}{Eqs.~(#3#1#4)--(#5#2#6)}
\Crefformat{equation}{Equation~(#2#1#3)}
\Crefmultiformat{equation}
  {Equation~(#2#1#3)}
  { and~(#2#1#3)}
  {, (#2#1#3)}
  { and~(#2#1#3)}
\Crefrangeformat{equation}{Equations~(#3#1#4)--(#5#2#6)}

\definecolor{limegreen}{HTML}{32CD32}
\definecolor{orangeX}{HTML}{F4AD43}

\defcitealias{Nibauer+:2023}{N23}
\newacronym{mle}{MLE}{maximum likelihood estimate}
\newacronym{com}{CoM position}{centre of mass position}
\newacronym{col}{CoL position}{centre of light position}
\newacronym{Q1}{Q1}{Quick Release 1}
\newacronym{hmc}{HMC}{Hamiltonian Monte Carlo}
\newacronym{DM}{DM}{dark matter}
\newacronym{SIDM}{SIDM}{self-interacting dark matter}
\newacronym{ICM}{ICM}{intracluster medium}
\newacronym{NFW}{NFW}{Navarro, Frenk, and White}
\newacronym{6D}{6D}{six-dimensional}
\newacronym{ARRAKIHS}{ARRAKIHS}{Analysis of Resolved Remnants of Accreted galaxies as a Key Instrument for Halo Surveys}
\newacronym{LSST}{LSST}{Legacy Survey of Space and Time}

\DeclareSIUnit\pixel{px}
\DeclareSIUnit\pix{pix}

\newcommand{\veckappa}{\mbs{\kappa}}
\newcommand{\uveckappa}{\hat{\veckappa}}
\newcommand{\qproj}{q_{\mathrm{proj}}}
\newcommand{\vecT}{\mbf{T}}
\newcommand{\uvecT}{\hat{\vecT}}
\newcommand{\vectheta}{\mbs{\theta}}

\begin{document}

   \title{Euclid Quick Data Release (Q1)}
   \subtitle{The geometry of dark matter halos from extragalactic streams}

%
%
%

\newcommand{\orcid}[1]{} 

\author{Euclid Collaboration: N.~Starkman\orcid{0000-0003-3954-3291}\thanks{\email{starkman@mit.edu}}\inst{\ref{aff1},\ref{aff2}}
\and J.~Nibauer\orcid{0000-0001-8042-5794}\inst{\ref{aff3}}
\and S.~Pearson\orcid{0000-0003-0256-5446}\inst{\ref{aff4}}
\and S.~Wu\orcid{0009-0003-4675-3622}\inst{\ref{aff4}}
\and M.~Walmsley\orcid{0000-0002-6408-4181}\inst{\ref{aff5},\ref{aff6}}
\and L.~Necib\orcid{0000-0003-2806-1414}\inst{\ref{aff2}}
\and J.~Bovy\orcid{0000-0001-6855-442X}\inst{\ref{aff5}}
\and F.~R.~Marleau\orcid{0000-0002-1442-2947}\inst{\ref{aff7}}
\and E.~Sola\orcid{0000-0002-2814-3578}\inst{\ref{aff8}}
\and D.~Scott\orcid{0000-0002-6878-9840}\inst{\ref{aff9}}
\and B.~Altieri\orcid{0000-0003-3936-0284}\inst{\ref{aff10}}
\and S.~Andreon\orcid{0000-0002-2041-8784}\inst{\ref{aff11}}
\and N.~Auricchio\orcid{0000-0003-4444-8651}\inst{\ref{aff12}}
\and C.~Baccigalupi\orcid{0000-0002-8211-1630}\inst{\ref{aff13},\ref{aff14},\ref{aff15},\ref{aff16}}
\and M.~Baldi\orcid{0000-0003-4145-1943}\inst{\ref{aff17},\ref{aff12},\ref{aff18}}
\and A.~Balestra\orcid{0000-0002-6967-261X}\inst{\ref{aff19}}
\and S.~Bardelli\orcid{0000-0002-8900-0298}\inst{\ref{aff12}}
\and P.~Battaglia\orcid{0000-0002-7337-5909}\inst{\ref{aff12}}
\and A.~Biviano\orcid{0000-0002-0857-0732}\inst{\ref{aff14},\ref{aff13}}
\and M.~Bolzonella\orcid{0000-0003-3278-4607}\inst{\ref{aff12}}
\and E.~Branchini\orcid{0000-0002-0808-6908}\inst{\ref{aff20},\ref{aff21},\ref{aff11}}
\and M.~Brescia\orcid{0000-0001-9506-5680}\inst{\ref{aff22},\ref{aff23}}
\and J.~Brinchmann\orcid{0000-0003-4359-8797}\inst{\ref{aff24},\ref{aff25},\ref{aff26}}
\and S.~Camera\orcid{0000-0003-3399-3574}\inst{\ref{aff27},\ref{aff28},\ref{aff29}}
\and V.~Capobianco\orcid{0000-0002-3309-7692}\inst{\ref{aff29}}
\and C.~Carbone\orcid{0000-0003-0125-3563}\inst{\ref{aff30}}
\and J.~Carretero\orcid{0000-0002-3130-0204}\inst{\ref{aff31},\ref{aff32}}
\and S.~Casas\orcid{0000-0002-4751-5138}\inst{\ref{aff33},\ref{aff34}}
\and M.~Castellano\orcid{0000-0001-9875-8263}\inst{\ref{aff35}}
\and G.~Castignani\orcid{0000-0001-6831-0687}\inst{\ref{aff12}}
\and S.~Cavuoti\orcid{0000-0002-3787-4196}\inst{\ref{aff23},\ref{aff36}}
\and K.~C.~Chambers\orcid{0000-0001-6965-7789}\inst{\ref{aff37}}
\and A.~Cimatti\inst{\ref{aff38}}
\and C.~Colodro-Conde\inst{\ref{aff39}}
\and G.~Congedo\orcid{0000-0003-2508-0046}\inst{\ref{aff40}}
\and C.~J.~Conselice\orcid{0000-0003-1949-7638}\inst{\ref{aff6}}
\and L.~Conversi\orcid{0000-0002-6710-8476}\inst{\ref{aff41},\ref{aff10}}
\and Y.~Copin\orcid{0000-0002-5317-7518}\inst{\ref{aff42}}
\and A.~Costille\inst{\ref{aff43}}
\and F.~Courbin\orcid{0000-0003-0758-6510}\inst{\ref{aff44},\ref{aff45},\ref{aff46}}
\and H.~M.~Courtois\orcid{0000-0003-0509-1776}\inst{\ref{aff47}}
\and M.~Cropper\orcid{0000-0003-4571-9468}\inst{\ref{aff48}}
\and H.~Degaudenzi\orcid{0000-0002-5887-6799}\inst{\ref{aff49}}
\and G.~De~Lucia\orcid{0000-0002-6220-9104}\inst{\ref{aff14}}
\and H.~Dole\orcid{0000-0002-9767-3839}\inst{\ref{aff50}}
\and F.~Dubath\orcid{0000-0002-6533-2810}\inst{\ref{aff49}}
\and C.~A.~J.~Duncan\orcid{0009-0003-3573-0791}\inst{\ref{aff40}}
\and X.~Dupac\inst{\ref{aff10}}
\and S.~Dusini\orcid{0000-0002-1128-0664}\inst{\ref{aff51}}
\and S.~Escoffier\orcid{0000-0002-2847-7498}\inst{\ref{aff52}}
\and M.~Fabricius\orcid{0000-0002-7025-6058}\inst{\ref{aff53},\ref{aff54}}
\and M.~Farina\orcid{0000-0002-3089-7846}\inst{\ref{aff55}}
\and R.~Farinelli\inst{\ref{aff12}}
\and S.~Ferriol\inst{\ref{aff42}}
\and F.~Finelli\orcid{0000-0002-6694-3269}\inst{\ref{aff12},\ref{aff56}}
\and P.~Fosalba\orcid{0000-0002-1510-5214}\inst{\ref{aff57},\ref{aff58}}
\and S.~Fotopoulou\orcid{0000-0002-9686-254X}\inst{\ref{aff59}}
\and M.~Frailis\orcid{0000-0002-7400-2135}\inst{\ref{aff14}}
\and E.~Franceschi\orcid{0000-0002-0585-6591}\inst{\ref{aff12}}
\and M.~Fumana\orcid{0000-0001-6787-5950}\inst{\ref{aff30}}
\and L.~Gabarra\orcid{0000-0002-8486-8856}\inst{\ref{aff60}}
\and S.~Galeotta\orcid{0000-0002-3748-5115}\inst{\ref{aff14}}
\and K.~George\orcid{0000-0002-1734-8455}\inst{\ref{aff61}}
\and B.~Gillis\orcid{0000-0002-4478-1270}\inst{\ref{aff40}}
\and C.~Giocoli\orcid{0000-0002-9590-7961}\inst{\ref{aff12},\ref{aff18}}
\and P.~G\'omez-Alvarez\orcid{0000-0002-8594-5358}\inst{\ref{aff62},\ref{aff10}}
\and J.~Gracia-Carpio\orcid{0000-0003-4689-3134}\inst{\ref{aff53}}
\and A.~Grazian\orcid{0000-0002-5688-0663}\inst{\ref{aff19}}
\and F.~Grupp\inst{\ref{aff53},\ref{aff54}}
\and S.~V.~H.~Haugan\orcid{0000-0001-9648-7260}\inst{\ref{aff63}}
\and W.~Holmes\inst{\ref{aff64}}
\and F.~Hormuth\inst{\ref{aff65}}
\and A.~Hornstrup\orcid{0000-0002-3363-0936}\inst{\ref{aff66},\ref{aff67}}
\and M.~Huertas-Company\orcid{0000-0002-1416-8483}\inst{\ref{aff39},\ref{aff68},\ref{aff69}}
\and K.~Jahnke\orcid{0000-0003-3804-2137}\inst{\ref{aff70}}
\and M.~Jhabvala\inst{\ref{aff71}}
\and B.~Joachimi\orcid{0000-0001-7494-1303}\inst{\ref{aff72}}
\and S.~Kermiche\orcid{0000-0002-0302-5735}\inst{\ref{aff52}}
\and A.~Kiessling\orcid{0000-0002-2590-1273}\inst{\ref{aff64}}
\and B.~Kubik\orcid{0009-0006-5823-4880}\inst{\ref{aff42}}
\and M.~K\"ummel\orcid{0000-0003-2791-2117}\inst{\ref{aff54}}
\and M.~Kunz\orcid{0000-0002-3052-7394}\inst{\ref{aff73}}
\and H.~Kurki-Suonio\orcid{0000-0002-4618-3063}\inst{\ref{aff74},\ref{aff75}}
\and A.~M.~C.~Le~Brun\orcid{0000-0002-0936-4594}\inst{\ref{aff76}}
\and S.~Ligori\orcid{0000-0003-4172-4606}\inst{\ref{aff29}}
\and P.~B.~Lilje\orcid{0000-0003-4324-7794}\inst{\ref{aff63}}
\and V.~Lindholm\orcid{0000-0003-2317-5471}\inst{\ref{aff74},\ref{aff75}}
\and I.~Lloro\orcid{0000-0001-5966-1434}\inst{\ref{aff77}}
\and M.~Magliocchetti\orcid{0000-0001-9158-4838}\inst{\ref{aff55}}
\and G.~Mainetti\orcid{0000-0003-2384-2377}\inst{\ref{aff78}}
\and O.~Mansutti\orcid{0000-0001-5758-4658}\inst{\ref{aff14}}
\and O.~Marggraf\orcid{0000-0001-7242-3852}\inst{\ref{aff79}}
\and M.~Martinelli\orcid{0000-0002-6943-7732}\inst{\ref{aff35},\ref{aff80}}
\and N.~Martinet\orcid{0000-0003-2786-7790}\inst{\ref{aff43}}
\and F.~Marulli\orcid{0000-0002-8850-0303}\inst{\ref{aff81},\ref{aff12},\ref{aff18}}
\and R.~J.~Massey\orcid{0000-0002-6085-3780}\inst{\ref{aff82}}
\and N.~Mauri\orcid{0000-0001-8196-1548}\inst{\ref{aff38},\ref{aff18}}
\and E.~Medinaceli\orcid{0000-0002-4040-7783}\inst{\ref{aff12}}
\and S.~Mei\orcid{0000-0002-2849-559X}\inst{\ref{aff83},\ref{aff84}}
\and M.~Melchior\inst{\ref{aff85}}
\and M.~Meneghetti\orcid{0000-0003-1225-7084}\inst{\ref{aff12},\ref{aff18}}
\and E.~Merlin\orcid{0000-0001-6870-8900}\inst{\ref{aff35}}
\and G.~Meylan\inst{\ref{aff86}}
\and P.~Monaco\orcid{0000-0003-2083-7564}\inst{\ref{aff87},\ref{aff14},\ref{aff15},\ref{aff13}}
\and A.~Mora\orcid{0000-0002-1922-8529}\inst{\ref{aff88}}
\and M.~Moresco\orcid{0000-0002-7616-7136}\inst{\ref{aff81},\ref{aff12}}
\and C.~Moretti\orcid{0000-0003-3314-8936}\inst{\ref{aff14},\ref{aff13},\ref{aff15}}
\and L.~Moscardini\orcid{0000-0002-3473-6716}\inst{\ref{aff81},\ref{aff12},\ref{aff18}}
\and R.~Nakajima\orcid{0009-0009-1213-7040}\inst{\ref{aff79}}
\and C.~Neissner\orcid{0000-0001-8524-4968}\inst{\ref{aff89},\ref{aff32}}
\and R.~C.~Nichol\orcid{0000-0003-0939-6518}\inst{\ref{aff90}}
\and S.-M.~Niemi\orcid{0009-0005-0247-0086}\inst{\ref{aff91}}
\and J.~W.~Nightingale\orcid{0000-0002-8987-7401}\inst{\ref{aff92}}
\and C.~Padilla\orcid{0000-0001-7951-0166}\inst{\ref{aff89}}
\and S.~Paltani\orcid{0000-0002-8108-9179}\inst{\ref{aff49}}
\and F.~Pasian\orcid{0000-0002-4869-3227}\inst{\ref{aff14}}
\and K.~Pedersen\inst{\ref{aff4}}
\and W.~J.~Percival\orcid{0000-0002-0644-5727}\inst{\ref{aff93},\ref{aff94},\ref{aff95}}
\and V.~Pettorino\orcid{0000-0002-4203-9320}\inst{\ref{aff91}}
\and A.~Pezzotta\orcid{0000-0003-0726-2268}\inst{\ref{aff11}}
\and S.~Pires\orcid{0000-0002-0249-2104}\inst{\ref{aff96}}
\and G.~Polenta\orcid{0000-0003-4067-9196}\inst{\ref{aff97}}
\and M.~Poncet\inst{\ref{aff98}}
\and L.~A.~Popa\inst{\ref{aff99}}
\and L.~Pozzetti\orcid{0000-0001-7085-0412}\inst{\ref{aff12}}
\and F.~Raison\orcid{0000-0002-7819-6918}\inst{\ref{aff53}}
\and A.~Renzi\orcid{0000-0001-9856-1970}\inst{\ref{aff100},\ref{aff51},\ref{aff12}}
\and J.~Rhodes\orcid{0000-0002-4485-8549}\inst{\ref{aff64}}
\and G.~Riccio\inst{\ref{aff23}}
\and I.~Risso\orcid{0000-0003-2525-7761}\inst{\ref{aff11},\ref{aff21}}
\and E.~Romelli\orcid{0000-0003-3069-9222}\inst{\ref{aff14}}
\and M.~Roncarelli\orcid{0000-0001-9587-7822}\inst{\ref{aff12}}
\and C.~Rosset\orcid{0000-0003-0286-2192}\inst{\ref{aff83}}
\and R.~Saglia\orcid{0000-0003-0378-7032}\inst{\ref{aff54},\ref{aff53}}
\and Z.~Sakr\orcid{0000-0002-4823-3757}\inst{\ref{aff101},\ref{aff102},\ref{aff103}}
\and D.~Sapone\orcid{0000-0001-7089-4503}\inst{\ref{aff104}}
\and B.~Sartoris\orcid{0000-0003-1337-5269}\inst{\ref{aff54},\ref{aff14}}
\and P.~Schneider\orcid{0000-0001-8561-2679}\inst{\ref{aff79}}
\and A.~Secroun\orcid{0000-0003-0505-3710}\inst{\ref{aff52}}
\and E.~Sihvola\orcid{0000-0003-1804-7715}\inst{\ref{aff105}}
\and P.~Simon\inst{\ref{aff79}}
\and C.~Sirignano\orcid{0000-0002-0995-7146}\inst{\ref{aff100},\ref{aff51}}
\and G.~Sirri\orcid{0000-0003-2626-2853}\inst{\ref{aff18}}
\and L.~Stanco\orcid{0000-0002-9706-5104}\inst{\ref{aff51}}
\and P.~Tallada-Cresp\'{i}\orcid{0000-0002-1336-8328}\inst{\ref{aff31},\ref{aff32}}
\and A.~N.~Taylor\inst{\ref{aff40}}
\and I.~Tereno\orcid{0000-0002-4537-6218}\inst{\ref{aff106},\ref{aff107}}
\and N.~Tessore\orcid{0000-0002-9696-7931}\inst{\ref{aff48}}
\and S.~Toft\orcid{0000-0003-3631-7176}\inst{\ref{aff108},\ref{aff109}}
\and R.~Toledo-Moreo\orcid{0000-0002-2997-4859}\inst{\ref{aff110}}
\and F.~Torradeflot\orcid{0000-0003-1160-1517}\inst{\ref{aff32},\ref{aff31}}
\and A.~Tsyganov\inst{\ref{aff111}}
\and I.~Tutusaus\orcid{0000-0002-3199-0399}\inst{\ref{aff58},\ref{aff57},\ref{aff102}}
\and E.~A.~Valentijn\inst{\ref{aff112}}
\and J.~Valiviita\orcid{0000-0001-6225-3693}\inst{\ref{aff74},\ref{aff75}}
\and T.~Vassallo\orcid{0000-0001-6512-6358}\inst{\ref{aff14},\ref{aff61}}
\and G.~Verdoes~Kleijn\orcid{0000-0001-5803-2580}\inst{\ref{aff112}}
\and A.~Veropalumbo\orcid{0000-0003-2387-1194}\inst{\ref{aff11},\ref{aff21},\ref{aff20}}
\and Y.~Wang\orcid{0000-0002-4749-2984}\inst{\ref{aff113}}
\and J.~Weller\orcid{0000-0002-8282-2010}\inst{\ref{aff54},\ref{aff53}}
\and A.~Zacchei\orcid{0000-0003-0396-1192}\inst{\ref{aff14},\ref{aff13}}
\and G.~Zamorani\orcid{0000-0002-2318-301X}\inst{\ref{aff12}}
\and F.~M.~Zerbi\orcid{0000-0002-9996-973X}\inst{\ref{aff11}}
\and E.~Zucca\orcid{0000-0002-5845-8132}\inst{\ref{aff12}}
\and M.~Ballardini\orcid{0000-0003-4481-3559}\inst{\ref{aff114},\ref{aff115},\ref{aff12}}
\and E.~Bozzo\orcid{0000-0002-8201-1525}\inst{\ref{aff49}}
\and C.~Burigana\orcid{0000-0002-3005-5796}\inst{\ref{aff116},\ref{aff56}}
\and R.~Cabanac\orcid{0000-0001-6679-2600}\inst{\ref{aff102}}
\and M.~Calabrese\orcid{0000-0002-2637-2422}\inst{\ref{aff117},\ref{aff30}}
\and A.~Cappi\inst{\ref{aff118},\ref{aff12}}
\and T.~Castro\orcid{0000-0002-6292-3228}\inst{\ref{aff14},\ref{aff15},\ref{aff13},\ref{aff119}}
\and J.~A.~Escartin~Vigo\inst{\ref{aff53}}
\and J.~Garc\'ia-Bellido\orcid{0000-0002-9370-8360}\inst{\ref{aff101}}
\and J.~Macias-Perez\orcid{0000-0002-5385-2763}\inst{\ref{aff120}}
\and R.~Maoli\orcid{0000-0002-6065-3025}\inst{\ref{aff121},\ref{aff35}}
\and J.~Mart\'{i}n-Fleitas\orcid{0000-0002-8594-569X}\inst{\ref{aff122}}
\and R.~B.~Metcalf\orcid{0000-0003-3167-2574}\inst{\ref{aff81},\ref{aff12}}
\and M.~P\"ontinen\orcid{0000-0001-5442-2530}\inst{\ref{aff74}}
\and V.~Scottez\orcid{0009-0008-3864-940X}\inst{\ref{aff123},\ref{aff124}}
\and M.~Sereno\orcid{0000-0003-0302-0325}\inst{\ref{aff12},\ref{aff18}}
\and M.~Tenti\orcid{0000-0002-4254-5901}\inst{\ref{aff18}}
\and M.~Tucci\inst{\ref{aff49}}
\and M.~Viel\orcid{0000-0002-2642-5707}\inst{\ref{aff13},\ref{aff14},\ref{aff16},\ref{aff15},\ref{aff119}}
\and M.~Wiesmann\orcid{0009-0000-8199-5860}\inst{\ref{aff63}}
\and Y.~Akrami\orcid{0000-0002-2407-7956}\inst{\ref{aff101},\ref{aff1}}
\and I.~T.~Andika\orcid{0000-0001-6102-9526}\inst{\ref{aff54}}
\and G.~Angora\orcid{0000-0002-0316-6562}\inst{\ref{aff23},\ref{aff114}}
\and S.~Anselmi\orcid{0000-0002-3579-9583}\inst{\ref{aff51},\ref{aff100},\ref{aff125}}
\and M.~Archidiacono\orcid{0000-0003-4952-9012}\inst{\ref{aff126},\ref{aff127}}
\and F.~Atrio-Barandela\orcid{0000-0002-2130-2513}\inst{\ref{aff128}}
\and L.~Bazzanini\orcid{0000-0003-0727-0137}\inst{\ref{aff114},\ref{aff12}}
\and D.~Bertacca\orcid{0000-0002-2490-7139}\inst{\ref{aff100},\ref{aff19},\ref{aff51}}
\and M.~Bethermin\orcid{0000-0002-3915-2015}\inst{\ref{aff129}}
\and F.~Beutler\orcid{0000-0003-0467-5438}\inst{\ref{aff40}}
\and A.~Blanchard\orcid{0000-0001-8555-9003}\inst{\ref{aff102}}
\and L.~Blot\orcid{0000-0002-9622-7167}\inst{\ref{aff130},\ref{aff76}}
\and M.~L.~Brown\orcid{0000-0002-0370-8077}\inst{\ref{aff6}}
\and S.~Bruton\orcid{0000-0002-6503-5218}\inst{\ref{aff131}}
\and A.~Calabro\orcid{0000-0003-2536-1614}\inst{\ref{aff35}}
\and B.~Camacho~Quevedo\orcid{0000-0002-8789-4232}\inst{\ref{aff13},\ref{aff16},\ref{aff14}}
\and F.~Caro\orcid{0009-0003-1053-0507}\inst{\ref{aff35}}
\and C.~S.~Carvalho\inst{\ref{aff107}}
\and F.~Cogato\orcid{0000-0003-4632-6113}\inst{\ref{aff81},\ref{aff12}}
\and A.~R.~Cooray\orcid{0000-0002-3892-0190}\inst{\ref{aff132}}
\and O.~Cucciati\orcid{0000-0002-9336-7551}\inst{\ref{aff12}}
\and F.~De~Paolis\orcid{0000-0001-6460-7563}\inst{\ref{aff133},\ref{aff134},\ref{aff135}}
\and G.~Desprez\orcid{0000-0001-8325-1742}\inst{\ref{aff112}}
\and A.~D\'iaz-S\'anchez\orcid{0000-0003-0748-4768}\inst{\ref{aff136}}
\and S.~Di~Domizio\orcid{0000-0003-2863-5895}\inst{\ref{aff20},\ref{aff21}}
\and J.~M.~Diego\orcid{0000-0001-9065-3926}\inst{\ref{aff137}}
\and P.-A.~Duc\orcid{0000-0003-3343-6284}\inst{\ref{aff129}}
\and V.~Duret\orcid{0009-0009-0383-4960}\inst{\ref{aff52}}
\and M.~Y.~Elkhashab\orcid{0000-0001-9306-2603}\inst{\ref{aff14},\ref{aff15},\ref{aff87},\ref{aff13}}
\and A.~Enia\orcid{0000-0002-0200-2857}\inst{\ref{aff12}}
\and Y.~Fang\orcid{0000-0002-0334-6950}\inst{\ref{aff54}}
\and A.~Finoguenov\orcid{0000-0002-4606-5403}\inst{\ref{aff74}}
\and A.~Franco\orcid{0000-0002-4761-366X}\inst{\ref{aff134},\ref{aff133},\ref{aff135}}
\and K.~Ganga\orcid{0000-0001-8159-8208}\inst{\ref{aff83}}
\and T.~Gasparetto\orcid{0000-0002-7913-4866}\inst{\ref{aff35}}
\and R.~Gavazzi\orcid{0000-0002-5540-6935}\inst{\ref{aff43},\ref{aff138}}
\and E.~Gaztanaga\orcid{0000-0001-9632-0815}\inst{\ref{aff58},\ref{aff57},\ref{aff139}}
\and F.~Giacomini\orcid{0000-0002-3129-2814}\inst{\ref{aff18}}
\and F.~Gianotti\orcid{0000-0003-4666-119X}\inst{\ref{aff12}}
\and E.~J.~Gonzalez\orcid{0000-0002-0226-9893}\inst{\ref{aff140},\ref{aff141}}
\and G.~Gozaliasl\orcid{0000-0002-0236-919X}\inst{\ref{aff142},\ref{aff74}}
\and A.~Gruppuso\orcid{0000-0001-9272-5292}\inst{\ref{aff12},\ref{aff18}}
\and M.~Guidi\orcid{0000-0001-9408-1101}\inst{\ref{aff17},\ref{aff12}}
\and C.~M.~Gutierrez\orcid{0000-0001-7854-783X}\inst{\ref{aff39},\ref{aff143}}
\and A.~Hall\orcid{0000-0002-3139-8651}\inst{\ref{aff40}}
\and C.~Hern\'andez-Monteagudo\orcid{0000-0001-5471-9166}\inst{\ref{aff143},\ref{aff39}}
\and H.~Hildebrandt\orcid{0000-0002-9814-3338}\inst{\ref{aff144}}
\and J.~Hjorth\orcid{0000-0002-4571-2306}\inst{\ref{aff4}}
\and L.~K.~Hunt\orcid{0000-0001-9162-2371}\inst{\ref{aff145}}
\and J.~J.~E.~Kajava\orcid{0000-0002-3010-8333}\inst{\ref{aff146},\ref{aff147},\ref{aff148}}
\and Y.~Kang\orcid{0009-0000-8588-7250}\inst{\ref{aff49}}
\and V.~Kansal\orcid{0000-0002-4008-6078}\inst{\ref{aff149},\ref{aff150}}
\and D.~Karagiannis\orcid{0000-0002-4927-0816}\inst{\ref{aff114},\ref{aff151}}
\and K.~Kiiveri\inst{\ref{aff105}}
\and J.~Kim\orcid{0000-0003-2776-2761}\inst{\ref{aff60}}
\and C.~C.~Kirkpatrick\inst{\ref{aff105}}
\and S.~Kruk\orcid{0000-0001-8010-8879}\inst{\ref{aff10}}
\and M.~Lattanzi\orcid{0000-0003-1059-2532}\inst{\ref{aff115}}
\and M.~Lembo\orcid{0000-0002-5271-5070}\inst{\ref{aff138}}
\and F.~Lepori\orcid{0009-0000-5061-7138}\inst{\ref{aff152}}
\and G.~Leroy\orcid{0009-0004-2523-4425}\inst{\ref{aff153},\ref{aff82}}
\and J.~Lesgourgues\orcid{0000-0001-7627-353X}\inst{\ref{aff33}}
\and T.~I.~Liaudat\orcid{0000-0002-9104-314X}\inst{\ref{aff154}}
\and S.~J.~Liu\orcid{0000-0001-7680-2139}\inst{\ref{aff55}}
\and A.~Loureiro\orcid{0000-0002-4371-0876}\inst{\ref{aff155},\ref{aff156}}
\and G.~Maggio\orcid{0000-0003-4020-4836}\inst{\ref{aff14}}
\and A.~Manj\'on-Garc\'ia\orcid{0000-0002-7413-8825}\inst{\ref{aff136}}
\and F.~Mannucci\orcid{0000-0002-4803-2381}\inst{\ref{aff145}}
\and C.~J.~A.~P.~Martins\orcid{0000-0002-4886-9261}\inst{\ref{aff157},\ref{aff24}}
\and L.~Maurin\orcid{0000-0002-8406-0857}\inst{\ref{aff50}}
\and M.~Miluzio\inst{\ref{aff10},\ref{aff158}}
\and G.~Morgante\inst{\ref{aff12}}
\and S.~Nadathur\orcid{0000-0001-9070-3102}\inst{\ref{aff139}}
\and K.~Naidoo\orcid{0000-0002-9182-1802}\inst{\ref{aff139},\ref{aff70}}
\and A.~Navarro-Alsina\orcid{0000-0002-3173-2592}\inst{\ref{aff79}}
\and S.~Nesseris\orcid{0000-0002-0567-0324}\inst{\ref{aff101}}
\and D.~Paoletti\orcid{0000-0003-4761-6147}\inst{\ref{aff12},\ref{aff56}}
\and F.~Passalacqua\orcid{0000-0002-8606-4093}\inst{\ref{aff100},\ref{aff51}}
\and K.~Paterson\orcid{0000-0001-8340-3486}\inst{\ref{aff70}}
\and L.~Patrizii\inst{\ref{aff18}}
\and C.~Pattison\orcid{0000-0003-3272-2617}\inst{\ref{aff139}}
\and A.~Pisani\orcid{0000-0002-6146-4437}\inst{\ref{aff52}}
\and D.~Potter\orcid{0000-0002-0757-5195}\inst{\ref{aff159}}
\and G.~W.~Pratt\inst{\ref{aff96}}
\and S.~Quai\orcid{0000-0002-0449-8163}\inst{\ref{aff81},\ref{aff12}}
\and M.~Radovich\orcid{0000-0002-3585-866X}\inst{\ref{aff19}}
\and K.~Rojas\orcid{0000-0003-1391-6854}\inst{\ref{aff160}}
\and W.~Roster\orcid{0000-0002-9149-6528}\inst{\ref{aff53}}
\and S.~Sacquegna\orcid{0000-0002-8433-6630}\inst{\ref{aff161}}
\and M.~Sahl\'en\orcid{0000-0003-0973-4804}\inst{\ref{aff162}}
\and D.~B.~Sanders\orcid{0000-0002-1233-9998}\inst{\ref{aff37}}
\and E.~Sarpa\orcid{0000-0002-1256-655X}\inst{\ref{aff14}}
\and A.~Schneider\orcid{0000-0001-7055-8104}\inst{\ref{aff159}}
\and D.~Sciotti\orcid{0009-0008-4519-2620}\inst{\ref{aff35},\ref{aff80}}
\and E.~Sellentin\inst{\ref{aff163},\ref{aff164}}
\and F.~Shankar\orcid{0000-0001-8973-5051}\inst{\ref{aff165}}
\and J.~Stadel\orcid{0000-0001-7565-8622}\inst{\ref{aff159}}
\and K.~Tanidis\orcid{0000-0001-9843-5130}\inst{\ref{aff166}}
\and F.~Tarsitano\orcid{0000-0002-5919-0238}\inst{\ref{aff167},\ref{aff49}}
\and G.~Testera\inst{\ref{aff21}}
\and R.~Teyssier\orcid{0000-0001-7689-0933}\inst{\ref{aff3}}
\and S.~Tosi\orcid{0000-0002-7275-9193}\inst{\ref{aff20},\ref{aff11},\ref{aff21}}
\and A.~Troja\orcid{0000-0003-0239-4595}\inst{\ref{aff14}}
\and M.~Urbano\orcid{0000-0001-5640-0650}\inst{\ref{aff168},\ref{aff129}}
\and C.~Valieri\inst{\ref{aff18}}
\and A.~Venhola\orcid{0000-0001-6071-4564}\inst{\ref{aff169}}
\and D.~Vergani\orcid{0000-0003-0898-2216}\inst{\ref{aff12}}
\and G.~Verza\orcid{0000-0002-1886-8348}\inst{\ref{aff170},\ref{aff171}}
\and P.~Vielzeuf\orcid{0000-0003-2035-9339}\inst{\ref{aff52}}
\and S.~Vinciguerra\orcid{0009-0005-4018-3184}\inst{\ref{aff43}}
\and N.~A.~Walton\orcid{0000-0003-3983-8778}\inst{\ref{aff8}}
\and J.~R.~Weaver\orcid{0000-0003-1614-196X}\inst{\ref{aff2}}
\and A.~H.~Wright\orcid{0000-0001-7363-7932}\inst{\ref{aff144}}}

\institute{CERCA/ISO, Department of Physics, Case Western Reserve University, 10900 Euclid Avenue, Cleveland, OH 44106, USA\label{aff1}
\and
MIT Kavli Institute for Astrophysics and Space Research, Massachusetts Institute of Technology, Cambridge, MA 02139, USA\label{aff2}
\and
Department of Astrophysical Sciences, Peyton Hall, Princeton University, Princeton, NJ 08544, USA\label{aff3}
\and
DARK, Niels Bohr Institute, University of Copenhagen, Jagtvej 155, 2200 Copenhagen, Denmark\label{aff4}
\and
David A. Dunlap Department of Astronomy \& Astrophysics, University of Toronto, 50 St George Street, Toronto, Ontario M5S 3H4, Canada\label{aff5}
\and
Jodrell Bank Centre for Astrophysics, Department of Physics and Astronomy, University of Manchester, Oxford Road, Manchester M13 9PL, UK\label{aff6}
\and
Universit\"at Innsbruck, Institut f\"ur Astro- und Teilchenphysik, Technikerstr. 25/8, 6020 Innsbruck, Austria\label{aff7}
\and
Institute of Astronomy, University of Cambridge, Madingley Road, Cambridge CB3 0HA, UK\label{aff8}
\and
Department of Physics and Astronomy, University of British Columbia, Vancouver, BC V6T 1Z1, Canada\label{aff9}
\and
ESAC/ESA, Camino Bajo del Castillo, s/n., Urb. Villafranca del Castillo, 28692 Villanueva de la Ca\~nada, Madrid, Spain\label{aff10}
\and
INAF-Osservatorio Astronomico di Brera, Via Brera 28, 20122 Milano, Italy\label{aff11}
\and
INAF-Osservatorio di Astrofisica e Scienza dello Spazio di Bologna, Via Piero Gobetti 93/3, 40129 Bologna, Italy\label{aff12}
\and
IFPU, Institute for Fundamental Physics of the Universe, via Beirut 2, 34151 Trieste, Italy\label{aff13}
\and
INAF-Osservatorio Astronomico di Trieste, Via G. B. Tiepolo 11, 34143 Trieste, Italy\label{aff14}
\and
INFN, Sezione di Trieste, Via Valerio 2, 34127 Trieste TS, Italy\label{aff15}
\and
SISSA, International School for Advanced Studies, Via Bonomea 265, 34136 Trieste TS, Italy\label{aff16}
\and
Dipartimento di Fisica e Astronomia, Universit\`a di Bologna, Via Gobetti 93/2, 40129 Bologna, Italy\label{aff17}
\and
INFN-Sezione di Bologna, Viale Berti Pichat 6/2, 40127 Bologna, Italy\label{aff18}
\and
INAF-Osservatorio Astronomico di Padova, Via dell'Osservatorio 5, 35122 Padova, Italy\label{aff19}
\and
Dipartimento di Fisica, Universit\`a di Genova, Via Dodecaneso 33, 16146, Genova, Italy\label{aff20}
\and
INFN-Sezione di Genova, Via Dodecaneso 33, 16146, Genova, Italy\label{aff21}
\and
Department of Physics "E. Pancini", University Federico II, Via Cinthia 6, 80126, Napoli, Italy\label{aff22}
\and
INAF-Osservatorio Astronomico di Capodimonte, Via Moiariello 16, 80131 Napoli, Italy\label{aff23}
\and
Instituto de Astrof\'isica e Ci\^encias do Espa\c{c}o, Universidade do Porto, CAUP, Rua das Estrelas, PT4150-762 Porto, Portugal\label{aff24}
\and
Faculdade de Ci\^encias da Universidade do Porto, Rua do Campo de Alegre, 4150-007 Porto, Portugal\label{aff25}
\and
European Southern Observatory, Karl-Schwarzschild-Str.~2, 85748 Garching, Germany\label{aff26}
\and
Dipartimento di Fisica, Universit\`a degli Studi di Torino, Via P. Giuria 1, 10125 Torino, Italy\label{aff27}
\and
INFN-Sezione di Torino, Via P. Giuria 1, 10125 Torino, Italy\label{aff28}
\and
INAF-Osservatorio Astrofisico di Torino, Via Osservatorio 20, 10025 Pino Torinese (TO), Italy\label{aff29}
\and
INAF-IASF Milano, Via Alfonso Corti 12, 20133 Milano, Italy\label{aff30}
\and
Centro de Investigaciones Energ\'eticas, Medioambientales y Tecnol\'ogicas (CIEMAT), Avenida Complutense 40, 28040 Madrid, Spain\label{aff31}
\and
Port d'Informaci\'{o} Cient\'{i}fica, Campus UAB, C. Albareda s/n, 08193 Bellaterra (Barcelona), Spain\label{aff32}
\and
Institute for Theoretical Particle Physics and Cosmology (TTK), RWTH Aachen University, 52056 Aachen, Germany\label{aff33}
\and
Deutsches Zentrum f\"ur Luft- und Raumfahrt e. V. (DLR), Linder H\"ohe, 51147 K\"oln, Germany\label{aff34}
\and
INAF-Osservatorio Astronomico di Roma, Via Frascati 33, 00078 Monteporzio Catone, Italy\label{aff35}
\and
INFN section of Naples, Via Cinthia 6, 80126, Napoli, Italy\label{aff36}
\and
Institute for Astronomy, University of Hawaii, 2680 Woodlawn Drive, Honolulu, HI 96822, USA\label{aff37}
\and
Dipartimento di Fisica e Astronomia "Augusto Righi" - Alma Mater Studiorum Universit\`a di Bologna, Viale Berti Pichat 6/2, 40127 Bologna, Italy\label{aff38}
\and
Instituto de Astrof\'{\i}sica de Canarias, E-38205 La Laguna, Tenerife, Spain\label{aff39}
\and
Institute for Astronomy, University of Edinburgh, Royal Observatory, Blackford Hill, Edinburgh EH9 3HJ, UK\label{aff40}
\and
European Space Agency/ESRIN, Largo Galileo Galilei 1, 00044 Frascati, Roma, Italy\label{aff41}
\and
Universit\'e Claude Bernard Lyon 1, CNRS/IN2P3, IP2I Lyon, UMR 5822, Villeurbanne, F-69100, France\label{aff42}
\and
Aix-Marseille Universit\'e, CNRS, CNES, LAM, Marseille, France\label{aff43}
\and
Institut de Ci\`{e}ncies del Cosmos (ICCUB), Universitat de Barcelona (IEEC-UB), Mart\'{i} i Franqu\`{e}s 1, 08028 Barcelona, Spain\label{aff44}
\and
Instituci\'o Catalana de Recerca i Estudis Avan\c{c}ats (ICREA), Passeig de Llu\'{\i}s Companys 23, 08010 Barcelona, Spain\label{aff45}
\and
Institut de Ciencies de l'Espai (IEEC-CSIC), Campus UAB, Carrer de Can Magrans, s/n Cerdanyola del Vall\'es, 08193 Barcelona, Spain\label{aff46}
\and
UCB Lyon 1, CNRS/IN2P3, IUF, IP2I Lyon, 4 rue Enrico Fermi, 69622 Villeurbanne, France\label{aff47}
\and
Mullard Space Science Laboratory, University College London, Holmbury St Mary, Dorking, Surrey RH5 6NT, UK\label{aff48}
\and
Department of Astronomy, University of Geneva, ch. d'Ecogia 16, 1290 Versoix, Switzerland\label{aff49}
\and
Universit\'e Paris-Saclay, CNRS, Institut d'astrophysique spatiale, 91405, Orsay, France\label{aff50}
\and
INFN-Padova, Via Marzolo 8, 35131 Padova, Italy\label{aff51}
\and
Aix-Marseille Universit\'e, CNRS/IN2P3, CPPM, Marseille, France\label{aff52}
\and
Max Planck Institute for Extraterrestrial Physics, Giessenbachstr. 1, 85748 Garching, Germany\label{aff53}
\and
Universit\"ats-Sternwarte M\"unchen, Fakult\"at f\"ur Physik, Ludwig-Maximilians-Universit\"at M\"unchen, Scheinerstr.~1, 81679 M\"unchen, Germany\label{aff54}
\and
INAF-Istituto di Astrofisica e Planetologia Spaziali, via del Fosso del Cavaliere, 100, 00100 Roma, Italy\label{aff55}
\and
INFN-Bologna, Via Irnerio 46, 40126 Bologna, Italy\label{aff56}
\and
Institut d'Estudis Espacials de Catalunya (IEEC),  Edifici RDIT, Campus UPC, 08860 Castelldefels, Barcelona, Spain\label{aff57}
\and
Institute of Space Sciences (ICE, CSIC), Campus UAB, Carrer de Can Magrans, s/n, 08193 Barcelona, Spain\label{aff58}
\and
School of Physics, HH Wills Physics Laboratory, University of Bristol, Tyndall Avenue, Bristol, BS8 1TL, UK\label{aff59}
\and
Department of Physics, Oxford University, Keble Road, Oxford OX1 3RH, UK\label{aff60}
\and
University Observatory, LMU Faculty of Physics, Scheinerstr.~1, 81679 Munich, Germany\label{aff61}
\and
FRACTAL S.L.N.E., calle Tulip\'an 2, Portal 13 1A, 28231, Las Rozas de Madrid, Spain\label{aff62}
\and
Institute of Theoretical Astrophysics, University of Oslo, P.O. Box 1029 Blindern, 0315 Oslo, Norway\label{aff63}
\and
Jet Propulsion Laboratory, California Institute of Technology, 4800 Oak Grove Drive, Pasadena, CA, 91109, USA\label{aff64}
\and
Felix Hormuth Engineering, Goethestr. 17, 69181 Leimen, Germany\label{aff65}
\and
Technical University of Denmark, Elektrovej 327, 2800 Kgs. Lyngby, Denmark\label{aff66}
\and
Cosmic Dawn Center (DAWN), Denmark\label{aff67}
\and
Universit\'e PSL, Observatoire de Paris, Sorbonne Universit\'e, CNRS, LERMA, 75014, Paris, France\label{aff68}
\and
Universit\'e Paris-Cit\'e, 5 Rue Thomas Mann, 75013, Paris, France\label{aff69}
\and
Max-Planck-Institut f\"ur Astronomie, K\"onigstuhl 17, 69117 Heidelberg, Germany\label{aff70}
\and
NASA Goddard Space Flight Center, Greenbelt, MD 20771, USA\label{aff71}
\and
Department of Physics and Astronomy, University College London, Gower Street, London WC1E 6BT, UK\label{aff72}
\and
Universit\'e de Gen\`eve, D\'epartement de Physique Th\'eorique and Centre for Astroparticle Physics, 24 quai Ernest-Ansermet, CH-1211 Gen\`eve 4, Switzerland\label{aff73}
\and
Department of Physics, P.O. Box 64, University of Helsinki, 00014 Helsinki, Finland\label{aff74}
\and
Helsinki Institute of Physics, Gustaf H{\"a}llstr{\"o}min katu 2, University of Helsinki, 00014 Helsinki, Finland\label{aff75}
\and
Laboratoire d'etude de l'Univers et des phenomenes eXtremes, Observatoire de Paris, Universit\'e PSL, Sorbonne Universit\'e, CNRS, 92190 Meudon, France\label{aff76}
\and
SKAO, Jodrell Bank, Lower Withington, Macclesfield SK11 9FT, UK\label{aff77}
\and
Centre de Calcul de l'IN2P3/CNRS, 21 avenue Pierre de Coubertin 69627 Villeurbanne Cedex, France\label{aff78}
\and
Universit\"at Bonn, Argelander-Institut f\"ur Astronomie, Auf dem H\"ugel 71, 53121 Bonn, Germany\label{aff79}
\and
INFN-Sezione di Roma, Piazzale Aldo Moro, 2 - c/o Dipartimento di Fisica, Edificio G. Marconi, 00185 Roma, Italy\label{aff80}
\and
Dipartimento di Fisica e Astronomia "Augusto Righi" - Alma Mater Studiorum Universit\`a di Bologna, via Piero Gobetti 93/2, 40129 Bologna, Italy\label{aff81}
\and
Department of Physics, Institute for Computational Cosmology, Durham University, South Road, Durham, DH1 3LE, UK\label{aff82}
\and
Universit\'e Paris Cit\'e, CNRS, Astroparticule et Cosmologie, 75013 Paris, France\label{aff83}
\and
CNRS-UCB International Research Laboratory, Centre Pierre Bin\'etruy, IRL2007, CPB-IN2P3, Berkeley, USA\label{aff84}
\and
University of Applied Sciences and Arts of Northwestern Switzerland, School of Engineering, 5210 Windisch, Switzerland\label{aff85}
\and
Institute of Physics, Laboratory of Astrophysics, Ecole Polytechnique F\'ed\'erale de Lausanne (EPFL), Observatoire de Sauverny, 1290 Versoix, Switzerland\label{aff86}
\and
Dipartimento di Fisica - Sezione di Astronomia, Universit\`a di Trieste, Via Tiepolo 11, 34131 Trieste, Italy\label{aff87}
\and
Telespazio UK S.L. for European Space Agency (ESA), Camino bajo del Castillo, s/n, Urbanizacion Villafranca del Castillo, Villanueva de la Ca\~nada, 28692 Madrid, Spain\label{aff88}
\and
Institut de F\'{i}sica d'Altes Energies (IFAE), The Barcelona Institute of Science and Technology, Campus UAB, 08193 Bellaterra (Barcelona), Spain\label{aff89}
\and
School of Mathematics and Physics, University of Surrey, Guildford, Surrey, GU2 7XH, UK\label{aff90}
\and
European Space Agency/ESTEC, Keplerlaan 1, 2201 AZ Noordwijk, The Netherlands\label{aff91}
\and
School of Mathematics, Statistics and Physics, Newcastle University, Herschel Building, Newcastle-upon-Tyne, NE1 7RU, UK\label{aff92}
\and
Waterloo Centre for Astrophysics, University of Waterloo, Waterloo, Ontario N2L 3G1, Canada\label{aff93}
\and
Department of Physics and Astronomy, University of Waterloo, Waterloo, Ontario N2L 3G1, Canada\label{aff94}
\and
Perimeter Institute for Theoretical Physics, Waterloo, Ontario N2L 2Y5, Canada\label{aff95}
\and
Universit\'e Paris-Saclay, Universit\'e Paris Cit\'e, CEA, CNRS, AIM, 91191, Gif-sur-Yvette, France\label{aff96}
\and
Space Science Data Center, Italian Space Agency, via del Politecnico snc, 00133 Roma, Italy\label{aff97}
\and
Centre National d'Etudes Spatiales -- Centre spatial de Toulouse, 18 avenue Edouard Belin, 31401 Toulouse Cedex 9, France\label{aff98}
\and
Institute of Space Science, Str. Atomistilor, nr. 409 M\u{a}gurele, Ilfov, 077125, Romania\label{aff99}
\and
Dipartimento di Fisica e Astronomia "G. Galilei", Universit\`a di Padova, Via Marzolo 8, 35131 Padova, Italy\label{aff100}
\and
Instituto de F\'isica Te\'orica UAM-CSIC, Campus de Cantoblanco, 28049 Madrid, Spain\label{aff101}
\and
Institut de Recherche en Astrophysique et Plan\'etologie (IRAP), Universit\'e de Toulouse, CNRS, UPS, CNES, 14 Av. Edouard Belin, 31400 Toulouse, France\label{aff102}
\and
Universit\'e St Joseph; Faculty of Sciences, Beirut, Lebanon\label{aff103}
\and
Departamento de F\'isica, FCFM, Universidad de Chile, Blanco Encalada 2008, Santiago, Chile\label{aff104}
\and
Department of Physics and Helsinki Institute of Physics, Gustaf H\"allstr\"omin katu 2, University of Helsinki, 00014 Helsinki, Finland\label{aff105}
\and
Departamento de F\'isica, Faculdade de Ci\^encias, Universidade de Lisboa, Edif\'icio C8, Campo Grande, PT1749-016 Lisboa, Portugal\label{aff106}
\and
Instituto de Astrof\'isica e Ci\^encias do Espa\c{c}o, Faculdade de Ci\^encias, Universidade de Lisboa, Tapada da Ajuda, 1349-018 Lisboa, Portugal\label{aff107}
\and
Cosmic Dawn Center (DAWN)\label{aff108}
\and
Niels Bohr Institute, University of Copenhagen, Jagtvej 128, 2200 Copenhagen, Denmark\label{aff109}
\and
Universidad Polit\'ecnica de Cartagena, Departamento de Electr\'onica y Tecnolog\'ia de Computadoras,  Plaza del Hospital 1, 30202 Cartagena, Spain\label{aff110}
\and
Centre for Information Technology, University of Groningen, P.O. Box 11044, 9700 CA Groningen, The Netherlands\label{aff111}
\and
Kapteyn Astronomical Institute, University of Groningen, PO Box 800, 9700 AV Groningen, The Netherlands\label{aff112}
\and
Caltech/IPAC, 1200 E. California Blvd., Pasadena, CA 91125, USA\label{aff113}
\and
Dipartimento di Fisica e Scienze della Terra, Universit\`a degli Studi di Ferrara, Via Giuseppe Saragat 1, 44122 Ferrara, Italy\label{aff114}
\and
Istituto Nazionale di Fisica Nucleare, Sezione di Ferrara, Via Giuseppe Saragat 1, 44122 Ferrara, Italy\label{aff115}
\and
INAF, Istituto di Radioastronomia, Via Piero Gobetti 101, 40129 Bologna, Italy\label{aff116}
\and
Astronomical Observatory of the Autonomous Region of the Aosta Valley (OAVdA), Loc. Lignan 39, I-11020, Nus (Aosta Valley), Italy\label{aff117}
\and
Universit\'e C\^{o}te d'Azur, Observatoire de la C\^{o}te d'Azur, CNRS, Laboratoire Lagrange, Bd de l'Observatoire, CS 34229, 06304 Nice cedex 4, France\label{aff118}
\and
ICSC - Centro Nazionale di Ricerca in High Performance Computing, Big Data e Quantum Computing, Via Magnanelli 2, Bologna, Italy\label{aff119}
\and
Univ. Grenoble Alpes, CNRS, Grenoble INP, LPSC-IN2P3, 53, Avenue des Martyrs, 38000, Grenoble, France\label{aff120}
\and
Dipartimento di Fisica, Sapienza Universit\`a di Roma, Piazzale Aldo Moro 2, 00185 Roma, Italy\label{aff121}
\and
Aurora Technology for European Space Agency (ESA), Camino bajo del Castillo, s/n, Urbanizacion Villafranca del Castillo, Villanueva de la Ca\~nada, 28692 Madrid, Spain\label{aff122}
\and
Institut d'Astrophysique de Paris, 98bis Boulevard Arago, 75014, Paris, France\label{aff123}
\and
ICL, Junia, Universit\'e Catholique de Lille, LITL, 59000 Lille, France\label{aff124}
\and
Laboratoire Univers et Th\'eorie, Observatoire de Paris, Universit\'e PSL, Universit\'e Paris Cit\'e, CNRS, 92190 Meudon, France\label{aff125}
\and
Dipartimento di Fisica "Aldo Pontremoli", Universit\`a degli Studi di Milano, Via Celoria 16, 20133 Milano, Italy\label{aff126}
\and
INFN-Sezione di Milano, Via Celoria 16, 20133 Milano, Italy\label{aff127}
\and
Departamento de F{\'\i}sica Fundamental. Universidad de Salamanca. Plaza de la Merced s/n. 37008 Salamanca, Spain\label{aff128}
\and
Universit\'e de Strasbourg, CNRS, Observatoire astronomique de Strasbourg, UMR 7550, 67000 Strasbourg, France\label{aff129}
\and
Center for Data-Driven Discovery, Kavli IPMU (WPI), UTIAS, The University of Tokyo, Kashiwa, Chiba 277-8583, Japan\label{aff130}
\and
California Institute of Technology, 1200 E California Blvd, Pasadena, CA 91125, USA\label{aff131}
\and
Department of Physics \& Astronomy, University of California Irvine, Irvine CA 92697, USA\label{aff132}
\and
Department of Mathematics and Physics E. De Giorgi, University of Salento, Via per Arnesano, CP-I93, 73100, Lecce, Italy\label{aff133}
\and
INFN, Sezione di Lecce, Via per Arnesano, CP-193, 73100, Lecce, Italy\label{aff134}
\and
INAF-Sezione di Lecce, c/o Dipartimento Matematica e Fisica, Via per Arnesano, 73100, Lecce, Italy\label{aff135}
\and
Departamento F\'isica Aplicada, Universidad Polit\'ecnica de Cartagena, Campus Muralla del Mar, 30202 Cartagena, Murcia, Spain\label{aff136}
\and
Instituto de F\'isica de Cantabria, Edificio Juan Jord\'a, Avenida de los Castros, 39005 Santander, Spain\label{aff137}
\and
Institut d'Astrophysique de Paris, UMR 7095, CNRS, and Sorbonne Universit\'e, 98 bis boulevard Arago, 75014 Paris, France\label{aff138}
\and
Institute of Cosmology and Gravitation, University of Portsmouth, Portsmouth PO1 3FX, UK\label{aff139}
\and
Departament de F\'{\i}sica, Universitat Aut\`onoma de Barcelona, 08193 Bellaterra (Barcelona), Spain\label{aff140}
\and
Instituto de Astronomia Teorica y Experimental (IATE-CONICET), Laprida 854, X5000BGR, C\'ordoba, Argentina\label{aff141}
\and
Department of Computer Science, Aalto University, PO Box 15400, Espoo, FI-00 076, Finland\label{aff142}
\and
Universidad de La Laguna, Dpto. Astrof\'\i sica, E-38206 La Laguna, Tenerife, Spain\label{aff143}
\and
Ruhr University Bochum, Faculty of Physics and Astronomy, Astronomical Institute (AIRUB), German Centre for Cosmological Lensing (GCCL), 44780 Bochum, Germany\label{aff144}
\and
INAF-Osservatorio Astrofisico di Arcetri, Largo E. Fermi 5, 50125, Firenze, Italy\label{aff145}
\and
Department of Physics and Astronomy, Vesilinnantie 5, University of Turku, 20014 Turku, Finland\label{aff146}
\and
Finnish Centre for Astronomy with ESO (FINCA), Quantum, Vesilinnantie 5, University of Turku, 20014 Turku, Finland\label{aff147}
\and
Serco for European Space Agency (ESA), Camino bajo del Castillo, s/n, Urbanizacion Villafranca del Castillo, Villanueva de la Ca\~nada, 28692 Madrid, Spain\label{aff148}
\and
ARC Centre of Excellence for Dark Matter Particle Physics, Melbourne, Australia\label{aff149}
\and
Centre for Astrophysics \& Supercomputing, Swinburne University of Technology,  Hawthorn, Victoria 3122, Australia\label{aff150}
\and
Department of Physics and Astronomy, University of the Western Cape, Bellville, Cape Town, 7535, South Africa\label{aff151}
\and
Departement of Theoretical Physics, University of Geneva, Switzerland\label{aff152}
\and
Department of Physics, Centre for Extragalactic Astronomy, Durham University, South Road, Durham, DH1 3LE, UK\label{aff153}
\and
IRFU, CEA, Universit\'e Paris-Saclay 91191 Gif-sur-Yvette Cedex, France\label{aff154}
\and
Oskar Klein Centre for Cosmoparticle Physics, Department of Physics, Stockholm University, Stockholm, SE-106 91, Sweden\label{aff155}
\and
Astrophysics Group, Blackett Laboratory, Imperial College London, London SW7 2AZ, UK\label{aff156}
\and
Centro de Astrof\'{\i}sica da Universidade do Porto, Rua das Estrelas, 4150-762 Porto, Portugal\label{aff157}
\and
HE Space for European Space Agency (ESA), Camino bajo del Castillo, s/n, Urbanizacion Villafranca del Castillo, Villanueva de la Ca\~nada, 28692 Madrid, Spain\label{aff158}
\and
Department of Astrophysics, University of Zurich, Winterthurerstrasse 190, 8057 Zurich, Switzerland\label{aff159}
\and
University of Applied Sciences and Arts of Northwestern Switzerland, School of Computer Science, 5210 Windisch, Switzerland\label{aff160}
\and
INAF - Osservatorio Astronomico d'Abruzzo, Via Maggini, 64100, Teramo, Italy\label{aff161}
\and
Theoretical astrophysics, Department of Physics and Astronomy, Uppsala University, Box 516, 751 37 Uppsala, Sweden\label{aff162}
\and
Mathematical Institute, University of Leiden, Einsteinweg 55, 2333 CA Leiden, The Netherlands\label{aff163}
\and
Leiden Observatory, Leiden University, Einsteinweg 55, 2333 CC Leiden, The Netherlands\label{aff164}
\and
School of Physics \& Astronomy, University of Southampton, Highfield Campus, Southampton SO17 1BJ, UK\label{aff165}
\and
Center for Astrophysics and Cosmology, University of Nova Gorica, Nova Gorica, Slovenia\label{aff166}
\and
Institute for Particle Physics and Astrophysics, Dept. of Physics, ETH Zurich, Wolfgang-Pauli-Strasse 27, 8093 Zurich, Switzerland\label{aff167}
\and
School of Physics and Astronomy, University of Nottingham, University Park, Nottingham NG7 2RD, UK\label{aff168}
\and
Space physics and astronomy research unit, University of Oulu, Pentti Kaiteran katu 1, FI-90014 Oulu, Finland\label{aff169}
\and
International Centre for Theoretical Physics (ICTP), Strada Costiera 11, 34151 Trieste, Italy\label{aff170}
\and
Center for Computational Astrophysics, Flatiron Institute, 162 5th Avenue, 10010, New York, NY, USA\label{aff171}}    


%
%
 \abstract{
    Wide-field surveys like \Euclid mark a new era of extragalactic stellar stream studies. With a large number of streams, it is now possible to constrain the dark matter halos of galaxies in a cosmological volume and draw comparisons to theoretical expectations for the geometry of dark matter halos. 
    This study combines \Euclid imaging with visual detection and segmentation annotations to analyse streams. %
    We use projected stream morphologies to constrain the shape and centre-of-mass position (CoM) of each host galaxy's potential, jointly probing baryonic and dark matter distributions. %
    These inferences complement weak lensing methods, with sensitivity to halo profile and geometry on sub-virial scales. %
    The method enables both stacked, population-level constraints on halo flattening and CoM position, and constraints on these quantities for individual halos. %
    We also present a novel method for transforming segmentation maps of stellar streams into smooth, curvature-preserving tracks optimised for fast and robust dynamical inference. %
    This approach enables rapid modelling of stream morphology, supports a statistically rigorous combination of constraints across multiple streams within a single galaxy, and enables joint inference across galactic hosts. %
    From our study of 13 galaxies with prominent tidal streams, we find agreement with spherical halos, albeit a mild preference for flattening with $q = 0.95^{+0.05}_{-0.10}$ at 68\% confidence. %
    This is promising early agreement with $\Lambda$CDM predictions. %
    With thousands more discovered streams expected across \Euclid's mission, our programme will enable precise measurements of halo shapes and CoM positions across large samples and redshifts, offering constraints on the geometry of dark matter halos.}
%
%
%
%

   \titlerunning{Geometry of dark matter halos with \Euclid}
   \authorrunning{Euclid Collaboration: N. Starkman et al.}
   
   \maketitle
%
%
%
%
   
\section{Introduction} \label{sec:intro}

    The \Euclid Space Telescope was launched in 2023 and is already delivering transformative imaging data across wide fields \citep{Q1-TP001}. %
    For the local Universe, \Euclid is revealing faint substructures in galaxies -- including tidal features -- with an unprecedented combination of resolution and depth \citep{Borlaff-EP16,Q1-SP047,Hunt+:2025:Euclid,Urbano+:2025}. %

    While most of the $1.5$ billion sources that \Euclid will detect are compact or marginally resolved, a substantial subset are extended and bright enough to permit detailed morphological analysis. %
    Recently, \citet{Q1-SP047} released the first \Euclid visual morphology catalogue, comprising \num{378000} extended galaxies from the 63.1\,deg$^2$ \gls{Q1} field. %
    These systems were annotated through a Galaxy Zoo campaign, where thousands of volunteers classified large-scale morphology and substructure in each galaxy \citep{Bamford+:2009}, and those annotations are used to fine-tune \texttt{Zoobot} \citep{Walmsley+:2023:zoobot} to automatically measure \Euclid morphology. %
    
    Among these measurements are many candidate stellar streams, shells, and other dynamical features. %
    The scale and large redshift range of the \gls{Q1} sample are offering the unprecedented opportunity to systematically study extragalactic tidal features across a cosmologically representative set of galaxies. %

    Stellar streams form from the tidal disruption of globular clusters and dwarf galaxies. %
    The resulting debris approximately traces the progenitor orbit and can retain coherent dynamical structure over billions of years. %
    This makes streams sensitive tracers of the host's gravitational potential \citep[e.g.,][]{Johnston+:1995, Helmi+:1999, Johnston+:2001}. %

    In the Milky Way, extensive \gls{6D} phase-space data have enabled detailed modelling of individual streams to constrain the mass distribution and shape of the \gls{DM} halo \citep{Koposov+:2010, Bovy+:2016:halo_shape, Bonaca+Hogg:2018, Erkal+:2019, Vasiliev+:2021, Ibata+:2024, Koposov+:2023}. %
    Even in the absence of full kinematic information, stream morphology is informative about the geometry of the host potential \citep{Price-Whelan+:2014, Pearson+:2015, Yavetz+:2023, Nibauer+:2023, Walder+:2024,NibPear:2025,Chemaly+:2026}. %
    Thin streams can only persist on regular or near-resonant orbits, and are easily disrupted by time dependence in the global potential or strong non-axisymmetries \citep{Price-Whelan+:2016:chaos, Price-Whelan+:2016:Ophiuchus, Pearson+:2017}. %

    Stellar streams in the Milky Way have enabled detailed studies of the Galaxy's \gls{DM} halo, owing to our ability to resolve individual stars and measure their \gls{6D} phase-space coordinates. %
    In contrast, analogous studies of \gls{DM} halos in external galaxies remain comparatively underexplored, although see \citet{Fardal+:2013}, \citet{Pearson+2022:CenA}, \citet{Nibauer+:2023}, \citet{Walder+:2024}, \citet{NibPear:2025}, and \citet{Wu+:2026}. %

    In external galaxies, stellar streams are typically observed as faint, extended features in deep imaging surveys. %
    These systems predominantly lack distance or radial velocity measurements, and few individual stream stars are resolved beyond the Local Group \citep[e.g.,][]{Crnojevic:2019,Fielder:2025}. %
    Consequently, the traditional methods that rely on \gls{6D} data are inapplicable. %
    Instead, most work focusses either on modelling isolated cases (e.g., the Giant Southern Stream in M31; \citealt{Fardal+:2013,Preston+:2025}) or on cataloguing stream frequency and morphology \citep[e.g.,][]{Martinez-Delgado+:2010,Miro-Carretero+:2024,Sola+:2025:CFHT}. %
    \citet{Sola+:2025:STRRINGS} present a sample of 35 galaxies with often long, narrow, and looping streams from the Siena Galaxy Atlas \citep{Moustakas+:2023} and residual images from the Dark Energy Spectroscopic Instrument (DESI) Legacy Survey \citep[LS;][]{Dey+:2019}. %
    Additionally, \citet{Sola+:2025:CFHT} quantitatively characterise 100 streams across a sample of 472 galaxies. %
    Ongoing systematic cataloguing efforts with \Euclid are expected to dramatically expand these samples (e.g., Mir\'o-Carretero et al., in prep.), and the streams in such catalogues can be analysed using our annotation tool. %
    Collectively, these findings and ongoing efforts highlight the immense potential for mapping low-surface-brightness tidal features with deep photometry. %

    Despite limitations of only observing a 2D image of each stream, the projected morphology of stellar streams retains dynamical information. %
    \citet[][hereafter \citetalias{Nibauer+:2023}]{Nibauer+:2023} demonstrate that stream morphology alone can constrain the gravitational potential of an external galaxy, including its flattening and orientation. %
    This enables stream-based inference even in the absence of kinematic data. %
    Forward-modelling approaches can already provide informative constraints in this regime \citep[e.g.,][]{Walder+:2024,NibPear:2025}, but with only sparse observables they necessarily rely on stronger modeling assumptions, underscoring the need for complementary methods with fewer priors to motivate. %

    The transition from Milky Way streams to extragalactic streams is therefore not just one of distance, but also of method. %
    While the Milky Way enables dynamical modelling with detailed kinematic information, external systems demand inference from sparser observables. %
    The need for scalable morphological techniques is particularly urgent in this present and upcoming era of wide-field, deep imaging surveys. %

    To test predictions for halo shapes from cosmological simulations, we need a statistical sample of galaxies. %
    In \gls{DM}-only simulations, halos are typically triaxial, with shapes that correlate with mass, redshift, and environment \citep{Frenk+:1988, Jing+Suto:2002:triaxial, Allgood+:2006:halo_shapes}. %
    However, the inclusion of baryons, especially the formation of a central stellar disc, tends to round out the inner halo, driving it toward a more oblate, disc-aligned configuration \citep{Kazantzidis+:2004, Debattista+:2008, Chua+:2019}. %
    This transformation is strongest within a few scale radii, where baryonic condensation deepens the potential and feedback isotropises the kinematic distribution function \citep{Bovy:2026:book}. %
    Observationally, this is reflected in the kinematics of galaxies, where oblate fast rotators match axisymmetric models, while more massive slow rotators are more triaxial \citep{Cappellari:2016:review}. %
    For Milky Way-like systems, simulations predict modest flattening and late-time preference for oblate geometries \citep{Allgood+:2006:halo_shapes, Debattista+:2008,Prada+:2019}, but verifying these predictions observationally -- particularly outside the Local Group -- remains a major challenge. %

    In extragalactic systems, halo geometry is typically inferred through indirect dynamical probes. %
    Weak lensing provides statistical constraints on axis ratios for galaxy ensembles, but is largely insensitive to individual halos \citep[e.g.,][]{Mandelbaum+:2006, Clampitt+:2016}. %
    For the subset of galaxies that act as strong lenses, lens modelling can constrain the axis ratio and orientation of the halo, and recent \Euclid analyses demonstrate this capability \citep{Q1-SP048, Q1-SP063}. %
    Globular cluster and satellite kinematics are used to model the potential in systems such as Centaurus~A \citep{Vesic+:2024}, but these analyses often assume equilibrium and require large spectroscopic data sets. %
    \ion{H}{i} kinematics can also probe the outer potential, though typically under strong modelling assumptions \citep{Bosma+:1978}. %

    Stellar streams provide a complementary approach. %
    Because they trace orbits over gigayear timescales, their geometry reflects the underlying potential, without requiring equilibrium or resolved kinematics \citep[e.g.,][]{Shipp+:2021,Koposov+:2023,Nibauer+Bonaca:2025}. %
    Assuming a forward model, this may be used to constrain the potential's flattening \citep[e.g.,][]{Chemaly+:2026} or radial profile \citep[e.g.,][]{NibPear:2025}.
    As shown in \citetalias{Nibauer+:2023} for the stream orbiting NGC\,5907, even the projected curvature direction of an unresolved stream can constrain halo flattening and alignment with the disc. %
    \citet{Wu+:2026} extend this approach to a sample of 15 streams from the Stellar Stream Legacy Survey \citep{Martinez-Delgado+:2010,Sola+:2025:STRRINGS}, applying curvature-based inference at the catalogue level in the local Universe and showing that the strongest constraints arise from edge-on wrapping loops and sharp turning points. %
    With current and upcoming data sets from \Euclid and upcoming data from Rubin's \acrlong{LSST}\glsunset{LSST} \citep[\gls{LSST};][]{Ivezic+:2019:LSST}, the \textit{Nancy Grace Roman} Space Telescope \citep[\textit{Roman}; see][]{Spergel+:2013:WFIRST, Spergel+:2015:WFIRST}, and \acrlong{ARRAKIHS}\glsunset{ARRAKIHS} \citep[\gls{ARRAKIHS};][]{Guzman+:2022:ARRAKIHS}, the curvature method from \citetalias{Nibauer+:2023} can be applied at scale. %
    The combination of large samples and curvature-based inference provides a new way for testing predictions of halo geometry across cosmological galaxy populations. %

    The number of extragalactic systems that could host detectable streams in wide-field surveys like \Euclid is already far beyond what can be reasonably inspected by a small group of people. %
    To scale up, we need a reliable way to identify and label stream candidates across millions of galaxies. %
    Machine learning will eventually help with this (e.g., \citealt{Walmsley+:2019:identify, Shih+:2022}) but detection of low-surface-brightness features like streams is challenging and sensitive to instrument specifics, making it hard to generalise search models. %
    High-quality training data are essential, and for streams, that means human-labelled examples. %
    This is where citizen science becomes crucial: large-scale visual classification provides both immediate discovery potential and the labelled data sets needed to train survey-specific detection pipelines. %
    This strategy is already demonstrated in practice by machine-learning frameworks trained on citizen science data (e.g., \citealt{Walmsley+:2019:identify,Walmsley+:2022:DECaLS,Gordon+:2024:DECaLS,Demirjian+:2020,Bickley+:2021,DominguezSanchez+:2023}). %

    In this paper, we present the first application of curvature-based halo inference to extragalactic stellar streams identified in \Euclid imaging. %
    A companion study, \citet{Wu+:2026}, develops a complementary application of the same curvature framework to a sample of 15 nearby streams from the Stellar Stream Legacy Survey; here we build on the framework with new tools for citizen-science annotation, full Bayesian propagation of track uncertainty, and joint inference of the halo \gls{com}, applied to a cosmological-volume sample drawn from Euclid \gls{Q1}. %
    We relied on expert annotations in this study to develop and illustrate our methodology, and in a future work will scale the approach to more tidal features using volunteer annotations from Zooniverse \citep{Lintott+:2008}. %
    The paper is organised as follows:  %
    In \cref{sec:the_data}, we describe the selection of galaxy candidates from the Euclid \gls{Q1} visual morphology catalogue and the identification of stellar stream candidates within these systems. %
    In \cref{sec:methods}, we introduce our method for fitting stream tracks and inferring the geometry of each host galaxy's gravitational potential from the projected stream morphology. %
    In \cref{sec:results}, we present the results of applying this method to the set of \Euclid streams shown in \cref{fig:streams-panel}, and perform a joint analysis that combines information across all galaxies. %
    In \cref{sec:discuss}, we discuss the robustness and limitations of this approach, explore its applicability to a broader range of features, and outline future directions. %


\section{Data}\label{sec:the_data}

    In this section, we describe the selection of galaxy candidates from the Euclid Quick Release 1 (Q1; \citealt{Q1-TP001}) morphology catalogue and the identification of stellar stream candidates within these galaxies. %
    \gls{Q1} is the first public data release from \Euclid, comprising calibrated VIS and Near-Infrared Spectrometer and Photometer (NISP) imaging and derived catalogues over $63\,\mathrm{deg}^2$ \citep{EuclidSkyOverview,EuclidSkyVIS}. %
    We use the candidate streams identified in this data set both to infer the geometry of each host galaxy's gravitational potential in \cref{sec:results} and to validate the scalable workflow for an ongoing citizen science campaign on larger \Euclid data sets. %

    \subsection{Galaxy selection}\label{sec:the_data:galaxy_selection}

        Our starting sample is the Euclid \gls{Q1} detailed morphology catalogue \citep{Q1-SP047}, which comprises \num{378000} galaxies selected to be bright or spatially extended. %
        These galaxies are drawn from the \Euclid VIS imaging pipeline \citep{Q1-TP002}, with morphological classification
        performed by the \texttt{Zoobot} foundation model \citep{Walmsley+:2023:zoobot}. %
        \texttt{Zoobot} is trained on over 2.9 million human classifications contributed by nearly \num{10000} Galaxy Zoo volunteers, enabling \texttt{Zoobot} to robustly identify features such as bars, spiral arms and morphological signs of mergers. %
        %
\begin{figure*}[ht!]
    \centering
    \includegraphics[width=1\textwidth]{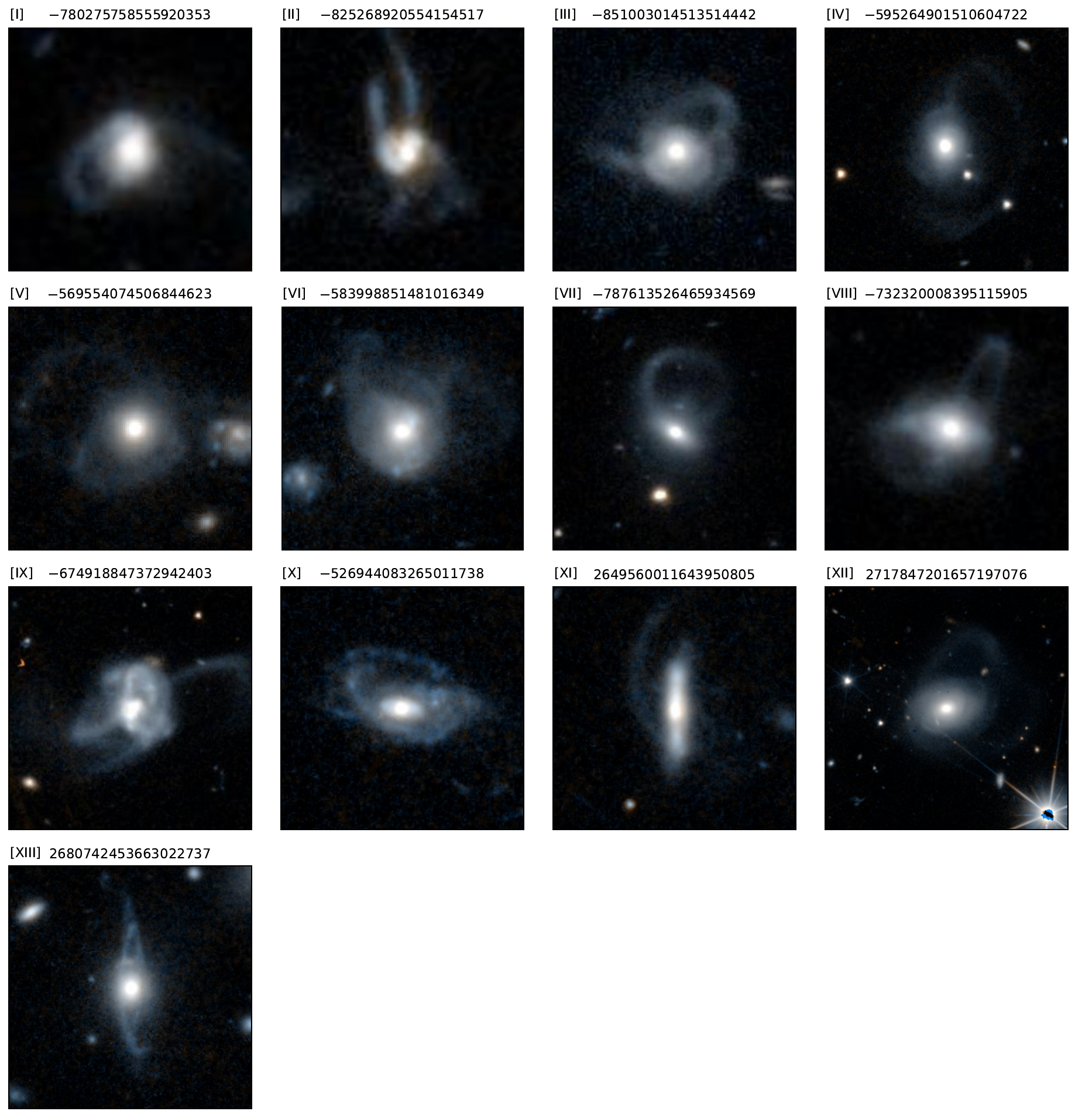}
    \vspace{-3.7cm} 

    \makebox[\textwidth][r]{%
      \begin{minipage}{0.73\textwidth} 
        \captionsetup{type=figure, justification=raggedright, singlelinecheck=false}%
        \caption{%
          Sample of extragalactic streams. %
          \Euclid image cutouts of galaxies selected for follow-up from our visual inspection of the sub-sample from the Euclid \gls{Q1} morphology catalogue (\cref{sec:the_data:galaxy_selection}). %
          These are the same \texttt{Zoobot} input cutouts, reproduced as described by \citet{Q1-SP047} using an arcsinh stretch. %
          Apparent perturbations are typically much smaller in physical amplitude than they visually appear. %
          The sample includes a range of stream morphologies, from prominent structures to more ambiguous features -- see \cref{tab:streams} for redshift information. %
        }%
        \label{fig:streams-panel}%
        \vspace{30pt}
      \end{minipage}%
    }
\end{figure*}
        %
        To build the visual morphology catalogue, \citet{Q1-SP047} applied conservative selection criteria aimed at isolating galaxies with morphological features. %
        Galaxies met the inclusion criteria either by having a large segmentation area (more than 1200 pixels), or by being sufficiently
        bright ($\IE < 20.5$) and moderately extended (segmentation area $> 200$ pixels). %
        The \IE magnitude is the \Euclid VIS broad-band flux measurement that serves as a proxy for apparent flux in the optical \citep{Q1-TP002}. %
        These cuts were designed to ensure high signal-to-noise detections and to exclude compact or marginally resolved systems, keeping only galaxies that are sufficiently extended to be reliably classified by the \texttt{Zoobot} model. %

        We select a subsample of \num{4000} galaxies from the full \gls{Q1} morphology catalogue\footnote{\href{https://zenodo.org/records/15106473}{https://zenodo.org/records/15106473}} by filtering by the column \texttt{merging=major\_disturbance} for systems exhibiting the largest visible structural asymmetries, which might be from tidal tails or merger signatures. %
        The image cutouts used for visual inspection were reproduced from the \texttt{Zoobot} procedure for \texttt{arcsinh\_vis\_y} images described by \citet{Q1-SP047}. %
        The arcsinh stretch magnifies the visual scale of faint features; perturbations are smaller in physical amplitude than they appear. %
        This selection used the \texttt{Zoobot} model's predicted vote fractions for signs of morphological disturbances, selecting the most disturbed systems not otherwise classified. %
        We also filter out problematic measurements, specifically columns \texttt{problem=zoom} and \texttt{smooth-or-featured=problem}). %
        We limited the search to the top \num{4000}, as morphological selection no longer identifies stream-containing galaxies beyond this threshold. By focusing on these selected galaxies, we aimed to increase the likelihood of identifying stellar streams. %
        For this study, this curated set provided a tractable and informative sample for detailed visual inspection and annotation. %

        Although our strategy of selecting visually disturbed systems increases the likelihood of identifying prominent stellar streams, it is a biased method for stream discovery. %
        Stellar streams originate from a range of progenitors -- from globular clusters to dwarf galaxies -- and their visibility in imaging depends on their mass, stellar content, distance from the host, and the survey depth. %
        While \Euclid is not expected to be sensitive to globular cluster streams in integrated light \citep{Pearson+:2019}, it can observe more massive streams, such as those from dwarf galaxies. %
        Many such streams are faint, extended, and well separated from their host galaxy, and therefore may not cause obvious morphological disturbances in the host light profile. %

        As a result, our sub-catalogue is inherently biased toward systems with more massive or recently disrupted progenitors located close to their host galaxies, where the dynamical impact is visually apparent. %
        Moreover, we are more likely to include non-stream structures -- such as disc warps or tidal arms -- that are not dynamically cold and thus are less good tracers of the host potential. %

        In later section \cref{sec:discuss:future_work} we discuss how future work might expand the search to less-disturbed galaxies. %
        For this first study, however, the goal is to demonstrate that stream structures can be detected and used to constrain halo shapes, even within a biased sample. %


    \subsection{Stream selection}\label{sec:the_data:stream_sel}

        \subsubsection{Finding candidate streams}\label{sec:the_data:stream_sel:finding}

            For this study we visually inspected all \num{4000} galaxies in our pre-selected subsample of morphologically disturbed systems
            (\cref{sec:the_data:galaxy_selection}), searching for substructures consistent with stellar streams. %
            We focused on identifying narrow, elongated features that appear offset from the galaxy and that exhibit coherent morphology suggestive of dynamically cold debris. %
            We excluded features associated with internal disc phenomena, such as rings, warps, and bars, as well as large-scale tidal structures likely produced by symmetric tidal interactions, e.g., Toomre--Toomre bridges and tails \citep{Toomre+Toomre:1972,Barnes:1988, Duc+:2015}. %
            Stream detectability also depends on the redshift of the host galaxy -- which affects both surface brightness sensitivity and angular resolution -- as well as on the orientation and viewing angle of the stream relative to the line of sight. %

            Our aim in this step was to identify candidate stellar streams that originate from satellite disruption and are not part of the host's internal structure. %
            Three expert astronomers each independently reviewed the full \num{4000}-galaxy sample. %
            A system was retained as a stream candidate only when all three classifiers agreed. %
            Classifications were performed on the Galaxy Zoo false-colour \Euclid image cutouts used throughout this work, which combine multiple bands to aid the identification of low-surface-brightness features. %
            The arcsinh stretch magnifies the visual scale of faint features; apparent perturbations are smaller in physical amplitude than they appear. %
            Future work replaces this expert visual inspection with a citizen science task in which multiple independent volunteers classify each image, and expert classification will be reserved for a final quality check. %
            This crowd-sourced approach enables probabilistic stream detection, provides redundancy against individual misclassifications, and generates the training set needed for machine-learning classifiers. %

            \Cref{fig:streams-panel} displays the systems selected for analysis from our visually identified candidates. %
            These galaxies span a range of morphologies and stream-like substructures, from clearly identifiable stellar streams to more ambiguous features whose nature is less certain and may represent tidal tails or disc disturbances. %
            This diversity allows us to test the robustness of our annotation and track-fitting procedures under different observational conditions. %
            From the larger set of visually identified candidates, we selected for analysis a non-exhaustive subset that samples a range of stream morphologies. %
            This was sufficient to develop and validate the method and demonstrate its scientific contribution. %
            A detailed treatment of selection effects is deferred to forthcoming larger-catalogue studies that will build on this work. %
            In the sections that follow, we analyse each system in turn, fitting ridge lines and evaluating the curvature-based constraints on the underlying gravitational potential. %
            For discussion of each system, see \cref{app:all_streams}. %


        \subsubsection{Stream annotation}\label{sec:the_data:finding:annotating_the_stream}

            \begin{figure}
                \centering
                \includegraphics[width=0.8\linewidth]{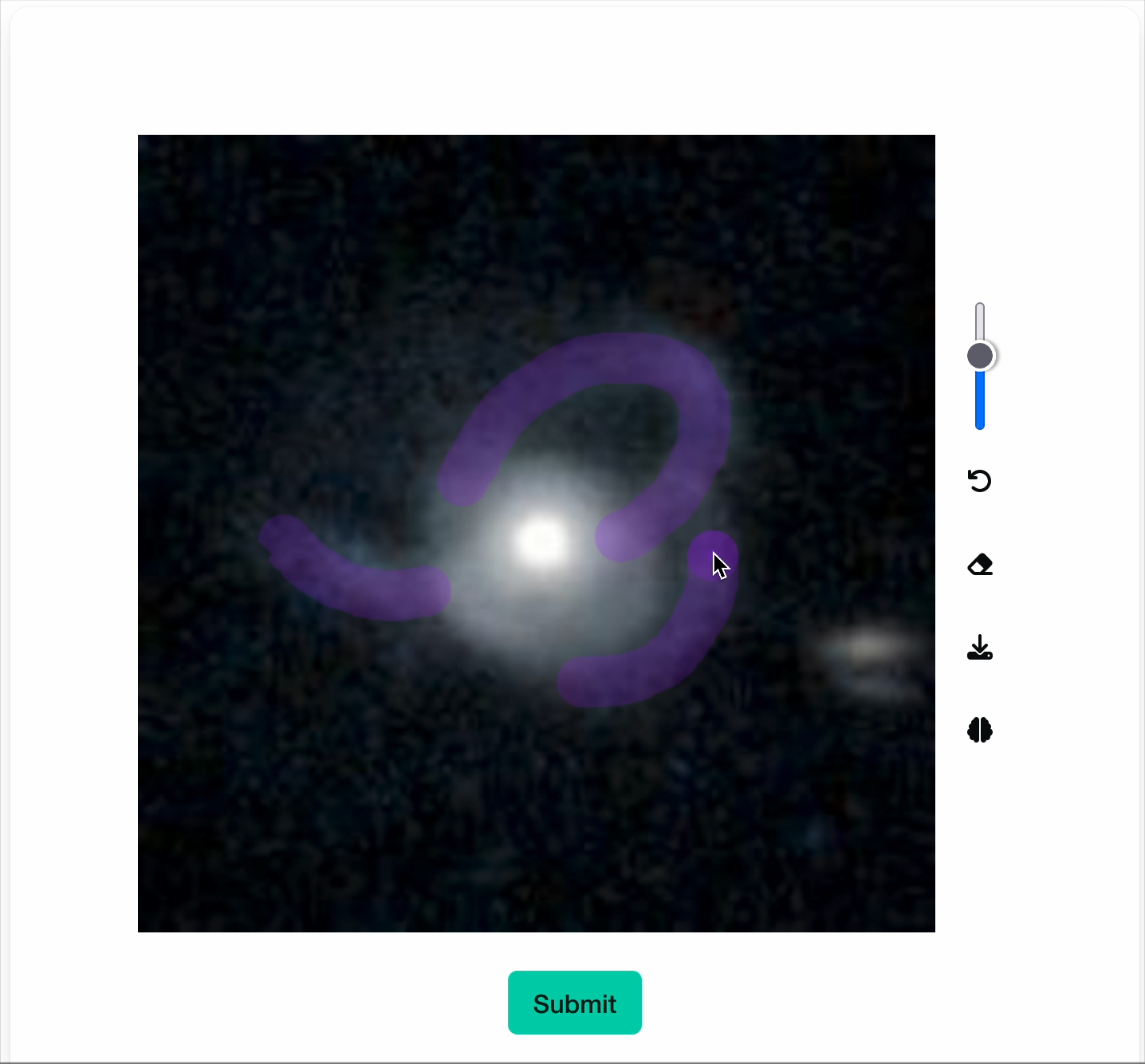}
                \caption{
                    Screen capture of the annotation process. %
                    Users are shown an anonymised \Euclid galaxy image. %
                    They can use a slider to select the brush width and then use their mouse to draw a path (purple overlay) tracing the stream-like feature. %
                    An eraser tool is also available to remove parts of the annotation. %
                    During the development phase, the annotation can also be directly downloaded as a \texttt{JSON} file. %
                    As \texttt{Zoobot} is further developed the `brain' option will provide assistance to the user in identifying stream-like features. %
                    When done, the user can submit their annotation, which is then stored in a database. %
                    \label{fig:zooniverse-annotatation-process}%
                } 
            \end{figure}
            %

            We apply a new web-based annotation tool that allows users to trace stellar streams directly on \Euclid images. %
            With a positively labelled set of galaxy images containing candidate stellar streams, we use this tool to separate the stream from the host galaxy, identifying and isolating the stream directly from its projected light distribution in the imaging. %

            Related annotation approaches aimed at low-surface-brightness structures, such as \texttt{Jafar}, allow experts to delineate tidal features with flexible geometries and varying widths and to derive their average tracks \citep{Sola+:2022,Sola+:2025:CFHT}. %
            Our tool differs in emphasis: it is designed for rapid volunteer marking of stream tracks and, crucially, records the annotation as an ordered path for downstream inference. %
            In this study, three astronomers acted as stand-in volunteers, collectively annotating the sample of stream-like features. %

            This interface presents users with an anonymised \Euclid image and provides a freehand brush tool for marking candidate stream segments. %
            During annotation, the brush width is adjustable between strokes, allowing users to select features with varying widths in the image plane. %
            Unlike an unordered segmentation, the web tool stores each annotation as an ordered path. %
            This ordered path provides crucial information for downstream ridge-line fitting, as discussed in \cref{sec:methods:track_fitting}. %

            \Cref{fig:zooniverse-annotatation-process} shows an example annotation from our study using the tool. %
            In the cutout, also shown in a panel of \cref{fig:streams-panel} (row 1, column 3), there is a central galaxy with a large
            stream wrapping around it. %
            In this case, the stream candidate is both spatially distinct from the host and morphologically coherent, making it a very promising tracer. %
            The purple overlay marks a hand-drawn annotation tracing this structure. %
            The three annotated segments are presumably parts of the same stream; however, the low surface brightness of the stream makes it difficult to distinguish from the host galaxy and in this annotation has been split into three segments. %
            Our curvature analysis method, from \citetalias{Nibauer+:2023} and explained subsequently, works on
            each segment and does not require them to be part of the same
            stream. %
            In other annotation sessions on this stream, the portion in front of the galaxy was also included, resulting in only two annotated segments. %
            This motivates our discussion of annotation confidence. %


        \subsubsection{Confidence in stream segments}\label{sec:the_data:finding:confidence}

            With this tool we show the same image to multiple volunteers
            (typically hundreds to thousands). %
            Once we have a set of annotations from multiple volunteers on the same image, we will estimate the pixel-wise probability that a given location contains a stellar stream by hierarchical modelling over the annotations. %
            Expertise level among volunteers may contribute weighting to this annotation fusion. %
            This vote-based approach is a mainstay of Galaxy Zoo \citep{Lintott+:2008} and has been demonstrated to be an effective means of citizen science classification in identifying real astrophysical structures. %

            Without a large ensemble of annotators, we cannot yet quantitatively assess the fidelity of this approach. %
            Our follow-up study (in preparation) will explore the reliability of pixel-wise stream identification from annotation agreement. %
            The base assumption is that errors in stream labelling are approximately uncorrelated across volunteers and are unbiased with respect to confusion classes (e.g.,\ Toomre--Toomre tidal features or asymmetric shells). %
            If this assumption holds at the population level, then vote aggregation should mitigate misclassifications. %
            If not, and there is systematic confusion, then post-processing will be required to distinguish between streams and other features and to correct the annotations accordingly. %

            In this first study we do not yet have a large corpus of volunteer annotations. %
            Instead, the 13 galaxies analysed here were annotated by a mix of three astronomers acting as stand-in volunteers. %
            For the dynamical analysis, each retained stream segment is based on a single expert annotation; segments judged to be low-confidence were reviewed by consensus among the three annotators. %
            This interim procedure is supported by the live Zooniverse programme, where repeated independent annotations are already recovering the same prominent stream candidates. %
            We nevertheless reproduce the annotation process as described in \cref{sec:the_data:finding:annotating_the_stream} and limit annotation times to 20 seconds per image, to avoid excessive pixel-level scrutiny and better demonstrate the volunteer experience. %
            We discuss in \cref{sec:discuss:advantages:tidal_tails_vs_streams} how sensitive we are to incorrect labelling of streams. %




\section{Methods}\label{sec:methods}

    In this section we describe the methods used to characterise stellar streams in \Euclid images and to infer the geometry of each host galaxy's gravitational potential from fits to the projected stream morphology. %

    The analysis proceeds in stages. %
    In \cref{sec:methods:track_fitting} we describe our procedure for fitting stream ridge lines from the \Euclid images using a constrained spline-based optimisation method. %
    In \cref{sec:methods:halo_geometry} we describe the parameterised halo model used to infer the host galaxy's gravitational potential from the stream morphology. %
    We define the curvature-based likelihood used to constrain host
    galaxy potentials in \cref{sec:methods:single_likelihood}. We then discuss how nuisance parameters are marginalised over in \cref{sec:methods:marginalisation_over_spline_parameters}, and derive how to combine constraints across multiple streams in \cref{sec:methods:combining_likelihoods}. %

    \subsection{Ridge-line fitting}\label{sec:methods:track_fitting}

        With annotated stream segments in hand, we turn to the task of fitting a smooth, representative ridge-line track for each stream. %
        These ridge lines are the basis for downstream inference, directly constraining the acceleration field of the host galaxy's gravitational potential \citepalias{Nibauer+:2023}. %
        In studies with resolved stars, stream tracks are often defined either as latent variables in a hierarchical Bayesian model \citep[e.g.,][]{Starkman+:2023}, or as a rolling summary statistic (e.g.,\ local means) of the observed data along the stream.
        Here, by contrast, we do not observe individual stars, but instead work with pixel-based annotations containing the stream segments. %
        Our eventual goal is to compute curvature vectors along the stream, which are related to the geometry of the host galaxy's gravitational potential \citepalias{Nibauer+:2023}. %
        We therefore define the optimal stream track as a curve within the annotated region that aligns with the stream's large-scale morphology, while avoiding spurious small-scale features from imprecise annotations.
        In this section, we describe our spline-based model for the track and detail the geometric loss function for fitting the ridge line; the underlying geometric formalism is given in \cref{app:math_details}. %
        The resulting curve, parameterised by the arc length, provides a robust representation of the stream's track and a basis for gravitational potential fitting. %

        \subsubsection{Ridge-line initialisation}\label{sec:methods:track_fitting:initialising}

            First we take the stream segment annotation. This is an ordered set of circles at pixel positions $\{\mbf{x}_i \}_{i=1}^{N}$ with radii $r_i$. %
            The pixels are in the Zooniverse web-defined pixel space, which can later be mapped to \Euclid's pixel space. %
            We compute the point-to-point distances $d_{i} = \|\mbf{x}_{i} - \mbf{x}_{i-1}\|$, where $d_1 \equiv 0$. %
            By construction $d_i$ is non-negative, but to ensure monotonicity in the arc length we filter out any $d_{i} = 0$, so that $d_{i}$ is strictly positive. %
            From here on we take $N$ to be the number of $d_i > 0$ points in the annotation and filter all quantities accordingly. %

            We take the running cumulative sum of the point-to-point distance as a proxy for the arc length $\ell_i = \sum_{j=1}^{i} d_j$. %
            The annotations are dense, so this is a good approximation of the true arc length. %
            Then we define the affine parameter $\gamma$ as the normalised arc length in the range $[-1, 1]$ by
            \begin{equation} \label{eq:gamma_from_arclength}
                \gamma_i = \frac{2 \ell_i}{L} - 1 \;, 
            \end{equation}
            where $L$ is the total arc length of the stream. %
            The advantage of this definition over other parameterisations, e.g.,\ an ordinal $\gamma = \{1, \dots, N\}$, is that the track `velocity' $\rm{d}\ell/\rm{d}\gamma$ (Eq.\,\ref{eq:speed}) is (numerically nearly) constant. %

            Using the data $\{(\gamma_i, \mbf{x}_i)\}_i$, we define a fiducial spline
            \begin{equation} \label{eq:spline}
                \mcal{S}(\gamma, \vectheta) : \mathbb{R}^1 \to \mathbb{R}^2 \;,
            \end{equation}
            where $\vectheta$ includes the set of spline knots at $\{(\gamma_k, \mbf{x}_k)\}_{k=1}^K$ (as well as any control points). %
            The tools for defining and optimising stream-related splines are built upon \citet{Wu+:2025:potamides}, the software backbone also used in the companion analysis of \citet{Wu+:2026}.
            This is an interpolating spline and $K \ll N $. %
            To reduce the density of points along the stream while preserving its large-scale structure, we chunk the stream into $K$ segments and compute the median $\gamma$ and $\mbf{x}$ in each chunk. %
            We also include the endpoints $(\gamma_1, \mbf{x}_1)$ and $(\gamma_N, \mbf{x}_N)$ to avoid shortening the track. %
            The total number of knots $K+2$ is a hyperparameter for the user to select. %
            Above a minimum number of knots needed to represent the curve, our procedure is robust to the choice of $K$. %
            Using fewer knots minimizes the computational cost, so a reasonable choice is one per end point plus one per about 50 pixels along the stream. %


        \subsubsection{Ridge-line optimisation}\label{sec:methods:track_fitting:ridge_line_optimisation}

            Here we provide a prescription for optimising the stream's ridge line against a user-drawn annotation. %
            The fiducial spline is just a rough running chunked median estimate and is thus not a good fit of the stream track. %
            Since we are using a spline to describe the track, the goal is to set the locations of the spline knots $\vectheta$. %
            We write our track definition as a loss function and optimise the spline knot locations. %
            The total cost function has four components:
            \begin{equation} \label{eq:cost_function}
                \mcal{C}_{\rm tot} =
                      w_{\mcal{R}} \, {\mathcal{C}_{\mathcal{R}}}
                    + w_{\Delta\kappa} \, {\mcal{C}_{\Delta\kappa}}
                    +  w_{\mrm{shape}} \, {\mcal{C}_{\mrm{shape}}}
                    + w_{\mrm{end}} \, {\mcal{C}_{\mrm{end}}} \;,
            \end{equation}
            where $w_{X}$ are weightings to determine the relative importance of the component terms.

            The first term $\mcal{C}_{\mathcal{R}}$ is the region cost, which enforces that the spline track lies entirely within the annotated segmentation map. %
            This acts as a hard spatial prior but does not otherwise constrain the track's shape. %
            The second term $\mcal{C}_{\Delta\kappa}$ is the concavity cost, which penalises spurious changes in the track's curvature. %
            Because potential inference relies on accurately capturing this curvature, the term is essential for preventing biases from the spline fit. %
            The third term $\mcal{C}_{\mrm{shape}}$ is the shape cost, which encourages the track to follow the overall morphology of the stream.
            This provides guidance in flat regions of the loss landscape, which arise frequently because the first two terms intentionally permit many spline configurations that match the annotation equally well. %
            The fourth term $\mcal{C}_{\mrm{end}}$ is the endpoint cost, which regularises the spline knots at the endpoints of the track. %
            This stabilizes the optimisation near boundaries, where the knots are less constrained by data. %
            In \cref{app:track_fitting} we describe in detail each component of the cost function and their weights. %

            While the loss function components are fixed, the relative weights and steepness parameters are user-defined hyperparameters. %
            For this study, we adopt a set of values that are well motivated by the mathematical properties of the loss terms and that perform robustly across all streams in our sample. %
            That said, the method is modular: each term serves a purpose, and the associated hyperparameters can be adjusted. %
            The cost function is not a one-size-fits-all solution, but a toolkit that can be tuned as needed for more complex systems. %


        \begin{figure}
                \hspace*{-0.5cm}
                \includegraphics[width=1.075\linewidth]{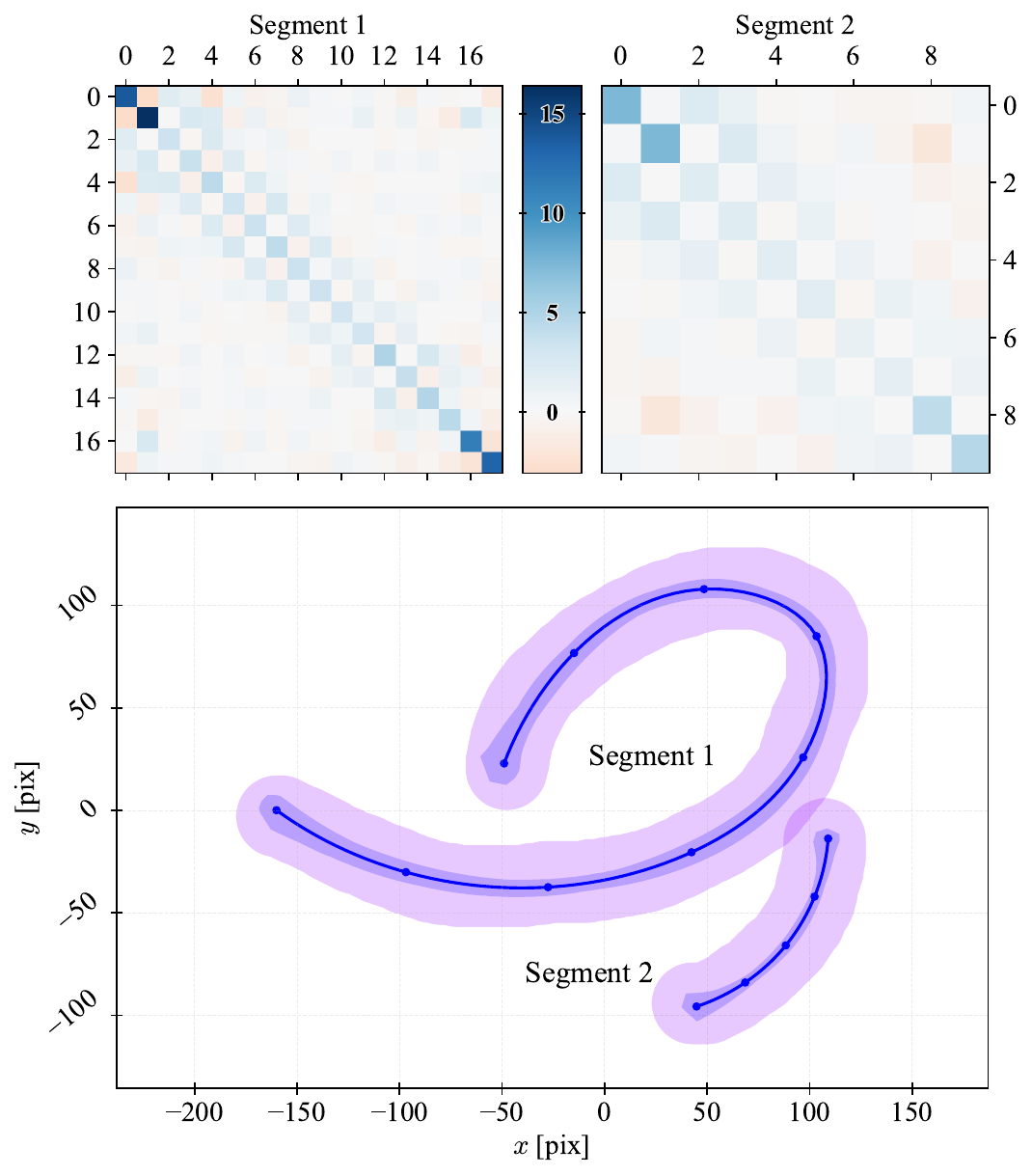}
                \caption{%
                    \emph{Top}: %
                    covariance matrices of the posterior samples of the spline knots. %
                    The $x$, $y$ positions are flattened into a single vector with adjacent elements corresponding to the $x$ and $y$ coordinates of each knot. %
                    The diagonal elements are the variances of the posterior samples, while the off-diagonal elements are the covariances between the knot positions. %
                    The band-diagonal structure indicates that the posterior samples are highly correlated, with the largest variances occurring at the end knots. %
                    \emph{Bottom}: %
                    sampling from the posterior of the spline knots. %
                    The blue lines are the MAP splines with the knot points shown as larger blue circles. %
                    The purple regions are the annotations. %
                    The blue shaded regions are the 90\% confidence intervals from the posterior of the spline knots, showing the uncertainty in the knot positions. %
                    Consistent with the covariance matrix, the interior knots are much better constrained than the end points. %
                    The track and posterior is contained and well aligned with the annotation, shown as the purple region. %
                \label{fig:ridge-line-posterior-samples}}
        \end{figure}

        \subsubsection{Posterior exploration}\label{sec:methods:track_fitting:ridge_line_optimisation:posterior_exploration}

            We optimise the spline track in two stages. %
            First, we apply gradient-based optimisation using the \texttt{Optax} \citep{optax2020github} library to minimise the total loss function and identify a good initialisation for the knot positions. %
            This step yields a \gls{mle} $\vectheta^{\rm MLE}$, which significantly reduces the burn-in time required for subsequent sampling. %
            We then explore the full posterior distribution over the spline parameters using \gls{hmc}, specifically No-U-Turn \citep{Hoffman+Gelman:2014} implemented with \texttt{BlackJAX} \citep{Cabezas+:2024:blackjax}. %
            This provides the {maximum a posteriori} (MAP) estimate $\vectheta^{\rm opt}$ and samples from the posterior distribution of the spline knots. %
            The uncertainty estimates are propagated through to the curvature-based potential inference. %

            In \cref{fig:ridge-line-posterior-samples}, we show an example posterior distribution over spline knot positions. %
            The bottom panel displays the MAP spline (blue line) overlaid on the stream annotation (purple region), with MAP knot positions marked as large blue circles. %
            The blue contour shows the $90\%$ posterior region, revealing the uncertainty in each knot's location. %
            The top panel summarises this structure via the covariance matrices of the spline knots, which have a clear band-diagonal pattern -- highlighting weak local correlations among neighbouring knots and increased correlation at the endpoints. %
            This locality is also consistent with our earlier point that, once the spline has enough knots to represent the stream, the precise choice of knot number is not especially important: if the spline were under-resolved, we would instead expect much stronger long-range covariance across many knots, rather than only the secondary off-diagonal structure seen here. %

            This behaviour is expected in a $C^2$ spline, where the degrees of freedom include both knot positions and two velocity-like control parameters. %
            The concavity loss strongly constrains how curvature changes, so knots surrounded by annotation data on both sides are tightly
            constrained. %
            Endpoints, which are only constrained on one side, show larger uncertainties. %
            In later sections, we marginalise over this posterior when propagating uncertainty from the fitted stream tracks to the curvature-based likelihood. %


        \subsubsection{Optimal track re-parametrisation} \label{sec:methods:track_fitting:reparameterisation}

            After optimisation, the knot positions are updated to $\mbf{x}^{\rm opt}$, but the original parameter values $\gamma$ remain unchanged. %
            As a result, the optimised spline $\mcal{S}(\gamma, \vectheta^{\rm opt})$ no longer has constant speed along $\gamma$. %
            To restore this construction, we redefine $\gamma$ as the normalised arc length along the optimised track using the continuous form of \cref{eq:gamma_from_arclength} which ensures that $\rm d \ell/\rm d \gamma$ is constant along the reparameterised track: $ \gamma^{\rm opt} \equiv \gamma \mapsto 2 \ell(\gamma)/L - 1 $. 
            We then define the optimal spline $\mcal{S}^{\rm opt}$ using knot positions $\{(\gamma^{\rm opt}_j,\mbf{x}^{\rm opt}_j)\}_{j=1}^{K}$, and hereafter refer to it as $\mcal{S}(\gamma, \vectheta)$. %

            Since the full posterior over $\vectheta$ was explored using \gls{hmc}, this reparameterisation is applied to each sample, not just the MAP estimate. %
            This ensures that all posterior realisations of the stream track are described using a consistent arc length parameterisation, as visualized in \cref{fig:ridge-line-posterior-samples}. %
            Because arc length is proportional to $\gamma$, this reparameterisation enables segment-wise quantities like the total length and likelihood contributions to be computed efficiently and consistently across samples. %
            It provides both a computational efficiency and a physically interpretable coordinate for hierarchical inference. %

    

    \subsection{Stream ridge lines to dark matter halos connection} \label{sec:methods:halo_geometry}

        Here we describe how to connect, via curvature, the smooth ridge lines from the preceding sections to a model for the gravitational potential of the host system. %
        The relationship between the curvature vector $\uveckappa$ of the stream and the local acceleration $\mbf{a}_\perp$ perpendicular to the stream is taken from equation 8 of \citetalias{Nibauer+:2023} and given by
        \begin{equation}\label{eq:nb23_curvature_acceleration}
            \mbf{a}_\perp = \left( \dot{v}_\kappa + v_{\mrm{T}} \left\| \frac{\mrm{d}\uvecT}{\mrm{d}t} \right\| \right) \uveckappa \;,
        \end{equation}
        where $v_{\mrm{T}}$ is the tangential velocity of the stream, $\uvecT$ is the unit tangent vector, and $\uveckappa$ is the unit curvature vector of the stream. %
        In \citetalias{Nibauer+:2023}, it is shown that the observed curvature of a stream is therefore connected to the perpendicular acceleration of the host halo (i.e., the stream's curve and local acceleration field must be within \ang{90}). %
        Different trial potentials generate distinct perpendicular accelerations, some of which can be ruled out by the observed stream geometry. %
        Simply put, the curvature vector provides a directional constraint on the local acceleration field. %

        \begin{figure}[htbp]
            \centering
            \includegraphics[width=1\linewidth]{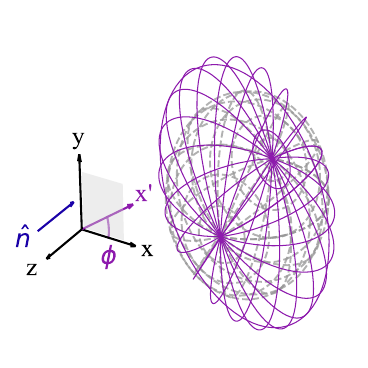}
            \caption{
                Geometry of the potential. %
                This visualisation represents how a gravitational potential may be rotated by some angle $\phi$ in the image plane, and flattened from the original scale $r_{\mrm{s}}$ by rescaling on the (rotated) axis by $q$. %
                In addition, a potential may be translated by $\vec{\Delta}$.
                The grey wire-frame globe represents a spherical potential with perfect symmetry, aligned with the observational axes. %
                The purple wire-frame globe applies these degrees-of-freedom (except any translation). %
                The `$x$-$y$ plane' represents the image plane, which for the observer sees under a flat-sky approximation, while the $z$-axis points toward the observer $\hat{\mbf{n}}$. %
            \label{fig:halo-geometry}}
        \end{figure}

        To test different mass models for a given stream, we must adopt a global gravitational potential for each system, $\Phi(\mbf{x}; \mbf{m})$, were $\mbf{m}$ are model parameters and $\mbf{x}$ is the position vector in the frame of the galaxy of interest. %
        Our coordinate system is illustrated in \cref{fig:halo-geometry}. %
        The $z$ axis is along the line-of-sight and the $x-y$ plane is the plane of the sky. %
        The precise placement of the origin is not important, and we take it based on the pixel positions in the image cutouts. %
        The streams in this study reside at a few scale radii from the host, where the dark matter halo dominates the potential. %
        At these radii the \gls{NFW} circular-velocity curve \citep{NFW:1997} is already approximately flat or gently declining -- and since the curvature-based method constrains only the {direction} of the local perpendicular acceleration, not its magnitude \citepalias{Nibauer+:2023}, the inference is robust to the precise radial form. %
        We therefore adopt the single-component logarithmic halo -- e.g., \citet{Law+Majewski:2010} -- with its symmetry axis aligned along the $z$-axis, allowing for an in-plane rotation by angle $\phi$ and flattening of the potential by $q$; further discussion appears in \cref{sec:discuss:modelling_bias} and \cref{sec:results:halo_geometry}. %
        We will also allow for the possibility that the \gls{com} differs from the \gls{col}. %
        The \gls{com} is labelled $\mbf{x}_0 = (\Delta x, \Delta y, 0)$. %
        The potential model is
        \begin{align} \label{eq:logarithmic_halo}
            \Phi(\mbf{x}; \mbf{m}) &= \frac{v_{\rm{c}}^2}{2} \ln\left(1 + \frac{\tilde{r}^2}{r_{\rm{s}}^2}\right) + \cancel{v_{\rm{c}}^2 \ln(r_{\rm{s}})}, \\
            \tilde{r}^2 &= \left\| \mbf{Q} \left(\mbf{x} -\mbf{x}_0 \right)\right\|^2, \quad
            \mbf{Q} = 
            \begin{bmatrix}
                \cos\phi & \sin\phi & 0 \\
                -\sin\phi/q & \cos\phi/q & 0 \\
                0 & 0 & 1
            \end{bmatrix} \;. \nonumber
        \end{align}
        where, $v_{\mrm{c}}$ is the circular velocity, $r_{\mrm{s}}$ is the core scale radius, and $\mbf{Q}$ is the rotation and flattening matrix. %
        This method relies on the curvature {direction}, not its magnitude, making it scale-free by construction \citepalias{Nibauer+:2023}. %
        Consequently, the choice of core scale-radius has little impact on the results, and we adopt $r_{\mrm{s}} = \qty{16}{\kilo\parsec}$ as a convenient galactic reference scale. %
        Its precise value is unimportant: the inference is carried out in pixel units due to poorly constrained distances. %
        Similarly, $v_{\mrm{c}}$ and any constant offset to the potential have no effect on the curvature direction, so it suffices to assign them fiducial values. %
        We therefore constrain the parameter set $\mbf{m} = (q, \Delta x, \Delta y, \phi)$, where the positional offsets are in pixel units -- a natural choice given that the curvature-based analysis is scale-free and yields purely geometrical constraints on the dark matter halo. %
        Two representative mass models from \cref{eq:logarithmic_halo} for different $\mbf{m}$ are shown in \cref{fig:halo-geometry}. %
        
        We note that in this formalism with flattening against $y$ in the image plane, an oblate halo may be rotated by $\phi$ into a prolate one. %
        We break this degeneracy by considering $q < 1$, though domain restrictions on $\phi$ are equivalent. %
        Additionally, the logarithmic potential admits negative densities for $q < 1/\sqrt{2}$, though still providing an accurate notion of oblate versus prolate potentials, particularly for the curvature-based metric adopted here. %
        Thus, even below this flattening the potential still provides a characterisation of dark matter halos far from the origin (10s of kpc), where most of the streams of interest reside. %
        Future work can consider adopting alternative mass models for the galaxies. %


    \subsection{Likelihood of a stream}\label{sec:methods:single_likelihood}

        In \cref{sec:methods:track_fitting}, we defined a stream track as a vector-valued spline $\mcal{S}(\gamma; \vectheta): \mapsto \mathbb{R}^1 \mapsto \mathbb{R}^2$, where $\gamma \in [-1, 1]$ is an affine parameter along the track. %
        The spline parameters $\vectheta$ include the knot positions, which have been optimised to align with the stream annotations. %
        We additionally parameterise $\gamma$ such that the track has approximately constant speed in arc length. %
        The resulting optimal spline $\mcal{S}(\gamma, \vectheta)$ follows the shape of the annotated stream, while maintaining a smoothly varying curvature profile. %

        To evaluate the likelihood of a stream track under a given host potential model, we sample a discrete set of $I$ points ${\gamma_i}$ along the continuous spline $\mcal{S}(\gamma)$ and compute the local unit-norm curvature vectors ${\uveckappa_i}$. %
        These curvatures are derived directly from the fitted spline, which is itself informed by user annotations (see \cref{sec:methods:track_fitting}). %
        For each point, we assume that the curvature measurement is conditionally independent, and adopt the likelihood form taken from equations (14) and (19) of \citetalias{Nibauer+:2023} and given by
        \begin{align} \label{eq:likelihood_single_independent}
            \mcal{L}(\{\uveckappa_i\} | \mbf{m}) &\propto \prod_{i=1}^{I} P(\uveckappa_i | \mbf{m}) \;,
            \\
            & P\left( \uveckappa_i \mid \mbs{m}, \vectheta \right)
            \propto
            \begin{cases}
            1, & |\theta_{\mrm{T},i}| < \pi/2,  \\
            0, & \text{else} \;,
            \end{cases}
        \end{align}
        where $\mbf{m}$ are the parameters of the gravitational potential model, and $P(\uveckappa_i\mid\mbf{m}, \vectheta)$ is the model-predicted probability of observing the curvature vector $\uveckappa_i$ given the difference $\theta_{\mrm{T},i}$ from the predicted acceleration direction at that location. %
        For the full derivation of the likelihood we refer the reader to \citet{Nibauer+:2023}, equations (14) to (19), and to \citet{Wu+:2025:potamides} and \citet{Wu+:2026} for its numerical implementation and an independent application to local-Universe streams. %
        Conceptually, this likelihood expresses the survival requirement that a stellar stream is sufficiently aligned with the local acceleration field. %
        At a single point \cref{eq:likelihood_single_independent} therefore quantifies the probability that an observed stream is dynamically consistent with a given gravitational potential. %

        We now consider the implications of moving from a continuous parameter $\gamma$ to discrete curvature measurements ${\uveckappa_i}$ sampled along the stream. %
        If the $\gamma_i$ points are poorly distributed, the likelihood may not accurately reflect the stream's overall geometry. %
        As an extreme case, if all the $\{\uveckappa_i\}$ measurements are concentrated at a single location, then only that part of the stream contributes to the likelihood -- clearly at odds with the goal of constraining $\mbf{m}$ using the {entire} stream. %
        Implicit in \cref{eq:likelihood_single_independent}, though not enforced, is the assumption that the full stream contributes. %
        This requires that measurements ${\uveckappa_i}$ be distributed across the entire track, i.e., that $\gamma_i$ spans the full range $[-1, 1]$. %
        Equivalently, there is a sampling distribution $P[\ell(\gamma)]$ over  the arc length along the stream. %
        A simple assumption is that this is uniformly distributed in $\ell$, i.e., $P[\ell(\gamma)] \sim U(-1, 1)$. %
        What this assumption fails to account for is the stream membership likelihood -- the probability that a given pixel contains stream stars. %
        In this first study where the sample size of annotations per stream segment is one, we take the single annotation as a binary classifier for stream membership, thus the membership for all stream segments is uniformly one, and the assumption $P[\ell(\gamma)] \sim U(-1, 1)$ is correct. %
        In \cref{sec:discuss:future_work:statistical_combination} we discuss upcoming work where the stream memberships are not uniform. %

        Therefore, while the measurements of each stream may be independent, their distribution along the stream is not. %
        Since we are processing user annotations and have a smooth representation $\mcal{S}(\gamma)$ of the stream we distribute our `measurements' evenly in $\ell(\gamma)$ over the stream.


     \subsection[Marginalising out spline parameters]{Marginalising out spline
     parameters $\vectheta$}\label{sec:methods:marginalisation_over_spline_parameters}

        From \cref{sec:methods:single_likelihood} we have a method to compute the likelihood of a stream track given a potential model $\mbf{m}$. %
        However, the spline parameters $\vectheta$ used to define the track are not known precisely -- they are inferred from the annotated data with associated uncertainty. %
        In \cref{sec:methods:track_fitting:ridge_line_optimisation:posterior_exploration} we explored the posterior distribution $P(\vectheta \mid \mrm{data})$ over these spline parameters using \gls{hmc}, yielding a set of samples $\{\vectheta_j\}_{j=1}^{N_{\vectheta}}$. %
        To propagate this uncertainty into our inference of the potential model, we marginalise over $\vectheta$ by integrating the likelihood with respect to its posterior %
        \begin{equation} \label{eq:marginal_likelihood}
            \mcal{L}_{\mrm{marg}}(\mbf{m}) = \int P(\{\uveckappa_i\} \mid \vectheta, \mbf{m}) \, P(\vectheta \mid \mrm{data}) \, \mrm{d}\vectheta \;.
        \end{equation}
        Since $P(\vectheta \mid \mathrm{data})$ is represented by posterior samples, we approximate this integral using Monte Carlo estimation %
        \begin{equation} \label{eq:mc_marginal_likelihood}
            \mcal{L}_{\mathrm{marg}}(\mbf{m}) \approx \frac{1}{N_{\vectheta}} \sum_{j=1}^{N_{\vectheta}} \mcal{L}\left(\{\uveckappa^{(j)}_i\} \mid \mbf{m} \right) \;,
        \end{equation}
        where $\{\uveckappa_i\}^{(j)}$ are the curvature vectors evaluated from the $j$-th sample of spline parameters $\vectheta_j$. %

        In principle, the number of Monte Carlo samples $N_{\vectheta}$ should be large to ensure convergence, but in practice we find that as few as 50 posterior samples suffice for the marginal likelihood estimate for our systems. %


     \subsection{Combining likelihoods}\label{sec:methods:combining_likelihoods}

        Thus far  we have focused on the likelihood calculation for a single stream segment. %
        To extend the framework to a set of streams (or stream segments), we introduce an index $s$ to label each system. %
        The full data set then consists of stream tracks $\{\mbf{x}^{(\mrm{s})}(\gamma^{(\mrm{s})}) \}_{\mrm{s}=1}^{N_\mrm{s}}$, where $N_\mrm{s}$ is the total number of streams. %

        We extend the thought example described in \cref{sec:methods:single_likelihood} to handle multiple streams around a galaxy. %
        The key insight is that if the arc length sampling distribution $P[\ell(\gamma)]$ is uniform across all streams, then the linear density of measurements will also be consistent. %
        One straightforward way to achieve this is to fix an arc length interval $\Delta \ell$ and place curvature measurements at regular intervals along each stream. %
        Now the total likelihood is given by
        \begin{equation} \label{eq:likelihood_many_constantdelta}
            \mcal{L}(\{\{\uveckappa_i\}_\mrm{s}\} | \mbf{m}) = \prod_{s=1}^{N_\mrm{s}} \mcal{L}(\{\uveckappa^{(\mrm{s})}_i\}) \propto \prod_{s=1}^{N_\mrm{s}} \prod_{i=1}^{I_{\mrm{\mrm{s}}}} P(\uveckappa^{(\mrm{s})}_i \,\mid\, \mbf{m}) \;,
        \end{equation}
        where $I_{\mrm{s}} = L_{\mrm{s}}/\Delta \ell$ for total arc length $L_{\mrm{s}}$ of the $s$-th stream. %

        There are two issues with the fixed-interval approach. %
        First, $L_{\mrm{s}}/\Delta \ell$ is not guaranteed to be an integer, as is required for the number of measurements $I_{\mrm{s}}$. %
        Therefore rounding (e.g., taking the floor $\left\lfloor L_{\mrm{s}}/\Delta \ell \right\rfloor$) introduces small inconsistencies in the linear density of measurements between streams. %
        This can cause some streams to be slightly under- or over-weighted relative to others in the combined likelihood. %
        While this source of error is small, it is preferable to avoid by construction. %

        The second issue concerns computational efficiency, and can manifest itself in two different ways, depending on numerical concerns. %
        In one way, a simple stream -- e.g., one following a great circle -- requires fewer points to characterise than a more informative stream with tight loops, varying curvature, or linear segments. %
        Ideally, we would vary the number of sampled points $I_{\mrm{s}}$ per stream based on the stream's complexity, adjusting for the change in $\Delta \ell$. %
        Alternatively, in a just-in-time compiled framework like \texttt{JAX} \citep{JAX:2021}, compilation occurs based on input shapes. %
        Keeping a fixed density of points across streams leads to varying length arrays, which is less efficient than fixed-length arrays at varying densities. %
        In either computational approach, it is beneficial to understand how to combine stream constraints with variable point density. %

        We introduce a weighting scheme that corrects for variations in point density across streams. %
        This ensures that streams sampled more densely do not disproportionately dominate the combined likelihood. %
        The exponential weight factor is
        \begin{equation} \label{eq:exponential_weight_factor}
            w_{\mrm{s}} = \frac{\Delta \ell}{\Delta \ell^{(\mrm{s})}} = \frac{I_{\rm tot} / L_{\mrm{tot}}}{I_{\mrm{s}} / L_{\mrm{s}}} \;,
        \end{equation}
        where $I_{\mrm{tot}} = \sum_{\mrm{s}}{I_{\mrm{s}}}$ is the total number of measurements and $L_{\mrm{tot}} = \sum_{\mrm{s}}{L_{\mrm{s}}}$ is the total arc length of all streams. %
        This exponential weight correctly goes to $1$ for $\Delta \ell^{(\mrm{s})} = \Delta \ell$ and will under/overweight contributions to the likelihood from streams measured above/below this sample density. %

        The curvature likelihood of \citetalias{Nibauer+:2023} is defined as a likelihood ratio relative to its maximum, and is therefore not fully normalised at low values. %
        For individual galaxies this does not affect our ability to identify incompatible potentials, but straightforward multiplication across systems would over-weight the under-normalised tail. %
        To avoid this artefact, we combine galaxies using the above arc-length–weighted scheme, adding an effective averaging across systems for a stable joint constraint, with full likelihood normalisation deferred to future work. %
        The more correct and general combined (ln-)likelihood is therefore %
        \begin{equation} \label{eq:lnlikelihood_many}
            \ln \mcal{L}(\{\{\uveckappa_i\}_{\mrm{s}}\} \mid \mbf{m}) = \frac{1}{N_{\mrm{s}}}\sum_{s=1}^{N_{\mrm{s}}} w_{\mrm{s}} \ln\mcal{L}(\{\uveckappa_{i}\}^{(\mrm{s})}) \;.
        \end{equation}



\section{Results}\label{sec:results}

    We apply our curvature-based inference method to the \Euclid-detected stellar streams shown in \cref{fig:streams-panel}, focusing on a selection of representative systems that showcase the method. %
    For each, we constrain the shape and \gls{com} of the host potential from stream morphology, identifying which segments most strongly inform the inference. %
    The full set of posterior results, including all annotated segments, is provided in \cref{app:all_streams}. %
    Finally, in \cref{sec:results:combining_information}, we combine constraints across the sample to demonstrate the feasibility of joint, population-level inference from stream morphology. %

    \subsection{Individual halo geometries} \label{sec:results:halo_geometry}

        \subsubsection{Procedure overview}\label{sec:results:procedure_overview}  

            Each trial potential is parameterised by the circular velocity $v_{\mrm{c}}$, scale radius $r_{\mrm{s}}$, axis ratio $q$, and in-plane orientation angle $\phi$. %
            We fix $v_{\mrm{c}} = \qty{250}{\kilo\metre\per\second}$ and $r_{\mrm{s}} = \qty{16}{\kilo\parsec}$ across all systems. %
            Because our method is sensitive only to the shape of the potential contours -- and not their normalisation -- these choices do not affect the inferred geometry for a single-component model (see \autoref{sec:methods:halo_geometry}). %
            This contour-based approach is a key strength of the curvature method of \citetalias{Nibauer+:2023}, allowing robust shape inference even in the absence of redshift or stellar mass information. %
            For each galaxy we consider two cases: one in which the \gls{com} is fixed to the galaxy's \gls{col}, so that only $q$ and $\phi$ are inferred, and one in which the \gls{com} is allowed to vary, so that $\Delta x$ and $\Delta y$ are inferred jointly with $q$ and $\phi$. %

            We sample the potential parameters using a Latin hypercube design \citep{McKay+:1979} with $10^6$ points spanning $q \in [0.1, 1.0]$, $\phi \in [\ang{-90},\ang{90}]$, and galaxy-specific search bounds on $\Delta x$ and $\Delta y$. %
            The $\Delta x,\Delta y$ ranges were chosen by visual inspection of the galaxy and stream scales, simply to exclude implausibly distant \gls{com} locations and keep the search volume reasonable. %
            They therefore act as a computational aid rather than as an informative prior, proven by how the solutions always lie comfortably within the adopted bounds. %
            Whenever we fix the halo centre to the galaxy light distribution, we use the centre-of-light defined as the brightest region in the cutout, identified from a pixel-intensity histogram, with the same procedure applied to every system (most relevant in \cref{sec:results:combining_information}). %
            This covers the full range of axisymmetric flattenings and their projected orientations; for example, an oblate halo rotated by \ang{90} becomes indistinguishable in projection from a prolate halo. %
            This large, uniform parameter coverage enables rapid evaluation and efficient posterior inference. %

            At the galactocentric radii probed by these streams, the dark matter halo dominates the total potential and baryonic contributions are subdominant, making a single-component description reasonable (see \cref{sec:methods:halo_geometry}). %
            Although more complex models -- including multi-component potentials -- can constrain additional parameters like mass ratios and scale radii (e.g., \citetalias{Nibauer+:2023}), our single-component setup provides a clean, interpretable testbed to demonstrate curvature-based inference on a galaxy's effective potential. %
            To illustrate how curvature measurements constrain host galaxy potentials, we present two example systems. %


        \subsubsection{Example constraint on circular streams}\label{sec:results:102018245_NEG851003014513514442}
\begin{figure*}[htbp]
                \hspace*{-0.7cm}
                \includegraphics[width=1.05\linewidth]{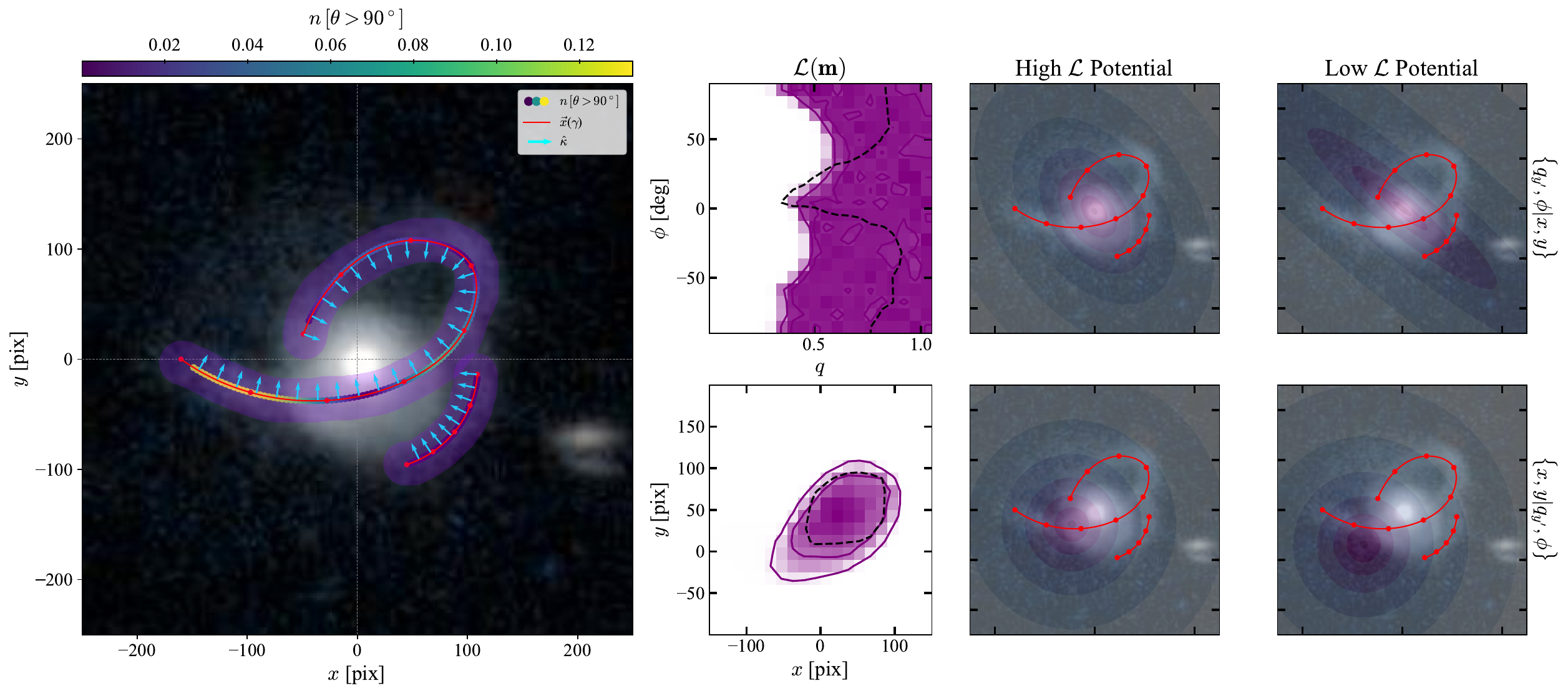}
                \caption{%
                    \emph{Left}: %
                        Galaxy \texttt{102018245:-851003014513514442}. %
                        The annotations for the stream segments are in purple, with the fitted splines $\vec{x}(\gamma)$ in red and with red dots marking the spline knots. %
                        The spline-based curvature vectors are cyan arrows. %
                        At each $\gamma$, the spline is coloured by the number of parameter grid points where the acceleration is misaligned with the curvature vector. The top colour bar shows this count, $n[\theta > \ang{90}]$. %
                        When the \gls{com} is fixed, the $x<100 \, \mrm{[pix]}$ region is the most constraining. %
                    \emph{Middle}: %
                        The marginal posteriors showing the constraints on the flattening $q$ and its axis angle $\phi$ for fixed \gls{com} (top) and the centre of the potential (bottom). %
                        The \gls{com} is strongly constrained to be $r\lesssim 50$ pixels from the \gls{col} of the galaxy. %
                        The flattening of the potential is $q>0.5$, which is consistent with the observed baryonic distribution. %
                        The dashed black lines show the same constraints, but from the alternative three-segment annotation in \cref{fig:zooniverse-annotatation-process}, in which the bridging portion across the galaxy is omitted. %
                        The constraints are notably tighter, caused by different curvature at the stream ends. %
                    \emph{Right}: %
                        Equipotential contours of the potential $\Phi$. %
                        The top row shows the fixed-\gls{com} case, while the bottom row shows the free-\gls{com} case. %
                        The left column shows contours of parameters with high likelihood for a flattened potential at fixed \gls{com} and spherical potential at free \gls{com}. %
                        The right column shows contours of parameters with low likelihood, with the same conditions. %
                        The curvature of the stream is compatible with angled and offset spherical or modestly oblate potentials. %
                \label{fig:102018245-851003014513514442-likelihood}}
            \end{figure*}

            Galaxy \texttt{102018245:$-$851003014513514442}, from \cref{fig:streams-panel} and detailed in
            \cref{fig:102018245-851003014513514442-likelihood}, is a bright, spheroidal system with a smooth, centrally concentrated light profile. %
            The pronounced bulge and lack of visible spiral structure suggest an early-type galaxy, such as an elliptical or face-on lenticular. %
            The broadband colour appears uniform across the galaxy, which can be seen in contrast to the redder neighbouring galaxy to the lower right. %
            These systems do not appear to be physically associated and are only close in projection. %
            While the host shows signs of past interaction -- from the faint and diffuse stellar material extending from the galaxy, in the upper right -- its inner morphology remains undisturbed. %
            Therefore, the tidal debris likely originates from a disrupted satellite that also sources the visible large-scale stream structure. %

            Encircling the host galaxy in \cref{fig:102018245-851003014513514442-likelihood} are several stream segments that have near circular morphologies. %
            The streams have a narrow width and smooth, continuous curvature, indicative of dynamically cold debris following a stream's orbit-like path. %
            The streams wrap around the host in projection, with the most prominent features including an arc on the left, a curved loop above the galaxy, and another arc to the lower right. %
            Where the longer segment passes  close to the central part of the galaxy, it does so below the central bulge, suggesting that the progenitor did not merge directly into the host core. %
            The streams have the same false-colour hue as the main galaxy, making separation by colour challenging and implying a similar stellar population or photometric age. %

            In \cref{fig:zooniverse-annotatation-process} the longer stream is split into two segments. %
            In contrast, the annotation in \cref{fig:102018245-851003014513514442-likelihood} connects the two segments, since the stream appears continuous in \cref{fig:streams-panel}. %
            We will explore below the effect of removing the tentative bridging portion where the stream overlaps the galaxy in projection. %
            The top middle panel of \cref{fig:102018245-851003014513514442-likelihood} shows the posterior distribution over the halo flattening $q$ and in-plane orientation angle $\phi$, fixing the \gls{com} to the photometric centre of the galaxy at $(x, y) = (-3, -4)$ pixels. %
            This corresponds to the peak of the baryonic light distribution, presumably near the true \gls{com}. %
            The posterior disfavours strongly flattened halos, with $q \lesssim 0.5$ excluded across all orientations. %
            Broadly similar constraints are obtained when allowing the centre of mass to vary freely (see \cref{app:all_streams:102018245_NEG851003014513514442}), though the fixed-centre case shown here yields more immediately interpretable bounds over the marginal distribution for the free-centre case. %

            The coloured stream tracks in the left panel highlight where the predicted acceleration is misaligned with the curvature, with colour indicating the number of potential models with angular separations greater than \ang{90} at each point. %
            This diagnostic indicates where the stream's morphology is most sensitive to the halo geometry, with yellow indicating a more constraining segment. %
            The lower-left arc contributes most strongly, being nearly orthogonal to the centre and thus most sensitive to the projected acceleration. %
            The lower-right segment, by contrast, is both short and nearly circular, offering little additional constraint beyond that of the more informative top arc. %

            In the free-centre case (bottom row of \cref{fig:102018245-851003014513514442-likelihood} panels, with $q, \phi$ shown in \cref{fig:102018245-NEG851003014513514442-constraints}), the posterior allows for a wider range of
            halo geometries due to the additional degrees of freedom. %
            Here, the long top loop is still the most constraining, allowing both spherical halos and highly flattened configurations. %
            The lower-right segment strongly favour flattened halos with $\phi \in [\ang{-80},\ang{0}]$. %
            It is the combination of these two segments that constrains the space of $q$ and $\phi$. %
            When combined, the constraints from the loop and the right arc yield a posterior consistent with the fixed-centre result in \cref{fig:102018245-851003014513514442-likelihood}, although the fixed-centre posterior is modestly tighter. %

            The stream's projected morphology, looping broadly and nearly tracing a great-circle path, prevents tighter constraints. %
            If this stream is viewed from a different angle, it might tighten the projected curvature of the top loop significantly, which would remove many allowed rotation angles $\phi$ and place stronger lower bounds on $q$. %
            However, at the current viewing angle, many halo geometries yield similar projected acceleration fields, and thus remain consistent with the stream's morphology. %
            This morphology-dependent sensitivity is also seen across the SSLS sample of \citet{Wu+:2026}, where edge-on wrapping loops and sharp turning points yield the strongest constraints while great-circle-like streams remain largely uninformative. %

\begin{figure*}[htbp]
\hspace*{-0.7cm}
\includegraphics[width=1.05\linewidth]{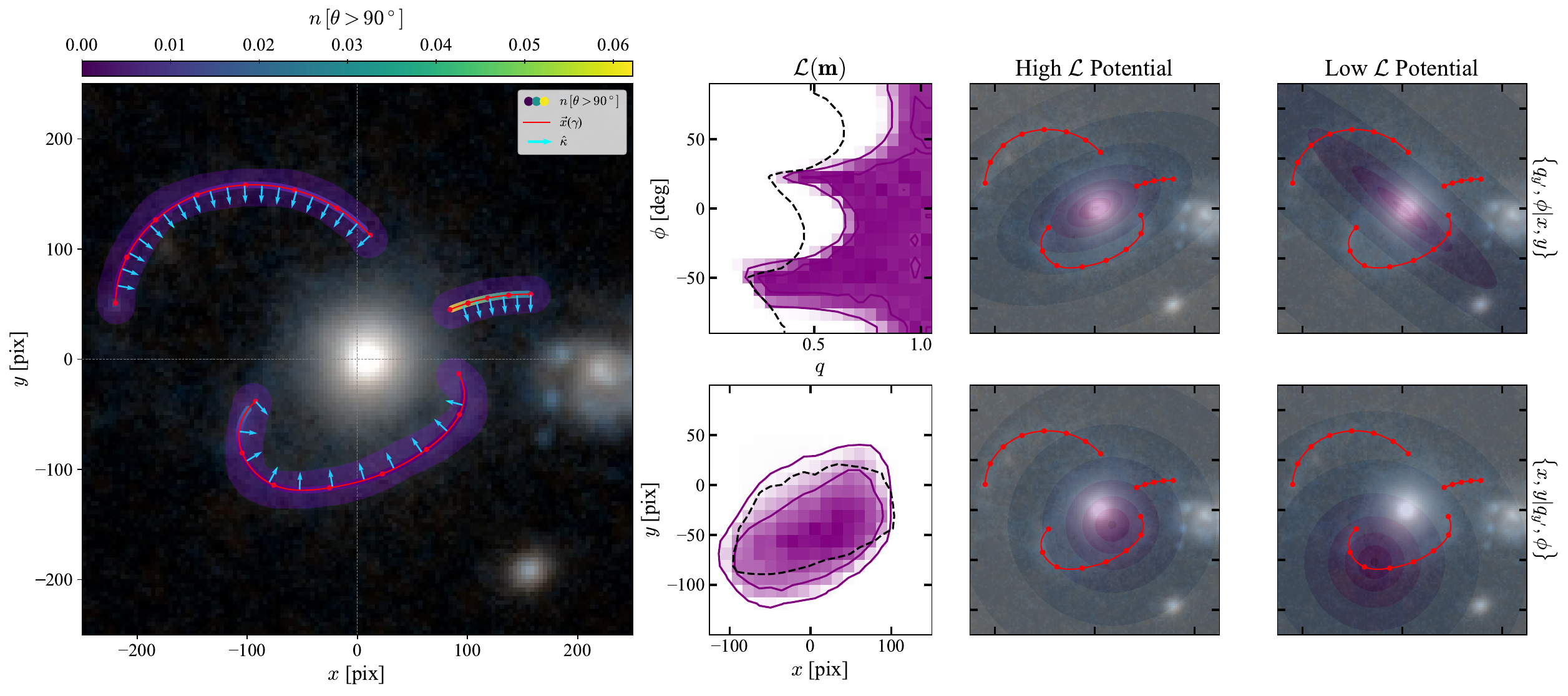}
\caption{%
    Same as \cref{fig:102018245-851003014513514442-likelihood}, but for galaxy \texttt{102019123:-569554074506844623}. %
    \emph{Left}: %
        With the \gls{com} as a free parameter, the centre of the first segment is the most constraining. %
        When the \gls{com} is fixed, the $x<100 \, \mrm{[pix]}$ region is the most constraining. %
    \emph{Middle}: %
        For $\phi$ of $[-80^\circ, 20^\circ]$, the potential is aligned with the system and the flattening is weakly constrained. %
        For other angles, the potential is more tightly constrained. %
        The constraints are systematically less tight for fixed \gls{com}, but tighter for the marginal distribution in \cref{fig:102019123-NEG569554074506844623-constraints}. %
        The dashed black lines show the constraints from a single continuous re-annotation connecting all three segments, with the portions crossing the galaxy excluded from the likelihood. %
    \emph{Right}: %
        The curvature of the stream is compatible with angled oblate potentials and offset spherical potentials. %
\label{fig:102019123-NEG569554074506844623-likelihood}}
\end{figure*}
%

            The \gls{com} is better constrained, with the highest-likelihood region enclosed within the top loop of the stream, just above the left arc. %
            As expected from the morphology of the stream, the projected marginal likelihood forms a broad plateau within the upper loop that contains the luminous centre of the galaxy at approximately $(x, y) = (0, 0)$, while also extending beyond it, and excludes much of the visible asymmetry in the lower half of the galaxy. %
            This breadth reflects the projection of a higher-dimensional likelihood in which $q$ and $\phi$ are free. %
            The inference is therefore consistent with the photometric centre. %

            The black dashed lines in the middle panels of \cref{fig:102018245-851003014513514442-likelihood} show constraints derived from an alternative set of stream annotations. %
            These annotations, shown in \cref{fig:zooniverse-annotatation-process}, include three separate segments rather than the two-segment annotation used for the main analysis in \cref{fig:102018245-851003014513514442-likelihood}. %
            In that three-segment annotation, the two outer arcs are still identified confidently, but the faint portion of the longest stream that overlaps the galaxy in projection is not marked, so the stream is split into separate segments. %

            In this example, the {tentative} segment is specifically that bridging portion across the galaxy, not the outer arcs themselves. %
            Faint sections projected against the galaxy are more easily missed or judged unrelated across repeated annotations, and are therefore assigned lower confidence. %
            The dashed black constraints are obtained by omitting this tentative bridging segment from the likelihood. %

            The two annotations are broadly consistent in their preferred region of parameter space, though the strength of the constraints differs. %
            Both favour a nearly spherical potential, in line with expectations from the baryonic light distribution (which is close to spherical) and with theoretical predictions that baryonic feedback drives halos toward rounder central shapes. %
            Excluding this tentative bridging segment yields noticeably tighter limits on $q$ -- by about $0.25$ for most $\phi$, converging only near $\phi \simeq 0$. %
            Although one might expect the longer annotation to be more informative, the shorter track has endpoints that curve inward more sharply, producing a stronger curvature signal. %
            This shows how including tentative, low-confidence segments can regularise the inferred track shape and thereby weaken curvature-based constraints.  %
            The difference between the two annotations thus reflects not a change in physical inference but a systematic uncertainty in the allowed track geometry. %
            Because marginalising over spline realisations captures only the uncertainty within a given annotation, a robust characterisation of the stream morphology ultimately requires an ensemble of annotations to average over heterogeneity in segment identification and reduce this source of systematic error. %

            The CoM positional constraints behave similarly. %
            The alternative annotation produces a slightly tighter posterior because the inward-curving endpoints strengthen the local curvature constraint. %
            Overall, both annotations give consistent results, indicating that the method is robust to modest annotation differences, especially in low-surface-brightness regions where segments are ambiguous. %
            The main difference is that the tentative segment helps regularise the curvature at the endpoints, highlighting that annotations should be treated as probabilistic measurements derived from many annotators rather than as ground truth. %

            The rightmost four panels of \cref{fig:102018245-851003014513514442-likelihood} illustrate example equipotential contours for both high- and low-likelihood models. %
            The top-left panel shows a good-fit solution with the \gls{com} fixed to the galaxy's photometric centre, allowing $q$ and $\phi$ to vary. %
            The potential is modestly oblate and rotated relative to the image frame. %
            The bottom-left panel shows a good fit for a fixed halo shape ($q \simeq 1$) but varying \gls{com}; for which high-likelihood solutions are confined within the stream loop. %
            By contrast, the adjacent low-likelihood solutions illustrate which configurations are ruled out: extremely flattened halos, or offsets in the \gls{com} that place the potential outside the stream loop, are disallowed -- even when those offsets lie near the visible galaxy. %
            These results demonstrate that we can robustly constrain both the shape and \gls{com} of the host galaxy's gravitational potential. %


        \subsubsection{Example constraint 
        with several streams around a single host}\label{sec:results:102019123_NEG569554074506844623}

            We now examine a second system, \texttt{102019123:} \texttt{$-$569554074506844623}, with a more complex stream morphology. %
            Unlike the previous example, this galaxy lies in close projected proximity to another system, making it a candidate for an interacting pair. %
            Such a configuration is particularly valuable for exploring potential \gls{com} offsets, since interactions can displace the dark matter halo relative to the stellar component. %

            This system, shown in \cref{fig:102019123-NEG569554074506844623-likelihood} as well as \cref{fig:102019123-NEG569554074506844623-constraints}, features an early-type galaxy with a bulge and no discernible internal structure such as spiral arms or bars. %
            The faint outer structure may indicate tidal disturbances, although it is likely associated with the large stream that wraps around the host. %
            The broadband colour appears uniform across the galaxy, suggesting a relatively quiescent and undisturbed stellar population. %
            To the right lies a second galaxy in close projected proximity. %
            This neighbouring system has significant substructure and colour variation, indicating a more complex star-formation history. %
            Without accurate redshift measurements on both systems, we are unable to confirm whether the two galaxies are physically associated or merely close in projection. %

            A stellar stream wraps around the host galaxy in \cref{fig:102019123-NEG569554074506844623-likelihood}, with multiple visible segments tracing a large loop. %
            The brightest clump in the stream lies on the far left, which may be the progenitor. %
            From there, the stream arcs over the galaxy, connecting at the top, then continues from the right, curves beneath the host, and appears to reconnect on the left side. %
            Another stream-like feature extends from the diffuse material on the upper right, pointing towards the nearby galaxy, but it does not appear to physically connect. %
            The stream is moderately thick, but remains morphologically coherent and consistent with tidal debris from a disrupted dwarf galaxy. %
            Its colour is similar to that of the host. %

            The top middle panel of \cref{fig:102019123-NEG569554074506844623-likelihood} shows the posterior distribution over the halo flattening $q$ and in-plane orientation angle $\phi$, inferred by fixing the \gls{com} to the galaxy's photometric centre. %
            The resulting posterior strongly constrains $q \gtrsim 0.8$ for $\phi \in [\ang{-90},\ang{-80}] \cup [\ang{30},\ang{90}]$. %
            At intermediate angles, especially around $\phi = \ang{-50}$, the posterior admits both spherical and significantly flattened halos. %

            The stream segments in this system also demonstrate the need for methodological improvement regarding segment connectivity when interpreting stream morphology. %
            Although the segments are probably part of the same stream, the separate annotations cause the spline fits to treat them independently. %
            As discussed in \cref{app:track_fitting:concavity_change_constraint}, the fitting procedure favours tracks with minimal curvature variation, often producing more circular arcs when allowed by the data. %
            This conservative behaviour ensures that the fits remain minimally informative, but can reduce curvature at segment endpoints when the segments are not connected. %

            To test the effect of connectivity, we re-annotated the system with a single continuous annotation connecting all three segments, then excluded the portions crossing the galaxy when computing the likelihood. %
            The resulting constraints (dashed black line in the middle panels of \cref{fig:102019123-NEG569554074506844623-likelihood}) are broadly consistent with the original annotation, but show some differences. %
            As expected, the constraints weaken when tentative segments are included, most notably for the curvature, where the inferred flattening $q$ can shift by about $0.25$. %
            This happens where the segment-wise curvature provides the tightest constraints on $\phi$. %
            In the two cases examined, the constraints become nearly identical at $\phi \simeq \ang{-50}$ and $\ang{25}$. %
            These differences arise because connecting the segments regularises the curvature at the endpoints, reducing sharp local curvature features that predominantly exclude potential orientations that likewise shift the alignment of the acceleration field. %
            Endpoint regularisation therefore broadens the range of potential orientations for which flattenings are consistent with the observed morphology. %

            A true connectivity condition would not override existing annotations, as in this illustrative test, but would instead guide the fit toward smoother, more physically plausible tracks. %
            We expect that including such a condition would produce results between the original annotation and the connected annotation, with improved curvature regularisation at the segment boundaries. %

            The bottom middle panel of \cref{fig:102019123-NEG569554074506844623-likelihood} shows the posterior over the \gls{com} $(x, y)$, now allowing for potential offset from the centre of the galaxy's bulge. %
            The highest-likelihood region remains close to the galaxy centre, consistent with the fixed-centre case, especially with the connectivity condition applied (black dashed line). %

            Off-centred potentials may reflect dynamical interaction between the two galaxies, if they are truly proximate. %
            \cref{app:all_streams:102019123_NEG569554074506844623} shows the full set of 2D marginalised posteriors. %
            The joint $q--(x, y)$ posterior reveals a degeneracy: off-centred potentials favour more spherical geometries, while more flattened halos are only permitted when the potential is tightly centred on the galaxy. %
            This degeneracy is consistent with expectations that when the dark matter halo's true centre is offset but forced in the model to align with the centre of luminosity, the model compensates via increased flattening to reproduce the stream curvature. %
            Although tentative, this system highlights how stream morphology can jointly constrain halo shape and \gls{com} and motivates follow-up studies of galaxies with possible dynamical offsets. %

            These examples demonstrate that the method is robust and well-suited to constraining halo properties from observed stream morphology. %
            The primary consideration, as in any stream analysis, is establishing which features belong to the stream. %
            Once stream membership is determined, the inference proceeds naturally and individual streams produce well-defined constraints on their host potential. %


        \begin{figure}[htbp]
            \centering
            \includegraphics[width=1\linewidth]{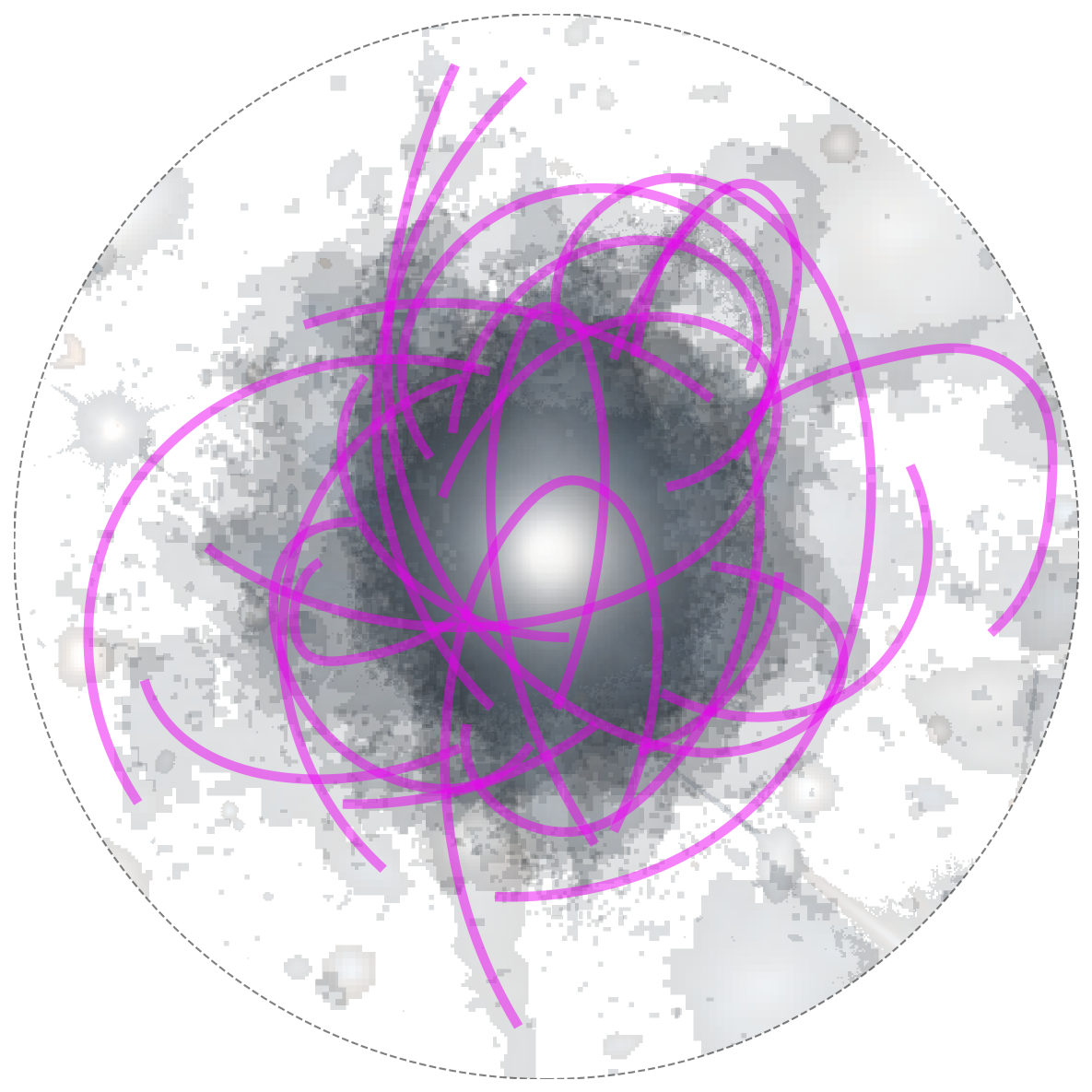}
            \caption{%
                Stacked view of galaxies and their streams. %
                Each host galaxy is shown in black-and-white, centred on its \gls{col}. %
                The stream tracks -- fitted ridge lines from our curvature analysis -- are overlaid in purple. %
                This visualisation demonstrates the stacking procedure used in our joint analysis: stream tracks from different galaxies are co-aligned as if they orbit a common host. %
                While each system has its own orientation and physical scale, stacking in this way allows us to combine curvature-based constraints across galaxies in a unified likelihood framework (Eq.\,\ref{eq:lnlikelihood_many}). %
            \label{fig:galaxy-stack}}
        \end{figure}

    \subsection{Cosmological model: combining information across galaxies}\label{sec:results:combining_information}

        Now we move from analysing individual systems to a joint, population-level analysis using the curvature-based inference framework. %
        Combining stellar stream constraints across galaxies, analogous to stacking in weak lensing, enables us to probe halo demographics, reduce statistical uncertainties, and test cosmological predictions. %
        By aggregating curvature-based inferences, we can infer the distribution of halo geometries and explore their dependence on properties such as mass, morphology, or redshift. %

        This first analysis serves as a critical proof-of-concept for joint inference from stellar stream morphology, despite the limited statistical power of our current sample to robustly address population-level questions. %
        With only a small number of galaxies, we cannot yet resolve dependencies on redshift, mass, or morphology. %
        Still, laying the methodological groundwork is essential. %
        Our framework is designed to scale up and supports precisely this kind of joint analysis. %

        \begin{figure*}[htbp]
            \centering
            \includegraphics[width=1\linewidth]{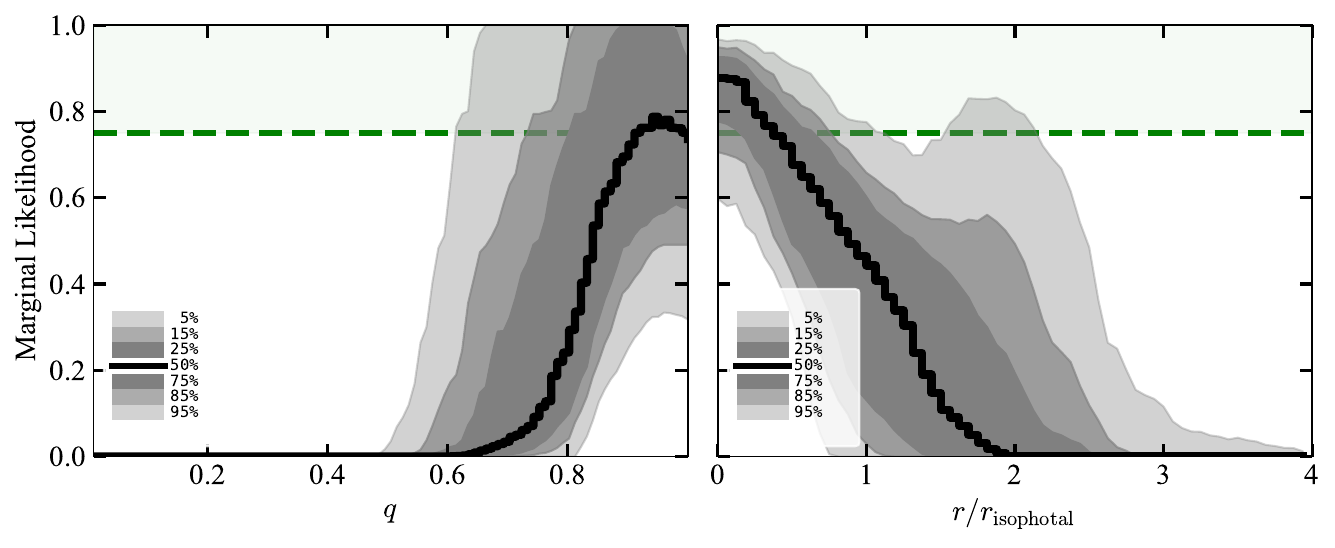}
            \caption{%
                \emph{Left}: %
                    Joint constraint on the projected flattening of the halo potential. %
                    The likelihood contours (grey) show the confidence intervals of the distribution of the projected flattening $q$ about the 50\% quantile (black line). %
                    The 75\% likelihood threshold is banded with a green dashed line. %
                    The data support potential flattenings $q \in [0.8, 1.0]$, with mild preference for flattened over spherical potentials. %
                \emph{Right Panel}: %
                    Joint constraint on the radial offset $r$ of the projected \gls{com} of the halo potential. %
                    The black line is the 50\% quantile, with grey shading for surrounding confidence intervals. %
                    The data constrains the total \gls{com} to be close to the luminous centre of the galaxy. %
            \label{fig:stacked-constraints}}
        \end{figure*}

        We combine curvature-based constraints from different galaxies by treating their streams as a single ensemble system. %
        Mathematically, this is analogous to combining multiple stream segments from a single host: we apply the same arc length weighting scheme described in \cref{eq:lnlikelihood_many}, ensuring that each stream contributes in proportion to its arc length. %
        \Cref{fig:galaxy-stack} illustrates this approach. %
        The stacked streams shown there are the same retained segments analysed individually in \cref{app:all_streams}, including tentative segments when they were retained in the per-galaxy analysis. %
        All 13 galaxies in our sample are included in this joint calculation. %
        Each stream, originally observed in a distinct host, is shown projected into a common frame positioned on the centre-of-light of its galaxy and treated as though it were part of a single, composite system. %
        The centre-of-light of each host is defined as the brightest region in the image, identified as the peak of the pixel intensity histogram of the galaxy cutout. %
        (The uncertainty on this estimate is negligible compared to the scale of the \gls{com} constraints, which are on the order of tens of pixels.) %
        This stacking enables a joint likelihood evaluation over shared halo shape parameters, aggregating curvature information across the sample. %
        Implicit in this approach is the assumption of a statistically shared halo geometry across the population of galaxies, while marginalising over the observer-dependent variations in orientation. %

        \subsubsection{Combining information on flattening} \label{sec:results:combining_information:flattening}

            Here we constrain the typical projected flattening of galaxy-scale dark matter halos by combining curvature-based likelihoods from multiple galaxies. %
            Each stream traces the shape of its host potential in projection, but the orientation of that potential on the sky is observer-dependent. %
            A direct stack of curvature likelihoods would therefore mix different orientations, washing out the true shape information. %

            We avoid this by statistically rotating each galaxy to a common frame before combining. %
            Under statistical isotropy, the observed orientation of any galaxy's potential is a random draw from a uniform distribution -- it carries no population-level information and can be treated as a nuisance parameter \citep{Ehlers+:1968, Wu+:1999, Saadeh+:2016}. %
            This assumption may break down on small scales due to environmental alignments, but is expected to hold in general. %
            Note that this is distinct from the relative alignment between the halo and its baryonic component, which does encode interesting physics but is not addressed here \citep{Tenneti+:2014,Velliscig+:2015}. %

            Practically, for each galaxy we sample the potential orientation $\phi$ from the high-likelihood region of its individual curvature likelihood, i.e.\ from the distribution of major-axis orientations consistent with the stream data. %
            Because these likelihoods often feature broad plateaus rather than sharply peaked maxima, sampling from this region is more robust than adopting a single \gls{mle} value. %
            For each sampled $\phi$ we evaluate the flattening likelihood $\mcal{L}(q;\phi,\mbf{m})$ over the full range of $q$; these curves are then stacked across galaxies. %
            This is equivalent to aligning each galaxy's potential along its inferred major axis before combining -- a statistical rotation to a common frame -- without committing to a single orientation estimate. %
            In principle, this stacked constraint could be built directly from the previously computed per-galaxy posteriors. %
            In practice, because the stacking step requires all systems to be evaluated on common grids once the stream tracks are fixed, it is expedient to recompute the likelihoods from scratch. %
            This recalculation is inexpensive, taking less than \qty{15}{\second} per stream segment and parallelising naturally, so the full stack can be regenerated in of order \qty{2}{\minute}; see \cref{sec:discuss:advantages:timing} for a broader discussion of run times. %

            The left panel of \cref{fig:stacked-constraints} shows the results of this stacked constraint procedure. %
            The grey contours are confidence intervals on the marginal likelihood for the projected flattening $q_{\mathrm{proj}}$, with shaded regions indicating the 90\% (5--95\%), 70\% (15--85\%), and 50\% (25--75\%) confidence intervals. %
            The solid black line traces the 50\% quantile, and green marks highlight regions where the likelihood exceeds a 75\% threshold. %
            The stacked constraint is consistent with spherical projected halos, with a mild preference for flattening and a median $q_{\mathrm{proj}} = 0.95^{+0.05}_{-0.10}$ at 68\% confidence. %
            Including the retained tentative segments does not materially change this result: any shift in the stacked flattening constraint remains well within the displayed confidence intervals. %

            This variation reflects uncertainty introduced by repeated sampling from the \gls{mle} plateau of each stream's orientation. %
            The full 2D likelihood surfaces in $(q, \phi)$ for each system are shown in \cref{app:all_streams}, where the \gls{mle} plateaus are apparent. %
            These stacks use the $(q, \phi)$ likelihoods with fixed potential centres, aligned with the centre-of-light of each galaxy, as shown in the top-right panels of the Appendix figures. %


        \subsubsection[Combining information on CoM Positions]{Combining information on \glspl{com}}\label{sec:results:combining_information:com}

            In addition to shape information, stellar stream morphology can also constrain the projected \gls{com} of the system. %
            By combining constraints across galaxies, we assess the typical alignment between the \gls{com} and baryonic components in the population. %
            This joint analysis provides a statistical probe of \gls{com} alignment, complementing the shape-based results and enabling tests of galaxy-halo co-location at kiloparsec scales. %

            For each galaxy, we compute the posterior over the \gls{com} $(x, y)$ relative to the centre-of-light defined in \cref{sec:results:combining_information}. %
            To combine these constraints across systems we use a scaled radial-slicing approach. %
            We convert each galaxy's 2D \gls{com} posterior into a set of 1D radial likelihoods by taking slices at 180 uniformly spaced azimuthal angles $\phi$. %
            At each angle, we evaluate the likelihood as a function of radial offset $r$ from the photometric centre, producing a set of radial likelihood curves conditioned on $\phi$. %
            The radius is scaled by the galaxy's isophotal radius, measured as a function of angle and interpolated to define a smooth normalisation profile (see \autoref{app:isophotal_radius}). %
            This allows us to characterise the angular variation in the \gls{com} constraints and compute azimuthally marginalised confidence intervals at each $r / r_{\mathrm{isophote}}$. %
            This preserves sensitivity to both tightly and broadly constrained posteriors. %

            This construction enables a statistically robust combination of \gls{com} information across systems. %
            While each galaxy constrains the location of its own halo centre, our goal is to measure the typical offset between the \gls{com} and light across the population. %
            The stacked radial likelihood reflects this distribution. %
            This analysis uses precomputed per-galaxy posteriors, making the computation highly efficient and modular. %

            The result of this stacking procedure is shown in \cref{fig:stacked-constraints}, right panel. %
            The grey contours indicate confidence intervals for the distribution of the projected radial offset $r$ between the \gls{com} and luminous centre, while the solid black line traces the 50\% quantile. %
            The \gls{com} is constrained to be close to the \gls{col} of the galaxy. %
            The \gls{com} is nicely constrained to lie near the galaxy's photometric centre, consistent with expectations that the dark matter halo and baryonic component are approximately co-located. %
            The median likelihood decreases monotonically with radius, further reinforcing that large displacements are disfavoured. %
            At $r/r_{\mathrm{isophote}} \gtrsim 0.75$, the contours widen substantially, and beyond $r \simeq 2 \ r_{\mathrm{isophote}}$, the $50\%$-quantile likelihood cuts off entirely. %
            This behaviour reflects the finite extent of the parameter space explored in the individual galaxy posteriors. %
            In the 70 and 90\% confidence intervals there is a secondary peak in the likelihood at $r \simeq 2 \ r_{\mathrm{isophote}}$, caused by off-centred distributions of the \gls{com} -- for example in \cref{fig:102014777-NEG780275758555920353-constraints}, which allows \gls{com} offsets in only some directions. %
            Taken together, these results show that even with a modest first sample, curvature-based inference provides informative constraints on halo centring, demonstrating both the power and scalability of the method. %
            



\section{Discussion}\label{sec:discuss}

    \subsection{Stream demographics}\label{sec:discuss:stream_demographics}

        The systems presented in \cref{fig:streams-panel} and individually detailed in \cref{app:all_streams} span a range of morphologies, environments, and photometric redshifts. %
        \cref{tab:streams} summarises their basic properties. %
        Based on visual morphology, the sample includes a mix of early- and late-type hosts. %
        Approximately half the galaxies are consistent with ellipticals or bulge-dominated spheroids, while the remainder exhibit disc-like or barred structures. %
        Despite these differences, the associated streams are morphologically coherent and show no overt evidence of ongoing star formation, suggesting that they likely originate from disrupted dwarf galaxies or major tidal disturbances of the galaxy. %
        This is supported by their low surface brightness, narrow widths, and lack of internal structure. %

        \begin{table*}[htbp]
            \centering
            \caption{%
                Selected \Euclid objects and their properties. %
            \label{tab:streams}}
            \normalsize
            \rowcolors{2}{}{gray!25}  
            \setlength{\tabcolsep}{0pt}
            \begin{tabular}{
                >{\columncolor{white}\rule{0pt}{4ex}\hspace{0.5em}}p{3em}
                >{\columncolor{white}\centering\arraybackslash}l
                >{\columncolor{white}\centering\arraybackslash}S[round-mode=places,round-precision=8,table-format=4.10,group-digits=false]<{\hspace{5pt}}
                >{\columncolor{white}\centering\arraybackslash}S[round-mode=places,round-precision=8,table-format=+3.10,group-digits=false]<{\hspace{5pt}}
                >{\columncolor{white}\centering\arraybackslash}c<{}
                >{\columncolor{white}\centering\arraybackslash}c @{\hspace{6pt}}
            }
                \specialrule{1pt}{0pt}{1pt}
                \specialrule{0.5pt}{0pt}{1pt}
                \textbf{No.} & \textbf{Tile/Object ID} & \textbf{RA [\unit{deg}]} & \textbf{Dec [\unit{deg}]} & \textbf{$\tilde{z}_{\mathrm{phot}}$} & \textbf{Companion?} \\
                \midrule
                I   & \texttt{102014777/$-$780275758555920353} & \phantom{0}78.027575812315940 & -55.59203530479052 & & No \\
                II  & \texttt{102014782/$-$825268920554154517} & \phantom{0}82.526892074520650 & -55.41545179813608 & & No \\
                III & \texttt{102018245/$-$851003014513514442} & \phantom{0}85.100301495076480 & -51.35144422980728 & & No \\
                IV  & \texttt{102018668/$-$595264901510604722} & \phantom{0}59.526490132129176 & -51.06047221648467 & 0.09 & No \\
                V   & \texttt{102019123/$-$569554074506844623} & \phantom{0}56.955407435921664 & -50.68446231373057 & 1.32 (0.32) & Yes \\
                VI  & \texttt{102021493/$-$583998851481016349} & \phantom{0}58.399885116931690 & -48.10163494386742 & 0.42 & Yes \\
                VII & \texttt{102022998/$-$787613526465934569} & \phantom{0}78.761352687759870 & -46.593456933048536 & & No \\
                VIII & \texttt{102030408/$-$732320008395115905} & \phantom{0}73.232000877101600 & -39.51159057298009 & & No \\
                IX  & \texttt{102032666/$-$674918847372942403} & \phantom{0}67.491884731685830 & -37.294240359865626 & & No \\
                X  & \texttt{102046112/$-$526944083265011738} & \phantom{0}52.694408338977340 & -26.501173834845325 & 0.38 & No \\
                XI & \texttt{102158269/2649560011643950805} & 264.956001137008570 & \phantom{+}64.39508057617573 & 0.08 & No \\
                XII & \texttt{102158893/2717847201657197076} & 271.784720103729060 & \phantom{+}65.71970761859983 & 0.06 & No \\
                XIII & \texttt{102159485/2680742453663022737} & 268.074245306568700 & \phantom{+}66.30227373902225 & 0.21 & Yes \\
                \specialrule{0.5pt}{0pt}{1pt}
                \specialrule{1pt}{0pt}{1pt}
            \end{tabular}
            \vspace{0.5em}
            \par\smallskip
            \raggedright\footnotesize
            \textbf{Note.} %
            {%
                The listed systems are drawn from the Zooniverse-\Euclid project and include their tile/object ID, right ascension (RA), declination (Dec), and median photometric redshift (Ph$z$ med). %
                Photometric redshifts are taken from \citet{Q1-SP047}, using the \texttt{phz\_median} value, except for \texttt{102019123/-569554074506844623}, for which we also report the \texttt{phz\_pp\_median\_redshift}.  %
                This exception is noted because the high \texttt{phz\_median} for that system is likely erroneous -- photo-$z$ outliers occur in about 10\% of cases and are typically biased high. %
            }
        \end{table*}

        The systems also show a range of colour variation. %
        Some exhibit uniform colours across both the host and the stream, consistent with an older stellar population. %
        Others display modest colour offsets between the stream and host, indicative of differences in stellar age or metallicity between the progenitor and the host galaxy. We do not exploit colour information in this study, though it could be utilised in future work. %

        In terms of halo geometry, almost all of the systems are consistent with spherical projected potentials (i.e., $q \simeq 1$), although many favour mildly flattened configurations with $q \lesssim 0.8$. %
        As illustrated by the system-by-system analyses in this section and in \cref{app:all_streams}, the degree of constraint varies with the observed projected stream morphology, including how many independent segments are visible, how strongly they curve, and how much azimuthal extent they span around the host. %

        Photometric redshifts fall in the approximate range $0.0 \lesssim z \lesssim 0.4$, corresponding to lookback times of up to $\qty{4}{\giga\year}$ and comoving distances of order \qty{1}{\giga\parsec}. %
        Most of the galaxies appear to be isolated, but several show signs of nearby (projected) companions with comparable flux, suggesting possible interactions. %

        Isolated systems offer cleaner conditions for halo modelling, where the central galaxy dominates the gravitational potential. %
        In contrast, galaxies with nearby companions present more complex dynamical environments, where interactions may produce \gls{com} offsets, increased triaxiality, and stronger projected flattening or rotations. %
        The present selection enables qualitative exploration of both relaxed and disturbed systems, each offering constraints on halo structure. %
        These examples together illustrate the flexibility of the curvature-based inference framework across varied astrophysical contexts and establish a foundation for joint analysis. %
        In the next section, we demonstrate how constraints from individual systems can be combined to infer population-level trends in halo shape and \gls{com} alignment. %


    \subsection{Dark matter halo statistics from stacking streams}\label{sec:discuss:combining_constraints}
        
        We have shown that while an individual stream might yield a low signal-to-noise constraint on the shape and offset of dark matter halos, when stacking 13 streams the signal-to-noise improves significantly, with a moderate preference for oblate halos (see \cref{sec:results:combining_information:flattening}) and no preference for halos that are offset from the \gls{col} (see \cref{sec:results:combining_information:com}). %
        While we have only considered 13 galaxies in this study, we expect at orders of magnitude more systems to be applicable to our modelling. 

        \subsubsection{Constraints on halo shapes}

            The joint modelling approach provides an independent framework to constrain population-level statistics of dark matter halos, particularly their flattening distribution. %
            Numerical simulations of structure formation provide a baseline for the expected distribution of halo shapes in a $\Lambda$CDM universe. %
            In dissipationless $N$-body simulations, halos are generically triaxial with minor-to-major axis ratios $s = c/a \simeq 0.5$ and only weak mass dependence \citep{Allgood+:2006:halo_shapes}. %
            Milky Way-mass halos tend to be moderately flattened, with axis ratios $a/c \approx 0.7$ and $b/c \approx 0.8$ \citep{Jing+Suto:2002:triaxial}, while more massive halos are typically more elongated.
            Even in projection, halos remain only mildly flattened, with projected ellipticity $\langle e \rangle \approx 0.24$ \citep{Bailin+Steinmetz:2005}. %
            However, these results pertain to dark matter-only simulations. %
            The inclusion of baryons, particularly the condensation of stars and gas into a central disc, significantly reshapes the inner halo. %
            Adiabatic contraction and feedback-driven potential fluctuations round out the mass distribution and reduce velocity anisotropy, transforming triaxial halos into more oblate, disc-aligned configurations \citep{Kazantzidis+:2004,Debattista+:2008,Chua+:2019,Prada+:2019,Bovy:2026:book}. %
            This sphericalisation is strongest within a few scale radii, where baryons dominate the gravitational potential. %
            As a result, the intrinsic shape distribution in hydrodynamical simulations shifts toward $q \simeq 1$ values, especially for fast-rotating disc galaxies \citep{Cappellari:2016:review}. %
            However, anisotropic mass accretion remains a factor \citep[e.g.,][]{Arora+:2025}, and halos rarely become fully spherical even in the presence of baryons. %
            These non-spherical geometries persist across a broad mass range. %
    
            Our stacked stream analysis measures the flattening of the gravitational potential across an ensemble of external galaxies. %
            We find that the data rule out very flattened projected potentials ($\qproj \lesssim 0.6$), disfavour flattened projected potentials ($\qproj \in [0.6,0.8)$), and favour mildly flattened potentials with $q_{\rm proj} \in [0.9,1.0)$, maximised at $q\simeq0.95$. %
            This is consistent with expectations from cosmological simulations (e.g., \citealt{Allgood+:2006:halo_shapes,Chua+:2019,Prada+:2019}).%
    
            Since the projected flattening sets an upper limit on the intrinsic potential flattening -- $\qproj \geq q_\Phi$ for randomly oriented halos -- our results imply that the underlying 3D potentials are also flattened. %
            At the low end of the 50\% likelihood in \cref{fig:stacked-constraints}, $\qproj \approx 0.8$ implies $q_\Phi < 0.8$, which is consistent with dark-matter-only simulations that find $q_\rho \simeq 0.5$ for \gls{NFW} halos. %
            The relation between $q_\Phi$ and $q_\rho$ arises from Poisson's equation, which links the Laplacian of the potential to the local density. %
            Because the Laplacian is a second-order spatial derivative, it smooths over sharp spatial features, making the potential inherently rounder than the density \citep[see][]{Binney+Tremaine:2008, Bovy:2026:book}. %
            This inequality -- $q_\Phi > q_\rho$ -- means that potential flattenings can reflect more elongated mass distributions. %

            At the more favoured higher end, $\qproj \approx 0.9 \text{--} 0.95$ corresponds to $q_\Phi \simeq 0.8 \text{--} 0.9$, consistent with the rounder shapes predicted when baryonic effects are included. %
            This is expected, since all the systems in \cref{fig:streams-panel} are within a few scale radii of the galaxy and close enough for baryonic effects to be important. %
            Thus, our measured range $\qproj \simeq 0.8$--$1$ spans the expected flattening from \gls{DM}-only to baryon-modified halos. %
            That our ensemble constraints for flattening favour $\qproj \gtrsim 0.8$ supports a cosmological picture in which galaxy-scale halos are mildly oblate: neither strongly triaxial nor perfectly spherical. %
            This aligns with the sphericalisation of halos seen in hydrodynamical simulations (e.g., \citealt{Chua+:2019,Prada+:2019}).
            Notably, \citet{Prada+:2019} find axis ratios of $b/a \simeq 0.8$--$0.9$ for Milky Way-mass halos in the Auriga magneto-hydrodynamical simulations, directly consistent with our preferred range of projected flattenings. %

            Observationally, the most direct comparison of our result on the stacked flattening of halos can be drawn to gravitational lensing. %
            Modern weak lensing surveys find ensemble-averaged halo ellipticities of $e = 0.2$--$0.4$ and axis ratios $q = 0.6$--$0.8$ on scales of 100--500 kpc. %
            For instance, \citet{vanUitert+:2012,vanUitert+:2017} measured $e_{\mrm{h}} \approx 0.38$ for early-type galaxies, while \citet{Schrabback+:2021} obtained $q \approx 0.82$ ($|\epsilon_\mrm{h}| \approx 0.17$). %
            Similar results have been reported for luminous red galaxies by
            \citet{Clampitt+:2016} and \citet{Robison+:2023}. %
            These values are consistent with our findings, preferring more oblate halos than spherical. %
            However, when extending our inference to hundreds of systems, our methodology will offer a test of halo geometries independent to lensing. %
        
        
        \subsubsection{Centre of mass (COM) position constraints}\label{sec:discuss:combining_constraints:com}

            Our first sample shows that the posterior distributions for the \gls{com} $(x,y)$ of individual galaxies are usually centred on their photometric centres. %
            Combining these posteriors with the radial-slicing approach described in \cref{sec:results:combining_information:com} produces an ensemble likelihood that declines steeply beyond the galaxies outer light radii. %
            Approximating galaxies as roughly $10 \,\unit{\kilo\parsec}$ in projected radius \citep{Shen+:2003}, our method achieves constraints on \gls{com} offsets at comparable scales across diverse systems.
                
            Galaxy clusters provide some of the clearest evidence for spatial separation between baryons and dark matter. %
            During major mergers, the dissipative \gls{ICM} can be displaced from the collisionless galaxies and dark matter, producing measurable offsets \citep{Clowe+:2006:bullet,Shan+:2010}. %
            Models of \gls{SIDM} predict an additional offset between the \gls{SIDM} and galaxies, with the magnitude depending on the interaction cross-section per unit mass \citep{Harvey+:2015,Kahlhoefer+:2015}. %
            Observationally, \citet{Cross+:2024} measured offsets between brightest cluster centres and X-ray centroids for 23 relaxed clusters in Dark Energy Survey \citep{2016MNRAS.460.1270D} and Sloan Digital Sky Survey \citep{2000AJ....120.1579Y}, finding non-zero offsets in nearly all clusters with a median projected separation of $(6\pm 5)$ kpc. %
            These results are broadly consistent with hydrodynamical simulations \citep{Roche+:2024}. %
            and may indicate a small but non-zero \gls{SIDM} cross-section. %
            In cosmological simulations, \citet{Roche+:2024} found that most offsets lie below the simulation's softening length, with extremely small median offsets. %
            Observational centroiding methods can overestimate these offsets by factors of 2--5. %
            Overall, these studies suggest that in relaxed clusters, true baryon-dark matter mis-centring is minimal and that many observed offsets may be driven by measurement systematics. %
    
            While our current sample is too small to identify robust correlations between \gls{com} offset and host properties, this study demonstrates the feasibility of such analyses, particularly for setting limits on the maximum offset inferred from stacking streams. %
            In order to do so more robustly, we will need to utilise redshift estimates for each galaxy to place its \gls{com} location constraint on a physical scale. %
            We defer this to future work.%


    \subsection{Comparison to relevant work}\label{sec:discuss:complementarity}

        Here we compare the methodology and results in this paper to recent work that utilised forward modelling (i.e., simulations) to constrain halo properties from stream images. %

        Several works have utilised simulations to connect the morphology of external stellar streams to the properties of the host halo. %
        \citet{Fardal+:2013} measured the mass of M31's dark matter halo using imaging and kinematics of the Giant Stellar Stream. %
        For each choice of a gravitational potential, they ran $N$-body simulations. %
        While this is the most accurate treatment of tidal tail formation and tidal stripping, their approach is computationally expensive, particularly in the absence of any kinematic or distance information to the host and along the stream. %
        More recently, \citet{Pearson+2022:CenA} used the particle-spray approach \citep{Fardal:2015} to model a stream around Centaurus A. %
        Their method explores a grid of models with different progenitor velocity vectors and halo masses. %
        \citet{NibPear:2025} improved upon their approach with a more exhaustive parameter search using Graphical Processing Unit (GPU) accelerated simulations. %
        However, their method fixes, among other assumptions, the on-sky location of the progenitor. %
        \citet{Chemaly+:2026} later developed a generative approach to constrain halo shapes at the population level from images of stellar streams. %
        Their model similarly fixes the on-track location of the progenitor in the plane of the sky and functional form of the stream formation. %
        Most of the streams we have analysed in this work have only tentative progenitors, if any, complicating the application of our streams to such modelling. %
        The curvature approach we have employed does not require any knowledge of the progenitor system, at the cost of not gaining likelihood sensitivity those assumptions impart. %
        Additionally, the runtime of the generative methods is on the order of hours, while ours works in seconds. %
        It is therefore difficult to scale fully generative methods at the level of thousands of galaxies, particularly over an ensemble of potential and progenitor models required to mitigate model bias. %
        An efficient forward model is developed by \citet{Walder+:2024}, based on orbit-fitting of tidal streams. %
        While their approach is less expensive, since each stream is represented by a single stellar orbit, streams do not trace perfect orbits and biases can arise under such an assumption \citep{Sanders:2013}. %

        While the above works have shown that there is substantial information to be obtained when using a generative model, it is at the cost of making additional assumptions about tidal stripping, the location of the progenitor, and the dynamical age of the stream. %
        In contrast, the curvature method developed by \citetalias{Nibauer+:2023} does not require a resolved progenitor, nor any prior on tidal stripping or the dynamical age of the stream. %
        We therefore view the constraints presented in this work as significantly less model dependent than forward modelling approaches. %
        Ideally, we hope to use the curvature method to rapidly constrain the allowed flattening of dark matter halos, and use the fits from our approach as priors for more expensive (and potentially less robust) forward models. %

        The closest methodological neighbour to this work is \citet{Wu+:2026}. %
        They apply the same curvature-based framework, built on the jointly-developed \texttt{potamides} software \citep{Wu+:2025:potamides}, to 15 nearby streams from the Stellar Stream Legacy Survey. %
        The two studies are complementary. %
        \citet{Wu+:2026} target local-Universe systems and use a grid-based maximum-likelihood evaluation over a flattened logarithmic halo with projected axis ratio and orientation $(q,\theta)$. %
        Here we instead couple the curvature likelihood to a citizen-science annotation pipeline tailored to \Euclid imaging, perform full posterior estimation over the spline track so that nuisance parameters can be marginalised over, jointly infer the halo \gls{com}, and stack constraints across galaxies. %
        The two analyses also probe different redshift regimes: local Universe versus $z \lesssim 0.4$. %
        Despite these differences, the two samples agree qualitatively. %
        In both, the streams that depart strongly from a great circle drive the inference, and the ensemble is consistent with halos no more flattened than $q \simeq 0.8$. %
        This convergence is evidence that the curvature method is robust to differences in track extraction, sampling strategy, and sample selection. %

        We also compare our work to that of \citet{Sola+:2025:STRRINGS,Sola+:2025:CFHT}, who have collectively constructed a large statistical sample of galaxies and streams using Canada-France-Hawaii Telescope data, DESI Legacy imaging, and the Sienna Galaxy Atlas. %
        They employ the \texttt{Jafar} tool \citep{Sola+:2022} to segment streams and subsequently fit the flux profiles of the segmented streams in angular bins using one or more Gaussians. %
        This work has been pioneering in the characterisation of streams from deep photometric data and in extracting their surface brightness information. %
        The \texttt{Jafar} tool is effective at deriving each stream's average track and width, although the curvature method requires smooth tracks. %
        In this work we have focused on developing a statistically rigorous formalism for fitting smooth stream tracks, although it is possible to utilise the approach of \citet{Sola+:2025:STRRINGS} as a starting point for initialising spline knots in our model (see \cref{sec:methods:track_fitting:initialising}). %
        We have also developed a scalable citizen science annotation platform that is accessible to the general public -- \url{https://ze.walmsley.dev}. %

        \subsection{Computational speed}\label{sec:discuss:advantages:timing}

            A major advantage of this method is its computational efficiency. %
            For almost all stream segments, the full analysis pipeline -- from stream detection through to posterior constraints -- completes in under 5 minutes on a laptop. %
            The steps for a single stream segment are as follows

            \begin{enumerate}[itemsep=0.5em, leftmargin=1.5em]
                \item Annotation -- A user marks the stream segment via a graphical interface. This is a lightweight step, typically taking less than 1 minute per segment. %
                \item Stream spline inference -- From the annotation, we estimate a ridge-line spline that captures the stream geometry. %
                We first find the \gls{mle} using gradient-based optimisation, which converges in 10--30 seconds. %
                We then refine this using NUTS to explore the spline's posterior distribution. Because we are primarily interested in marginalisation, the chain can be short and burn-in minimal. %
                This step typically takes \qty{20}{\second} to approximately \qty{1}{\minute}. %
                \item Potential likelihood evaluation -- Finally, we evaluate the curvature likelihood over a dense $10^6$-point grid of parameters while also marginalising over about $50$ realisations of the spline. %
                This step takes only around \qty{15}{\second} per segment. %
            \end{enumerate}

            For images with multiple stream segments, the total time scales accordingly (e.g., three segments take $\lesssim\,15$ minutes end-to-end). %
            These benchmarks are based on a 2023 M2 MacBook Pro with 32GB RAM and 10 Central Processing Unit (CPU) cores (six performance, four efficiency). %
            The pipeline is also GPU-compatible -- tests show runtimes for steps 2 and 3 combined $\lesssim\,1$ minute -- and fully parallelisable. %
            On a cluster with 100 GPU cores, \num{1000} stream segments could be analysed in under 10 minutes. %

            This performance is enabled by several technical features. %
            First, the curvature likelihood is nearly analytic, which is a distinct advantage over other methods. %
            Second, the pipeline is implemented in \texttt{JAX} \citep{JAX:2021}, using the custom libraries \texttt{galax} \citep{galax:2025:zenodo}, \texttt{coordinax} \citep{coordinax:2025:zenodo}, and \texttt{unxt} \citep{unxt+:2025}, for high performance and auto-differentiation features. %
            Through this framework it is simple to distribute compute across cores or devices; and since the method is inherently parallelisable, to accumulate results over chunked, vectorised computations, avoiding memory overload when performing dense evaluations of the likelihood. %
            This architecture enables the analysis to scale efficiently to large data sets and supports flexible extensions. %
            For example, we can rapidly evaluate parameter grids for a range of halo models -- including \gls{NFW} \citep{NFW:1997}, Einasto \citep{Einasto:1965}, baryon-modified profiles \citep[e.g.,][]{Dutton+Maccio:2014}, radially dependent potentials \citep{Allgood+:2006:halo_shapes}, figure rotation \citep{Bailin+Steinmetz:2005}, and even different cosmological contexts \citep[e.g.][]{Tollet+:2016} -- at minimal additional cost. %

        \subsection{Stellar streams or other tidal features?}\label{sec:discuss:advantages:tidal_tails_vs_streams}

            One of the major challenges when characterising and detecting extragalactic stellar streams is to know whether the features are indeed stellar streams, or stars ejected from the host galaxy's disc, e.g., from a tidal interaction \citep{2026A&A...707L...1S}. %
            
            Stellar streams from dwarf galaxies are partially produced through Lagrange point stripping during merger, and this type of tidal stripping occurs when the progenitor's orbital velocity around its host galaxy is much higher than the internal velocities of the progenitor's stars, i.e., when the system is dynamically cold. %
            As disc galaxies merge, on the other hand, they undergo quasi-resonant stripping episodes, where the orbital velocities of the disc galaxies around each other are similar to the orbital velocities of stars within the discs themselves \citep{Toomre+Toomre:1972, Barnes:1988, Hernquist+Ostriker:1992:SCF, DOnghia+:2010, Barnes+Hibbard:2009}. %
            These interactions give the largest distortion if the mergers are prograde, and such mergers can produce long tidal tails extending from the disc galaxies at various orientations and scales, depending on the mass ratio, encounter geometry, and viewing angle of the interaction. %
            
            Stellar streams are thus easier to classify if they are detected in a host halo far away from the disc and have a different population of stars than its host disc galaxy. %
            Throughout this paper we have investigated galaxies with disturbed morphologies, and we are therefore biased towards galaxies with brighter features close to the discs. %
            Stellar streams from dynamically cold systems could in some cases be confused with material stripped off of a disc galaxy, since many stellar streams cross the disc of their host galaxy in projection. %
            With the future release of stacked observations from the Euclid Deep Fields, and the corresponding automated morphology measurements via Galaxy Zoo and \texttt{Zoobot}, we can select streams that are fainter and that do not connect directly to their discs. %



    \subsection{Method limitations}\label{sec:discuss:limitations}
        
        In this section, we outline the main caveats and limitations of the
        methodology. %

        \subsubsection{Close encounters}
        \label{sec:discuss:limitations:close_encounters}

            A close encounter with a massive perturber, such as a satellite galaxy or large dark matter subhalo, can deflect a stellar stream and bias the inferred potential (e.g., \citealt{Erkal+:2019, Shipp+:2021, Dillamore+:2022, Lilleengen2023, Koposov+:2023}). %
            For the dwarf galaxy streams studied here, only a relatively massive perturber would produce a substantial change in the stream's trajectory \citep{Yoon+:2011}. %
            Because the method does not use stellar velocities, where perturbation signatures are most evident, it is less sensitive to such effects \citep{Carlberg:2009, Erkal+Belokurov:2015}. %
            Nevertheless, a sufficiently strong or recent encounter effectively alters the gravitational potential during part of the stream's orbit, making this limitation closely related to the assumption of a static potential. %


        \subsubsection{Potential time dependence}
        \label{sec:discuss:limitations:time_dependence}

            We assume that the host galaxy's gravitational potential is effectively static over the stream's $1$--$5$ Gyr timescale. %
            In reality, processes such as minor mergers, bar or spiral-arm growth, and disc or halo mass redistribution will evolve the potential. %
            Thus our curvature measurements probe a time-averaged field rather than the instantaneous potential. %
            Fully capturing this evolution would demand time-dependent dynamical models (e.g., \citealt{Hattori+:2016}; \citealt{Price-Whelan+:2016:chaos}; \citealt{Banik+Bovy:2019}; \citealt{Pearson+:2017}), which we leave for future work. %

            Another form of time dependence -- close-flyby perturbations -- appear to be a minor concern for our first sample. %
            Most of our hosts are early-type, bulge-dominated galaxies; however, at least one (galaxy~VII, \texttt{102022998}) shows an elongated light profile consistent with a bar or inclined disc (see \cref{app:all_streams:102022998_NEG787613526465934569}), and two others show faint disc-like features. %
            For such systems, bar or disc growth could introduce time-dependence beyond the discussed halo evolution, though at present the structures appear dynamically unimportant at the radii probed by the streams. %
            Moreover, we see no tell-tale signatures of recent strong encounters in any of our target streams. %
            Even the candidate interacting system in \cref{sec:results:102019123_NEG569554074506844623} shows no localised `kinks' or density gaps that would betray a recent perturbation \citep[e.g.,][]{Erkal+:2019}. %
            Thus though a general caveat for stream analyses, close encounters do not measurably impact our curvature-based constraints for this set of early-type systems. %


        \subsubsection{Choice of potential model}
        \label{sec:discuss:modelling_bias}

            We adopted a single-component, flattened logarithmic potential for inference \citep{Law+Majewski:2010}. %
            Real galaxies comprise multiple mass components (bulge, disc, and halo) and may exhibit radial or morphological variations that our model does not capture. %
            If the true potential deviates significantly from our assumed form, the curvature measurements may not accurately reflect the potential's geometry. %
            The curvature measurements themselves remain correct, but the inferred acceleration field is model-dependent. %
            \citet{Wu+:2026} reach consistent conclusions with a general flattened logarithmic potential, indicating that our results are not driven by the oblate-by-construction parameterisation used here. %
            The method constrains only the {direction} of the perpendicular acceleration, not its radial normalisation \citepalias{Nibauer+:2023}; potentials agreeing on the acceleration orientation at stream locations are observationally equivalent. %
            At a few scale radii the \gls{NFW} rotation curve is approximately flat and therefore directionally equivalent to a logarithmic halo. %
            The inference is thus robust to the choice of radial form by construction, not by approximation. %
            Closer to and further from the host, the model dependence is more significant. %
            Fortunately, one of the method's key strengths is its computational speed, which allows us to rapidly explore alternative potential families -- such as \gls{NFW} \citep{NFW:1997}, Einasto \citep{Einasto:1965}, or baryon-modified profiles -- and seamless integration of \texttt{Zoobot}-derived baryonic mass models. %


        \subsubsection{Endpoints and derivatives of the track}
        \label{sec:discuss:limitations:derivatives}

            Our curvature inference depends on the second derivative of the fitted spline, which inherently amplifies noise, particularly at the segment boundaries. %
            We mitigated this with a concavity regulariser, shape-matching term, and soft endpoint penalties, as described in \cref{sec:methods:track_fitting:ridge_line_optimisation}. %

            Nevertheless, endpoints remain the weakest points. %
            Interior knots draw support from annotated data on both sides, but endpoints have only one-sided information. %
            This asymmetry makes curvature sensitive to small shifts at the ends. %
            Disconnected segments compound the issue, since each endpoint floats independently when continuity is broken. %
            These effects broaden credible intervals near the ends and can weaken potential constraints when boundary regions dominate. %
            To limit this impact, we evaluate the curvature likelihood only over $\gamma \in [-0.95, 0.95]$, discarding 5\% of arc length at each end. %

            To improve robustness, combinations of multiple independent annotations can be used to sharpen endpoint distributions and introduce hierarchical priors that enforce smooth curvature continuity across the population. %
            These steps would suppress edge-driven noise and strengthen curvature-based inferences. %



    \subsection{Future directions}\label{sec:discuss:future_work}
    
        \subsubsection{Statistical combination of annotations}\label{sec:discuss:future_work:statistical_combination}

            As the citizen science program scales up, many streams will receive annotations from multiple independent volunteers, enabling analysis of streams from systematic \Euclid catalogues (e.g., Mir\'o-Carretero et al., in prep.). %
            We aim to combine these independent measurements of the stream's morphology into a probabilistic model of the stream track. %

            Before combining annotations, we must first determine whether two annotations correspond to the same stream segment. %
            In implementation this is a form of classification task that assigns annotations to segment classes. %
            One practical approach, used by \citet{Sola+:2022} for tidal features in MATLAS, is to group annotations by overlap: pairs sharing more than 25\% of their area are assigned to the same structure \citep[Sect.\,4.8]{Sola+:2022}. %
            Matched annotations are combined probabilistically; unmatched ones contribute independently to the total likelihood. %

            Instead of a hard segmentation mask, we define a soft probability map in which each pixel's weight reflects annotator agreement. %
            A similar per-pixel weighting strategy is used by \citet{Richards+:2024} for cirrus segmentation, where annotator consensus determines the training target and loss weight at each pixel \citep[Sect.\,2.4]{Richards+:2024}. %
            This modifies the track-fitting procedure described in \cref{sec:methods:track_fitting}. %
            For example, the region constraint $\mathcal{C}_\mathcal{R}$ (\cref{app:track_fitting:region_constraint}) is updated to penalise distance from high-probability regions using a continuous distance metric, such as the Jensen--Shannon divergence \citep{Lin:1991:shannon_divergence}. %
            Similarly, the shape cost $\mathcal{C}_{\rm shape}$ (\cref{app:track_fitting:central_distance}) becomes probabilistic, incorporating uncertainty in the stream's true location. %
            This weighting also propagates into the likelihood (Eq.\,\ref{eq:likelihood_single_independent}), analogously to the discrete per-segment factors in \cref{sec:methods:combining_likelihoods} but varying continuously with $\gamma$. %

            As an illustrative case, in \cref{sec:results:102018245_NEG851003014513514442} we combined two annotations of the same stream, one with two segments and another with three, weighting overlapping regions by annotator agreement. %
            This allows the tentative connecting segment to regularise the endpoints and improve curvature estimates, while still representing the uncertainty in that segment. %
            This uncertainty propagates into the posterior over track parameters $\vectheta$, broadening the fitted spline and weighting the curvature likelihood by annotator confidence at each point along the stream. %

            Alternatively, because the track-fitting and potential-inference pipeline is computationally cheap (\cref{sec:discuss:advantages:timing}), annotations can be combined in post-processing: fitting the full potential posterior independently for each annotation and then merging the resulting posteriors. %
            This sidesteps the grouping problem entirely and naturally propagates per-annotation uncertainty into the final constraints. %

        
        \subsubsection{Tentative segments}\label{sec:discuss:future_work:tentative_segments}

            Here, a segment is termed {tentative} only through disagreement across repeated annotations of the same system, whether those annotations come from the same person or from different annotators. %
            Rather, tentativeness is inferred only by comparing repeated annotations. %
            Building on the discussion in \cref{sec:results:102018245_NEG851003014513514442}, we compare two annotations of the same stream, one with two segments and one with an additional connecting segment (for a total of three). %
            The previous subsection (\cref{sec:discuss:future_work:statistical_combination}) described how to combine multiple annotations of the same segment to improve robustness to endpoint curvature. %
            Here, we discuss two related types of tentative segments: disconnected and connected. %

            Disconnected tentative segments are identified independently and are not physically connected to any other segment. %
            As in \cref{sec:discuss:combining_constraints}, where streams were weighted by projected arc length, we can assign each segment an additional weight corresponding to its probability of being real. %
            Although this factor is not yet incorporated into \cref{sec:methods:single_likelihood}, it will be important for future analyses. %

            Connected tentative segments are physically attached to high-confidence segments but appear in only a subset of annotations, making them lower-probability features of the overall segment. %
            The example in \cref{sec:results:102018245_NEG851003014513514442} illustrates this case, where a tentative segment bridges two well-identified segments. %
            A similar situation arises when different annotations trace the same segment to different lengths. %
            If the length distribution is bimodal, the extra portion beyond the shorter annotations can be treated as a tentative extension. %
            Such cases can be handled using the probabilistic combination approach described in \cref{sec:discuss:future_work:statistical_combination}. %

            Bridging segments like the one in \cref{sec:results:102018245_NEG851003014513514442} are more challenging because some annotations treat the segments as connected while others do not. %
            The most robust approach is to run track fitting and curvature analysis separately for the connected and disconnected cases, then combine the results weighted by the probability that the tentative bridge is real. %
            This roughly doubles the computational cost, but speaks to the advantages of this method, since it remains orders of magnitude faster than other methods, even with this additional complexity. %
            Also, with this approach, the resulting curvature constraints are more robust, especially near the segment endpoints. %

            These considerations highlight how tentative segments can be systematically incorporated into curvature-based inference. %
            By treating both disconnected and connected cases probabilistically, we can capture the additional information they provide without overstating their significance. %
            As the number of annotations grows, such approaches will ensure that endpoint constraints and curvature estimates remain robust while retaining computational efficiency. %


        \subsubsection{Baryon modelling}\label{sec:discuss:future_work:baryon_Modelling}

            Curvature-based inference constrains the direction of the local acceleration vectors. %
            Consequently, the overall normalisation of the gravitational potential is unconstrained, and for a single-component model our inferences depend only on the potential's geometry. %
            This makes a single-component logarithmic halo model appealing: it is simple, interpretable, and avoids degeneracies with baryonic mass components. %
            However, as we scale to higher precision and deeper samples, incorporating a realistic model for the baryonic contribution becomes essential. %

            In future work, we plan to integrate empirical models of the baryonic mass distribution derived from the \texttt{Zoobot} morphological framework. %
            These models will use imaging-based morphology classifications to construct parametric forms, e.g., \citet{Sersic:1963} discs or exponential discs \citep{Freeman:1970:exponential_disks} for the stellar mass distribution. %
            The resulting baryonic components can then be added to the total gravitational potential. %

            Incorporating baryons introduces both a technical challenge and a scientific opportunity. %
            First, deprojecting the stellar mass distribution from 2D imaging requires marginalising over the unknown inclination and intrinsic shape of the luminous component. %
            This type of deprojection is common in dynamical modelling of elliptical galaxies, where both Schwarzschild orbit-superposition models and Jeans anisotropic models (JAMs) use photometric profiles to construct 3D mass models \citep{Cappellari+:2007:SAURON, Cappellari:2016:review}. %
            In particular, the multi-Gaussian expansion (MGE) technique is often used to deproject the surface brightness under assumed viewing angles and to build axisymmetric or triaxial models of the stellar components \citep[e.g.,][]{Emsellem+:2004, Cappellari:2008:JAM}. %
            In our framework, morphological class information from \texttt{Zoobot} can provide informative priors for this deprojection. %

            Second, incorporating baryons changes the interpretation of the flattening parameter $q$. %
            Rather than describing the total potential, $q$ becomes a constraint on the residual dark matter geometry once the baryonic component is subtracted. %
            This separation enables more direct comparisons with predictions from $\Lambda$CDM and hydrodynamical simulations, where the dark halo is expected to become rounder and more disc-aligned in the presence of a dissipative stellar component \citep{Debattista+:2008, Chua+:2019, Bovy:2026:book}. %

            Modelling the baryonic contribution also enables tests of the alignment between baryonic and dark matter distributions, which are not expected to exactly coincide \citep{Shao+:2016}. %
            This allows us to identify systems where baryons dominate the local acceleration and to track how this dominance evolves with radius. %
            In those regions, curvature constraints remain valid but reflect the baryonic potential instead, offering a way to probe the transition to dark matter dominance. %



    \subsection{Forecasting stream detections}\label{sec:discuss:number_of_streams}

        Even in the early Euclid \gls{Q1} data, our analysis already identifies a scientifically useful sample of extragalactic streams and demonstrates that this approach can be applied to real survey-quality imaging. %
        Looking ahead, the forthcoming Galaxy Zoo and \texttt{Zoobot} morphology measurements of \Euclid images, combined with systematic stream cataloguing efforts (e.g., Mir\'o-Carretero et al., in prep.), will greatly expand this sample. %
        To estimate the scale on which this approach may operate, we consider a simple back-of-the-envelope forecast. %

        The catalogue from \citet{Q1-SP047} contains \num{378000} bright sources, of which \num{4000} of the most morphologically perturbed galaxies were selected for visual inspection (see \cref{sec:the_data:galaxy_selection}). %
        From this subset, we identified $13$ highly promising stream systems (see \cref{sec:the_data:stream_sel}), already showing that \Euclid contains a nontrivial population of streams suitable for our analysis, though additional candidates were present. %
        Importantly, streams often have low surface brightness and lie far from the host galaxy, so they may not trigger automated perturbedness classifiers. %
        The arcsinh stretch used in our cutouts also magnifies the visual scale of faint features; their physical amplitude is typically smaller than it appears. %
        Thus, these $13$ detections likely represent a lower bound on the number of streams present. %

        We can use these $13$ galaxies to derive approximate predictions for the expected number of streams detected by \Euclid that will be applicable to our analysis. %
        The Euclid \gls{Q1} field is approximately \qty{60}\,deg$^2$, while the wide field will cover \qty{14000}\,deg$^2$, around $230$ times the area. %
        Thus a conservative bound is 2300 more streams. %
        Note, however, that we have presented only a subset of tidal features in \gls{Q1}. %
        If the true number is even only a few times larger, for example $50$ streams, then there could be over $10^4$ streams suited for our analysis in the \Euclid wide field. %



\section{Summary and conclusions}\label{sec:conclusions}

    We have applied our curvature-based inference method to a first sample of \Euclid-detected stellar streams (\cref{fig:streams-panel}). %
    While we highlight a few illustrative examples in the main text, the full set of results -- including posteriors for all annotated segments -- is provided in \cref{app:all_streams}. %
    These examples demonstrate the range of halo geometries that can be constrained from stream morphology and showcase the diagnostic power of our approach across diverse systems. %

    In \cref{sec:results:halo_geometry}, we examined individual host galaxies, using the curvature of their stellar streams to infer halo shape and \gls{com} location. %
    We highlighted systems with especially informative stream geometries, such as \cref{sec:results:102018245_NEG851003014513514442,sec:results:102019123_NEG569554074506844623}, to provide tight constraints on both the projected flattening $\qproj$ and \gls{com} offsets at about $0.5 \, r_{\mrm{isophotal}}$ or $5\,{\rm kpc}$ precision. %
    We identified how and which stream segments dominate the inference, showing that long, highly curved segments provide the strongest constraints. %

    In \cref{sec:results:combining_information}, we combined constraints across galaxies to infer the population-level distribution of halo flattening and \gls{com} alignment. %
    Stacking the individual likelihoods yields median projected flattening $\qproj = 0.95^{+0.05}_{-0.10}$ at 68\% confidence, with 90\% confidence interval (CI) at $0.7$--$1.0$, and no evidence for systematic offsets between light and total-mass centres. %
    These results are consistent with predictions for $\Lambda$CDM halos of mass around $10^{12}\,M_\odot$ at $z<0.5$ \citep[e.g.][]{Allgood+:2006:halo_shapes}. %

    Our sample spans a broad range of galaxy morphologies, photometric redshifts, and environments, with both disc and early-type hosts represented. %
    All stellar streams are significantly more massive than globular cluster streams and have not massively disturbed their host galaxies, suggesting that they are likely the remnants of dwarf galaxy accretion events. %
    The diversity in stream colours and curvatures highlights the variety of accretion histories accessible in \Euclid imaging. %
    Several systems show signs of recent accretion or possible companions, raising the prospect of isolating environmental effects in future work. %

    Throughout, our method provides robust geometric constraints with minimal assumptions. %
    By avoiding explicit modelling of the progenitor and being by design robust to much of the tidal disruption process, the curvature-based likelihood is less sensitive to model misspecification than traditional forward modelling approaches. %
    This makes it particularly well-suited for stacking analyses and ensemble inference, where heterogeneity makes joint modelling challenging. %
    Stream morphology thus complements other techniques: lensing probes statistical halo shapes at large scales, while satellite kinematics require equilibrium assumptions and spectroscopy. %
    Streams directly constrain the local acceleration field on tens to hundreds of kiloparsec scales, with few dynamical assumptions. %

    This study establishes the viability of curvature-based inference using real \Euclid data and lays the foundation for scale-up. %
    Improvements to the method -- e.g.,\ better segment-connectivity models, refined curvature theory, and integration of stream membership probabilities -- will further enhance its power. %
    Future \Euclid releases will expand the sample by orders of magnitude. %
    With thousands of streams anticipated in the citizen science program, this framework will enable precise, population-level constraints on dark matter halo geometry across cosmic time, providing a new dynamical test of $\Lambda$CDM. %

    Our main findings are as follows:
    \begin{enumerate}
        \item Methodological advances: We developed an optimal track-fitting approach for segmented stream annotations, integrated with a curvature-based likelihood that is robust to progenitor modelling and many complexities of tidal debris, enabling both individual and ensemble inference. %
        \item Constraints from individual streams: For the most informative systems, we obtain $\qproj \gtrsim 0.8$ (90\% CI) and \glspl{com} within $r \lesssim 0.5\;r_{\mrm{isophotal}} \simeq 5\,{\rm kpc}$ of the photometric centre. %
        \item Population-level inference: Stacking posteriors across galaxies yields a median $\qproj = 0.95^{+0.05}_{-0.10}$ at 68\% confidence (90\% CI $0.7$--$1.0$) and no evidence for $r>0.5r_{\rm isophotal}$ systematic baryon-dark matter offsets, consistent with $\Lambda$CDM predictions. %
    \end{enumerate}


\vspace{30pt}
\begin{acknowledgements}\label{sec:acknowledgments}

\AckQone
\,\AckEC

    Support for this work was provided by The Brinson Foundation through a Brinson Prize Fellowship grant to author Nathaniel Starkman. %

    This work was supported by a research grant (VIL53081) from VILLUM FONDEN. %
    This work was co-funded by the European Union (ERC, BeyondSTREAMS, 101115754). %
    Views and opinions expressed are however those of the author(s) only and do not necessarily reflect those of the European Union or the European Research Council. %
    Neither the European Union nor the granting authority can be held responsible for them. %
    
    JB received partial support from NSERC (funding reference number RGPIN-2020-04712) and from an Ontario Early Researcher Award (ER16-12-061; PI Bovy). %
    The Dunlap Institute is funded through an endowment established by the David Dunlap family and the University of Toronto. %

    This publication uses data generated via the Zooniverse.org platform, development of which is funded by generous support, including a Global Impact Award from Google, and by a grant from the Alfred P. Sloan Foundation. %
    The ZooTasks interface used to annotate galaxy images was supported by the Data Sciences Institute at the University of Toronto via grant number DSI-RSDY3R1P02. %

    This research was supported by the International Space Science Institute (ISSI) in Bern, through ISSI International Team project \#23-584. %

    We thank Patrick Simon, Annette Ferguson, Ariane Lancon, Crescenzo Tortora, Denis Erkal, Elham Saremi, Federico Lelli, Jose Maria Diego, Maarten Baes, and Pascale Jablonka for reading the manuscript and providing helpful comments. %

    The data in this paper are the result of the efforts of the Galaxy Zoo volunteers, without whom none of this work would be possible. Their efforts are individually acknowledged at \url{http://authors.galaxyzoo.org}.   %

\end{acknowledgements}

%
%
\bibliography{main.bib}
%
%

%
%

\begin{appendix}

\section{Geometric properties of 2D curvilinear paths}\label{app:math_details}

    Let $\mcal{S}(\gamma;\vectheta)$ be a spline parameterisation of a curvilinear path. %
    The spline is the mapping from an affine parameter $\gamma$ to the 2D plane, i.e., $ \mcal{S}(\gamma, \vectheta) : \mathbb{R}^1 \to \mathbb{R}^2 $, (Eq.\,\ref{eq:spline}). %
    The spline parameters $\vectheta$ includes the set of spline knots at $\{(\gamma_k, \mbf{x}_k)\}_{k=1}^K$ (as well as any control points). %

    Points along the spline are trivially defined as
    \begin{equation}
        \mbf{x}_i \equiv \mcal{S}(\gamma_i, \vectheta) \;,
    \end{equation}
    where $\mbf{x}_i$ is the 2D position of the point at affine parameter $\gamma_i$. %
    
    The arc length along the spline is defined as

    \begin{equation} \label{eq:arc_length}
        \ell(\gamma, \vectheta) = \int_{\min(\gamma)}^{\gamma} \left\| \frac{\mrm{d}}{\mrm{d}\gamma} \mcal{S}(\gamma, \vectheta) \right\| \, \mrm{d}\gamma \;.
    \end{equation}
    Henceforth we drop requiring explicit inclusion of $\vectheta$, the spline parameters. %

    A common parameterisation of stream splines is to use the affine parameter $\gamma$ as the arc length along the spline, normalised to the range $[-1, 1]$. %
    $\gamma = 0$ is the centre of the spline and is set to the progenitor location, if known. %
    The following are true regardless of the affine parameterisation. %

    The tangent vector $\vecT(\gamma)$ is the $\gamma$-derivative of the spline
    \begin{equation} \label{eq:tangent}
        \vecT(\gamma;\vectheta) = \frac{\rm{d}}{\rm{d}\gamma}\mcal{S}(\gamma;\vectheta), \qquad \uvecT = \frac{\vecT}{\|\vecT\|} \;.
    \end{equation}
    Therefore the arc length along the spline can be expressed as
    \begin{equation} \label{eq:arc_length_tangent}
        \ell(\gamma) = \int_{\min(\gamma)}^{\gamma} \|\vecT(\gamma')\| \, \mrm{d}\gamma' \;.
    \end{equation}
    The track `speed' is the $\gamma$-derivative of this quantity
    \begin{equation} \label{eq:speed}
        v(\gamma) = \frac{\rm{d}\ell(\gamma)}{\rm{d}\gamma} = \|\vecT(\gamma)\| \;.
    \end{equation}
    The curvature of the spline track is thus
    \begin{equation} \label{eq:curvature}
        \veckappa(\gamma) = \dfrac{\rm{d}\uvecT}{\rm{d}\ell}(\gamma) = \left[\frac{1}{v} \dfrac{\rm{d}\uvecT}{\rm{d}\gamma} \right](\gamma) \equiv \kappa(\gamma) \; \hat{\mbf{N}}\;.
    \end{equation}
    where $\kappa(\gamma) = \|\veckappa(\gamma)\|$ is the scalar curvature and $\hat{\mbf{N}}$ the normal vector. %
    This normal is defined from the curve. %
    Rotate $\vecT(\gamma)$ by $\ang{90}$ counter-clockwise to get the left-handed normal: %
    \begin{equation} \label{eq:left_handed_normal}
        \mbf{N}_{\rm left}(\gamma) = \left( -T_y, T_x \right)(\gamma) \;.
    \end{equation}

    The signed curvature is a vector that exists only in two dimensions, and is defined as the projection of the curvature vector onto the left-handed normal (see Eq.\,\ref{eq:left_handed_normal}):
    \begin{equation}
        \kappa_{\pm}(\gamma) = \veckappa(\gamma) \cdot \mbf{N}_{\rm left}(\gamma) = \frac{x'(\gamma) y''(\gamma) - y'(\gamma) x''(\gamma)}{[x'(\gamma)^2 + y'(\gamma)^2]^{3/2}} \;,
    \end{equation}
    where $x(\gamma)$ and $y(\gamma)$ are the $x$ and $y$ coordinates of the spline track $\mcal{S}(\gamma)$ and prime denotes differentiation with respect to $\gamma$. %
    The dot product form is generally more numerically stable than the coordinate form. %
    In higher dimensions, the signed curvature is not defined, but is similarly described by a torsion vector. %


\section{Track optimisation details}\label{app:track_fitting}

    This appendix provides detailed definitions and motivations for the components of the track-fitting cost function introduced in \cref{sec:methods:track_fitting:ridge_line_optimisation}. %
    There, we describe how the stream ridge line is modeled by optimising a spline track to match a user-provided annotation. %
    The total cost function comprises four terms: a region constraint $\mcal{C}_\mcal{R}$, a concavity regulariser $\mcal{C}_{\Delta\kappa}$, a shape loss $\mcal{C}_{\rm{text}}$, and an endpoint penalty $\mcal{C}_{\rm{end}}$. %
    Each term is designed to enforce specific properties of the optimal spline track, and we describe them in detail below. %

    \subsection[Region constraint]{Region $\mcal{R}$ constraint $\mcal{C}_\mcal{R}$}\label{app:track_fitting:region_constraint}

        With this component of the cost function we ensure the optimal spline lies within the annotation region $\mcal{R}$. %
        Let $d_\mcal{R}(\mbf{x})$ be the region distance field given by
        \begin{equation} \label{eq:region_distance}
            d_{\mcal{R}}(\mbf{x}) =
            \begin{cases}
                0, & \quad \text{if } \mbf{x} \in \mcal{R}, \\
                \min\limits_{\mbf{y} \in \partial \mcal{R}} \| \mbf{x} - \mbf{y} \|, & \quad \text{if } \mbf{x} \notin \mcal{R} \;,
            \end{cases}
        \end{equation}
        where $\partial \mcal{R}$ is the boundary of the region $\mcal{R}$. %
        The spline and underlying set of annotation points are $\{(\gamma_i, \mbf{x}_i)\}_{i=1}^N$. %

        Using this region distance function we define the loss function as 
        \begin{equation} \label{eq:region_cost}
            \mcal{C}_{\mcal{R}}(\vectheta) = \sum_{i=1}^n \arctan\left[ \lambda_{\mcal{R}} \, d_\mcal{R}(\mcal{S}[\gamma_i; \vectheta]) \right] \;,
        \end{equation}
        where $\lambda_{\mcal{R}}$ is the `steepening' hyperparameter for the slope of the derivative of this cost function. %
        $\arctan(\cdot)$ is a smooth surrogate for the Heaviside function \citep{Heaviside:1893}, which does not have a derivative and cannot be used in gradient descent optimisation. %
        Like the Heaviside, inside the region, $\arctan(0) = 0$ there is no contribution from the cost function and no solution is penalised. %
        Outside the region $\arctan(\lambda_{\mcal{R}} d_\mcal{R})$ grows quickly. %
        Larger values of $\lambda_{\mcal{R}}$ force gradient-based optimisers to rapidly converge to solutions with $d_{\mcal{R}}=0$. %
        In this work we adopt $w_{\mcal{R}} = 10^{12}$.

        This region cost term $\mcal{C}_\mcal{R}$ is both incredibly restrictive and permissive. %
        The restrictive aspect is that it forces the solution to be within the region $\mcal{R}$, which is the annotation region. %
        The permissive part is that it allows any solution within that region. %
        Thus we require further constraints to find an optimal solution.%

    \subsection[Concavity change constraint]{Concavity change constraint $\mcal{C}_{\Delta \kappa}$}\label{app:track_fitting:concavity_change_constraint}

        In this contribution to the loss function we discourage the track from having spurious changes in the concavity. %
        The curvature of the track is related to the geometry of the host galaxy's gravitational potential. %
        Spurious changes in the curvature of the track are significant problems for the inference of the halo geometry. %
        We build to a cost function that penalises these concavity changes in terms of the interpolating spline $\mcal{S}$. %

        We project the curvature vector of the spline onto the spline's left-handed normal to get the signed curvature
        \begin{equation} \label{eq:signed_curvature}
            \kappa_{\pm}(\gamma) = \veckappa(\gamma) \cdot \mbf{N}_{\rm left},
        \end{equation}
        with terms derived in \cref{app:math_details}. %
        This signed curvature $\kappa_{\pm}(\gamma)$ indicates the local concavity of the stream and is only defined in two dimensions, which is appropriate here given the lack of line-of-sight (depth) information in the \Euclid image cutouts. %

        From the signed curvature we compute $\widetilde{\mrm{sgn}}(\kappa_\pm)$, the curvature sign indicator, 
        \begin{equation} \label{eq:curvature_sign_indicator}
            \widetilde{\mrm{sgn}}(\kappa_\pm) = \arctan\left[ \lambda_{\kappa}\, \kappa_{\pm}(\gamma) \right] \;,
        \end{equation}
        where $\lambda_{\kappa}$ controls how closely the indicator approximates the $\mrm{sgn}(\cdot)$ function (see \cref{app:sensitivity_analysis_curvature_sign} for further details).
        The purpose of using an indicator function for the sign is to maintain smooth differentiability for gradient-based optimisation. %
        
        The derivation up to \cref{eq:signed_curvature} established how to detect a change in concavity at a specific point $\gamma$ along the spline. %
        To construct a global cost function, we integrate over the derivative of the curvature sign indicator along the spline, thereby penalizing the total variation in concavity across the entire track,
        \begin{equation} \label{eq:concavity_cost}
            \mcal{C}_{\Delta\kappa}(\vectheta)
            = \!\int_{0}^{L} \!\!\! \left( \mrm{d}_\ell \, \widetilde{\mrm{sgn}} \right)^2 \! \mrm{d}\ell
            = \!\int_{-\!1}^{1} \!\! \left( \frac{1}{v}\frac{\rm{d}\widetilde{\mrm{sgn}}(\kappa_\pm)}{\mrm{d}\gamma} \right)^2 \!\! \mrm{d}\gamma \;.
        \end{equation}
        Conceptually, this cost function is the smooth equivalent of counting the number of sign changes in the curvature along the track. %

        Note that we need to take a $\gamma$-derivative of the $\widetilde{\mrm{sgn}}$, which in turn is proportional through $\kappa$ (Eq.\,\ref{eq:curvature}) to the second derivative of the spline. %
        Therefore we require that the spline $\mcal{S}$ (Eq.\,\ref{eq:spline}) is at least $C^2$ continuous. %
        In general we use a $C^2$ continuous cubic spline everywhere. %

        This cost function penalises changes in concavity while remaining differentiable, enabling gradient-based optimisation. %
        It is designed to be insensitive to variations in curvature that do not alter the concavity. %
        A sufficiently large $\lambda_\kappa$ is required to achieve both of these goals and is of particular importance if the stream has any nearly-straight portions (see \cref{app:sensitivity_analysis_curvature_sign} for explanation). %
        We adopt $w_{\Delta\kappa} = 10^7$; the choice of $\lambda_\kappa = 10^3$ is discussed in \cref{app:sensitivity_analysis_curvature_sign}. %

        Like the region constraint \hyperref[eq:region_cost]{$\mathcal{C}_{\mathcal{R}}$}, $\mcal{C}_\kappa$ is both a very restrictive and permissive constraint.
        For a $C^2$ spline, the degrees of freedom include the knot positions and two associated velocity-like control parameters. %
        Small adjustments to knot positions can significantly alter the curvature, making the concavity penalty highly sensitive and thus strongly constraining. %
        However, any parameters that avoid changes in concavity is treated nearly equivalently by the cost function and the solution space can be very degenerate. %
        As with $\mathcal{C}_{\mathcal{R}}$, additional constraints are needed for an optimal fit. %

    \subsection[track shape cost]{Track shape constraint $\mcal{C}_{\rm shape}$}\label{app:track_fitting:central_distance}

        This component of the loss function encourages the fitted spline to follow the overall shape of the annotation. %
        The previous terms constrain the spline to remain within the annotated region ({$\mcal{C}_\mcal{R}$}) and discourage spurious changes in concavity ({$\mcal{C}_{\Delta \kappa}$}), but neither term ensures that the spline reproduces the large-scale shape of the annotation. %
        As a result, the solution space may contain splines that are geometrically valid but qualitatively inconsistent with the observed stream. %

        For example in \cref{fig:zooniverse-annotatation-process} there is an annotated segment which is short, but noticeably curved. %
        With only $\mcal{C}_\mcal{R}$ and $\mcal{C}_{\Delta \kappa}$, a straight-line spline would be equally valid (and actually may be preferred since $\arctan$ is an imperfect sign indicator). %
        This term encourages the spline to reflect the shape of the annotation. %

        The functional form of the cost function is a simple weighted least squares,
        \begin{equation} \label{eq:shape_cost}
            \mcal{C}_{\rm shape}(\vectheta) = \frac{1}{N} \sum_{i=1}^{N} \left(\frac{\mbf{x}_i - \mcal{S}(\gamma_i)}{\sigma_i}\right)^2 \;,
        \end{equation}
        where $\sigma_i$ is an inverse weight on the point $(\gamma_i, \mbf{x}_i)$. %
        This tunable hyperparameter defaults to twice the half-width of the annotation region, i.e.\ $\sigma_i = 2\,r_\mcal{R}$, lending a gentle gradient towards the annotation's shape. %

        It is important that $\mcal{S}$ is at least $C^1$ continuous to ensure a well-defined derivative at all points. %
        For consistency with the concavity term, which involves second derivatives, we adopt a $C^2$ continuous cubic spline here as well. %
        We normalise this loss by $N$ to define a mean squared error (MSE), making it independent of the number of annotation points. %
        This scaling is atypical for Bayesian-style loss functions, but it ensures consistency since the other terms do not depend on $N$. %

        We give this shape term a lower weight than the others, adopting $w_{\mrm{shape}} = 10^3$.
        However, because the region and curvature constraints often lead to broad solution spaces with nearly flat gradients once satisfied, this term helps guide toward a more representative solution. %
        In particular, it encourages the spline to follow the overall shape of the stream, even when many solutions are otherwise equivalent. %

    \subsection[End point fixing]{End-point fixing $\mcal{C}_{\mrm{end}}$}\label{app:track_fitting:end_point}

        The final term in the loss function optionally constrains the locations of the track's end-points. %
        Splines, especially their curvature, are sensitive to boundary conditions, and small changes to the end-point location can have an outsized impact \citep{Lucas:1974:ErrorBoundsInterpolating}. %
        While annotations typically begin where the stream becomes visible, the stream likely extends beyond the annotated region. %
        We cannot constrain its shape outside this range, but we can approximate such a constraint by softly penalizing endpoints that deviate too far from the start or end of the annotation. %
        This regularisation helps stabilize the fit without enforcing hard boundaries. %

        The distance between the annotation endpoints and the end knots are
        \begin{equation}
            d_j = \left\| \mcal{S}(\gamma_{j}) - \mbf{x}_j \right\|, \quad j \in \{1, N\} \;.
        \end{equation}

        \noindent Then the total endpoint cost is
        \begin{equation} \label{eq:endpoint_cost}
            \mcal{C}_{\mrm{end}}(\vectheta) = \frac{1}{\lambda_{\mrm{end}}} \!\sum_{j \in \{1, N\}} \!\! \ln\left[1 + \mrm{e}^{\left(\lambda_{\mrm{end}} [d_j \!-\! d_{\max} + \epsilon] \right)}\right] \;,
        \end{equation}
        where $\lambda_{\mrm{end}}$ controls the steepening of the $\text{softplus}$ and $d_{\max}$ is the maximum distance. %
        $\epsilon$ is a small softening term that helps with numeric stability when the distance is in the denominator. %
        We adopt $w_{\mrm{end}} = 10^5$ and $d_{\max} = 2\,r_{\mrm{isophote}}$.

        While this term could technically be merged into the shape loss $\mathcal{C}_{\rm shape}$ we separate it to isolate the specific role of endpoint regularisation. %
        Tying the width parameter $\sigma$ in $\mathcal{C}_{\rm shape}$ to the annotation makes sense for most of the track, but endpoints are often less reliable and more variable across annotations. %
        Treating endpoints separately ensures better control over boundary behavior and allows us to tune this component separately without affecting the shape-fitting behavior. %

    \subsection{Sensitivity analysis of the curvature sign indicator}\label{app:sensitivity_analysis_curvature_sign}

    \begin{table*}[t]
        \centering
      \caption{%
        Outer isophotal radii (in pixels) measured at eight azimuthal angles for each galaxy, going counter-clockwise from the positive $x$-axis.
      \label{tab:isophotal-radii}}
        \renewcommand{\arraystretch}{1.3}
        \begin{tabular}{>{\ttfamily}l@{\hskip 0.5em}*{8}{S@{\hskip 0.2em}}}
          \toprule
          Tile Index / Object ID &
          \multicolumn{8}{c}{$r$ (pix) at $\{0^\circ, 45^\circ, \ldots, 315^\circ\}$} \\
          \cmidrule(lr){2-9}
          \begin{filecontents*}{figures/stack/isophotal_radii.csv}
Galaxy ID,0,45,90,135,180,225,270,315
102014777/-780275758555920353,67.2,64.1,74.2,70.1,65.7,79.6,77.4,71.7
102014782/-825268920554154517,65.6,59.9,62.2,58.1,64.7,64.0,74.9,71.2
102018245/-851003014513514442,82.9,68.4,55.3,62.5,66.9,72.2,84.4,95.6
102018668/-595264901510604722,57.6,65.7,78.5,68.1,62.7,75.6,77.3,56.4
102019123/-569554074506844623,76.7,83.4,75.0,83.2,85.8,76.5,75.7,72.4
102021493/-583998851481016349,80.8,65.9,67.4,97.7,106.3,98.7,107.0,90.3
102022998/-787613526465934569,69.8,62.2,62.2,72.4,58.9,49.2,56.3,69.4
102159485/2680742453663022737,54.6,55.7,58.9,57.0,57.2,58.6,58.7,57.4
102030408/-732320008395115905,83.5,76.2,68.3,97.0,119.7,88.4,85.0,83.2
102032666/-674918847372942403,51.2,55.1,44.6,42.6,48.7,49.6,51.1,51.4
102044826/-539137978275277599,52.0,50.3,50.5,53.4,42.8,28.7,33.3,41.8
102046112/-526944083265011738,76.6,58.1,51.3,64.8,66.5,44.9,37.6,58.0
102158269/2649560011643950805,38.5,51.5,101.3,42.0,30.1,40.1,112.9,51.4
102158893/2717847201657197076,75.7,67.7,58.1,60.0,85.1,78.4,64.7,61.5
\end{filecontents*}
\csvreader[
            head to column names,
            late after line=\\
          ]{figures/stack/isophotal_radii.csv}{}%
          {\csvcoli & \csvcolii & \csvcoliii & \csvcoliv & \csvcolv &
           \csvcolvi & \csvcolvii & \csvcolviii & \csvcolix}
          \bottomrule
        \end{tabular}
    \end{table*}

        In \cref{app:track_fitting:concavity_change_constraint}, we required a count of concavity changes along the stream track, which is inherently a discontinuous function. %
        For numerical purposes we introduced a smooth proxy function, defined by $\widetilde{\rm{sgn}}(\kappa_\pm)$, as given in \cref{eq:curvature_sign_indicator}. %
        Here $\lambda_\kappa$ is a tunable hyperparameter that controls the sensitivity of the curvature sign indicator to the curvature magnitude. %

        Setting $\lambda_\kappa$ is not trivial. %
        To inform this choice, we perform a sensitivity analysis of the curvature sign indicator. %
        Taking the derivative with respect to $\kappa_\pm$ gives
        \begin{equation}
            \frac{\rm{d}}{\rm{d}\kappa_\pm} \widetilde{\rm{sgn}}(\kappa_\pm) = \frac{\lambda_\kappa}{1 + (\lambda_\kappa \kappa_\pm)^2} \;.
        \end{equation}
        This expression reveals two key features of the curvature sign indicator: %

        First, it shows that in the regime $|\lambda_\kappa \kappa_\pm| \gg 1$, the denominator dominates, and the derivative tends toward zero. %
        In this case $\widetilde{\rm{sgn}}$ effectively saturates to $\pm {\pi}/{2}$, depending on the sign of $\kappa_\pm$. %
        Therefore, the indicator becomes sensitive to the concavity and not the curvature magnitude. %

        Second, when $\kappa_\pm$ is near zero, the indicator becomes approximately linear and $\widetilde{\rm{sgn}}$ is not sensitive to changes in concavity unless $\lambda_\kappa$ is large. %
        More clearly, the Taylor expansion of $\widetilde{\mrm{sgn}}(\kappa_\pm)$ about $\kappa_\pm = 0$ is
        \begin{equation}
            \arctan(\lambda_\kappa \kappa_\pm) = \lambda_\kappa \kappa_\pm + \mathcal{O}([\lambda_\kappa \kappa_\pm]^3) \;,
        \end{equation}
        so that for small $\kappa_\pm$ the curvature sign indicator is approximately $\widetilde{\rm{sgn}}(\kappa_\pm) \approx \lambda_\kappa \kappa_\pm$ with slope $\lambda_\kappa$. %

        To detect a small deviation $\epsilon$ from zero curvature as a change in concavity, we require the indicator to exceed some threshold $|\widetilde{\rm{sgn}}(\epsilon)| > \delta$. %
        Using the approximation above, this implies the condition
        \begin{equation}
            \lambda_\kappa > \frac{\tan(\delta)}{|\epsilon|} \;.
        \end{equation}

        For example, detecting a curvature deviation $\epsilon = 0.01$ with a threshold $\delta = \pi/6$ requires $\lambda_\kappa \gtrsim 577$. %
        This illustrates that in nearly straight regions, identifying small changes in concavity requires a relatively large value of $\lambda_\kappa$. %

        In this implementation, we set $\lambda_\kappa = 10^3$, though as a user-adjustable hyperparameter this can be changed as needed. %


\section{Isophotal radius} \label{app:isophotal_radius}

    This appendix supports the \gls{com} analysis in \cref{sec:results:combining_information:com}. %
    In that section, we compare dynamically inferred centres to the luminous centres of the galaxies. %
    Because the galaxies lie at different and uncertain distances, pixel-based offsets are not physically meaningful. %
    Moreover, due to their unknown intrinsic sizes, converting these offsets into real-space units is deferred to future work. %
    Instead, we normalise distances by an outer-light radius, defined here as the radius at which the galaxy light ceases to be visibly distinguishable from the local background, allowing us to express offsets as fractional, dimensionless quantities. %

    To define this normalisation scale, we estimate the outer isophotal radius of each galaxy at eight azimuthal angles, spaced every $\qty{45}{degree}$, proceeding counter-clockwise from the positive $x$-axis. %
    These radii are obtained from the galaxy annotations by marking, along each direction, the approximate point where the visible galaxy light ends; they are listed in \autoref{tab:isophotal-radii} for each system. %
    This angular profile allows us to normalise any radial measurement by the corresponding isophotal radius at that angle. %
    In particular, we interpolate the isophotal radius as a smooth, periodic function of angle to define a continuously varying normalisation profile. %
    This enables consistent, angle-dependent scaling of offset measurements throughout the analysis. %


\onecolumn
\section{Details on each selected system}\label{app:all_streams}

    In this appendix, we provide additional information on the systems selected in \cref{sec:the_data:stream_sel}, which are used in this work. %
    We examine each system individually, discussing its luminous host, associated stellar stream(s), and the constraints we infer on the geometry of its total matter distribution. %
    These systems are drawn from the study for a Zooniverse-\Euclid project -- a citizen science initiative that catalogues and annotates stream-like features in \Euclid imaging. %
    This section supplements the main-text results presented in \cref{sec:results:102018245_NEG851003014513514442} and \cref{sec:results:102019123_NEG569554074506844623}, and provides additional detail on all systems contributing to the stacked constraints discussed in \cref{sec:results:combining_information}. %
    Each system is displayed as a false-colour composite constructed from \Euclid VIS and \YE-band imaging using an inverse hyperbolic sine stretch (\texttt{vis\_y\_arcsinh}). %
    The arcsinh stretch magnifies the visual scale of faint features; apparent perturbations are smaller in physical amplitude than they appear. %
    The image preparation follows the procedure described in \citet{Q1-SP047}; we refer the reader to that work for a full account of the cutout construction and calibration. %


    \subsection[I) 102014777:-780275758555920353]{I) \texttt{102014777:$-$780275758555920353}}\label{app:all_streams:102014777_NEG780275758555920353}

        \begin{figure}[htbp]
            \centering
            \includegraphics[width=0.8\linewidth]{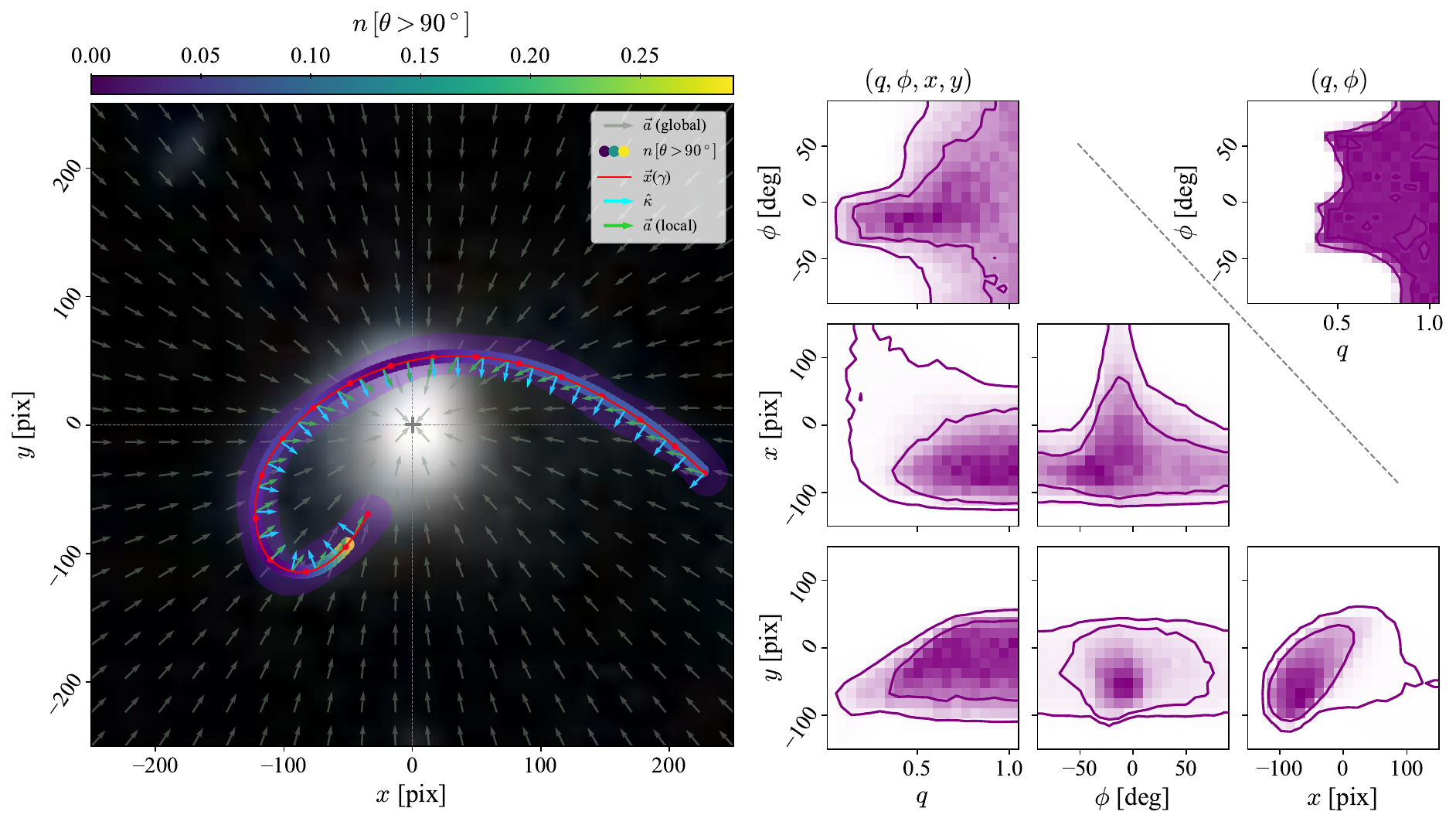}
            \caption{%
                Cutout~I: \texttt{102014777:$-$780275758555920353} and geometry
                constraints. %
                \emph{Left}: %
                Galaxy cutout centred on a host galaxy and a stellar stream (annotated in purple). %
                The stream connects to the galaxy's outer regions and loops in front of it in projection. %
                The optimal stream track is red with cyan blue unit vectors showing the curvature. %
                The acceleration field for a best-fit potential is grey with green vectors at the stream's location. %
                The stream is colour-coded by the fraction of misaligned curvature-acceleration vector pairs across the $(q, \phi)$ parameter grid, highlighting which regions of the track are most constraining. %
                \emph{Right}: %
                Posterior constraints on the potential geometry. %
                The lower-left panel shows the full parameter set $(q, \phi, x, y)$; the upper-right shows the $(q, \phi)$ subspace with \gls{com} fixed to the galaxy's \gls{col}, used to colour the stream in the left panel. %
                The \gls{com} is well constrained in the lower-left panel. %
                In the upper-right panel, where the \gls{com} is fixed, the flattening $q$ is constrained to be $q > 0.8$ for $\phi\in[-110^\circ,-40^\circ]$ and $q \gtrsim 0.5$ for $\phi\in[-40^\circ,70^\circ]$. %
            \label{fig:102014777-NEG780275758555920353-constraints}}
        \end{figure}

        The host in \cref{fig:102014777-NEG780275758555920353-constraints} has a bright concentrated core and a smooth, roughly symmetric light profile consistent with an early-type elliptical. %
        Broadband colour is uniform across the galaxy with no star-forming knots, indicating a quiescent stellar population. %
        Faint asymmetric outer material, with no visible disc or lens component, points to a post-merger origin and recent tidal interaction. %
        A single coherent stream arcs from the lower left to the upper right of the galaxy in \cref{fig:102014777-NEG780275758555920353-constraints}, partially overlapping the host in projection. %
        It remains smooth and narrow along its visible length, consistent with dynamically cold debris from a dwarf-galaxy satellite. %
        In projection, the stream wraps around the host in an approximately full \ang{360} spiral at an average radius of about $1.5$ times the galaxy's outer light radius. %

        The corner plot in \cref{fig:102014777-NEG780275758555920353-constraints} shows the posterior distribution over the host halo parameters inferred from the stream's morphology. %
        The \gls{com} is relatively well constrained, with the highest-likelihood region forming a compact blob within approximately $[-100, 20]$ pixels in both $x$ and $y$, inside the stream's concavity. %
        This is consistent with the galaxy's \gls{col}, as expected for an apparently relaxed system. %
        The flattening $q$ and orientation angle $\phi$ of the projected potential are more weakly constrained, and most of the parameter space remains consistent with a spherical halo ($q \approx 1$). %
        Very flattened halos are allowed only at certain orientations, particularly between \ang{-50} and \ang{0}, where the tight lower-left loop provides the strongest curvature information. %
        The colouring of the track highlights this region as the most constraining, while many other halo shapes still produce acceleration vectors aligned with the observed stream path. %
        
        Fixing the \gls{com} to the galaxy's \gls{col} gives a posterior over halo flattening and orientation that differs in shape but remains broadly consistent with the free-centre case (see upper right panel of \cref{fig:102014777-NEG780275758555920353-constraints}). %
        With the \gls{com} fixed, flattening is bounded from below across all orientation angles, disfavouring $q \lesssim 0.5$. %
        Spherical configurations remain allowed, but strongly flattened models are no longer supported, especially for orientation angles $\phi \in [\ang{-90},\ang{-40}], [\ang{70},\ang{90}]$, where they are difficult to reconcile with the observed loop morphology. %


    \subsection[II) 102014782:-825268920554154517]{II) \texttt{102014782:$-$825268920554154517}}\label{app:all_streams:102014782_NEG825268920554154517}

        \begin{figure}[htbp]
            \centering
            \includegraphics[width=0.8\linewidth]{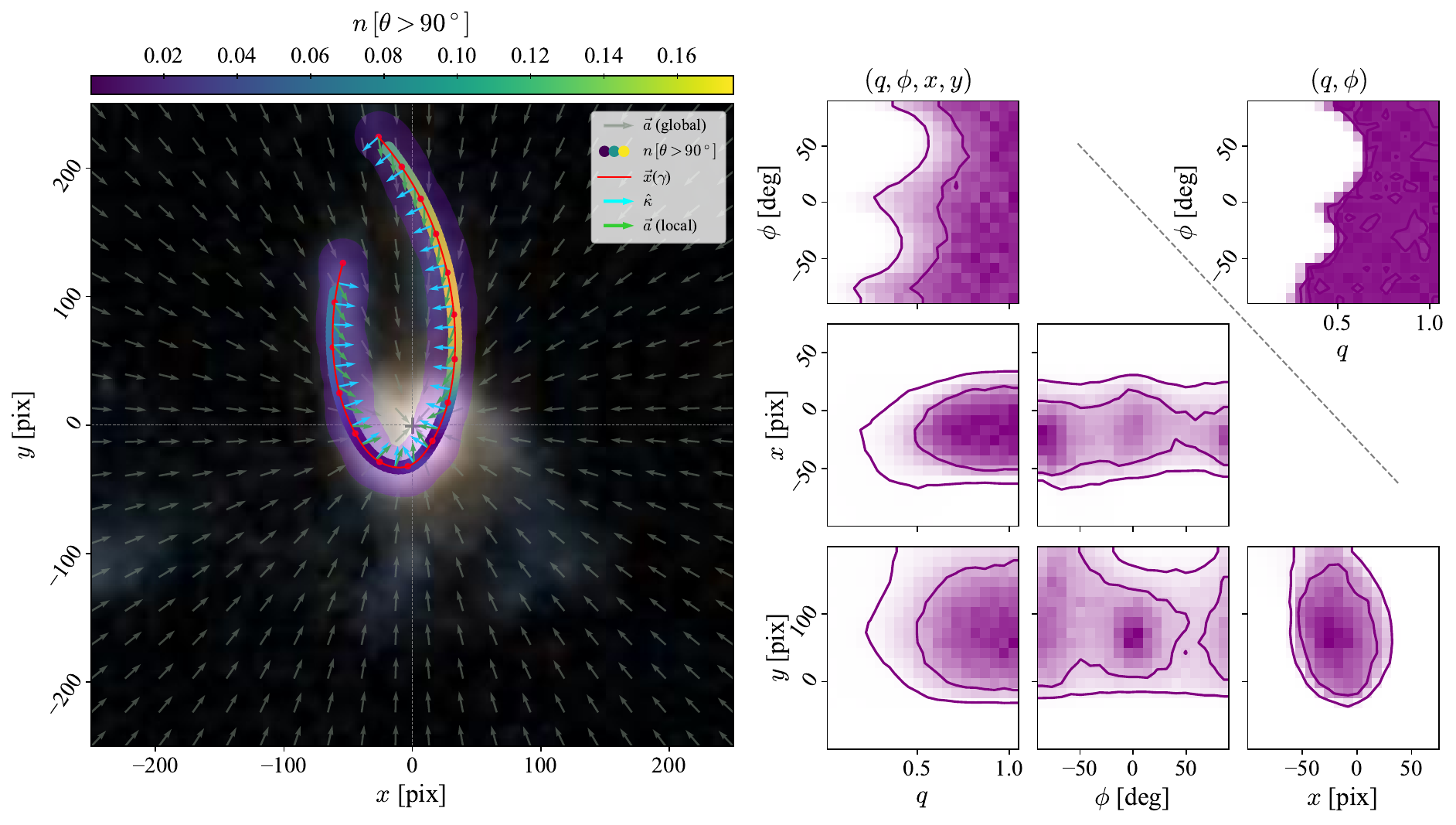}
            \caption{%
                Cutout~II: \texttt{102014782:$-$825268920554154517} and geometry
                constraints. %
                \emph{Left}: %
                Same as \cref{fig:102014777-NEG780275758555920353-constraints}. %
                \emph{Right}: %
                Posterior constraints on the potential geometry. %
                The lower-left panel shows the full parameter set $(q, \phi, x, y)$; the upper-right shows the $(q, \phi)$ subspace with \gls{com} fixed to the galaxy's \gls{col}, used to colour the stream in the left panel. %
                The \gls{com} is well constrained. %
                When fixed, the flattening $q$ is constrained to be $q \gtrsim 0.5$. %
            \label{fig:102014782-NEG825268920554154517-constraints}}
        \end{figure}

        The host in \cref{fig:102014782-NEG825268920554154517-constraints} has a bright compact core and an irregular, asymmetric outer light profile typical of a morphologically disturbed system. %
        Broadband colour varies from redder central regions to bluer diffuse outer material, indicating a mixed stellar population with no visible disc or spiral arms. %
        Diffuse light extending asymmetrically below the host points to recent disruption from a minor merger; the arcsinh stretch amplifies these features, and the likelihood constraints suggest a less disturbed system than it appears. %
        \texttt{102014782:$-$825268920554154517} features one or possibly two narrow, vertically extended stream-like arms emerging from the top side of the galaxy in \cref{fig:102014782-NEG825268920554154517-constraints}. %
        These arms curve upward from the host and remain spatially coherent over much of the image plane. %
        Despite the disturbed host morphology, they are smooth and well defined, consistent with dynamically cold debris. %
        The stream also appears bluer than the host galaxy in the false-colour image, suggesting an origin distinct from the main stellar population, for example in a merging dwarf galaxy. %

        The corner plot in \cref{fig:102014782-NEG825268920554154517-constraints} shows the posterior distribution over the host halo parameters inferred from the stream's morphology. %
        The \gls{com} is relatively well constrained, with the highest-likelihood region at approximately $x \in [-50, 20]$ and $y \in [-10, 150]$\,\unit{\pix}, within the concave interior defined by the upward-curving stream arms. %
        This posterior excludes most of the galaxy's luminous extent, but still includes the \gls{col} at $(x, y) = (5, -5)$\,\unit{\pix}, consistent with a roughly aligned baryonic and dark matter distribution. %
        The flattening $q$ and orientation angle $\phi$ are also constrained, with the posterior remaining consistent with a spherical halo ($q \approx 1$) across most angles. %
        For $\phi \in [\ang{20},\ang{90}]$, the constraints tighten and disfavour significantly flattened potentials. %
        This is driven primarily by the long, straighter rightward stream arm, whose length, proximity to the host, and low curvature make it especially constraining for those halo orientations. %

        When the \gls{com} is fixed to the galaxy's \gls{col}, the posterior over flattening and orientation closely resembles the free-centre case, but with slightly sharper bounds (see upper right panel of \cref{fig:102014782-NEG825268920554154517-constraints}). %
        The posterior then places a stronger lower bound on the flattening, broadly disfavouring values of $q \lesssim 0.5$, while spherical configurations remain allowed across all orientations. %


    \subsection[III) 102018245:-851003014513514442]{III) \texttt{102018245:$-$851003014513514442}}\label{app:all_streams:102018245_NEG851003014513514442}

        \begin{figure}[htbp]
            \centering
            \includegraphics[width=0.8\linewidth]{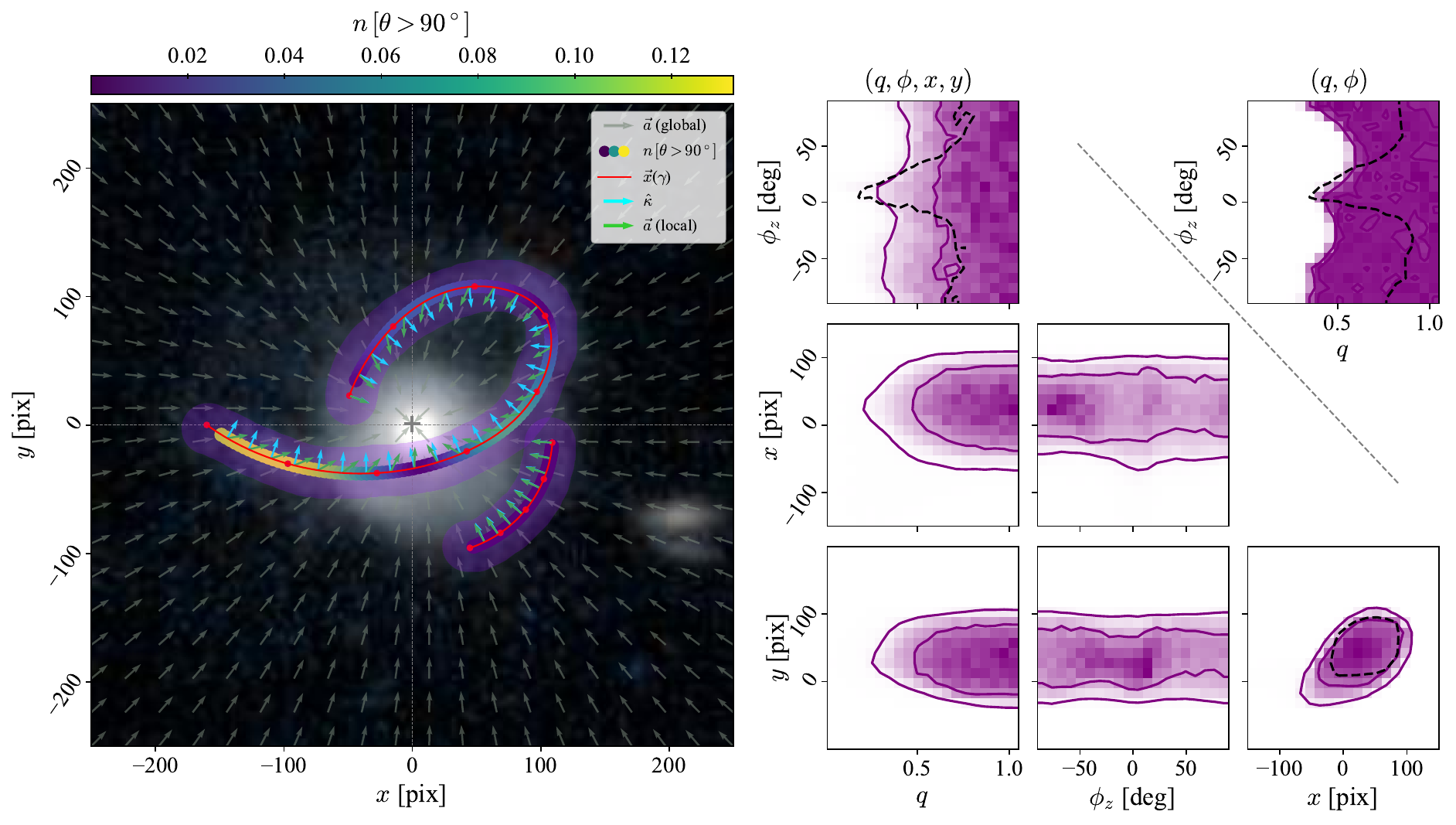}
            \caption{%
                Cutout~III: \texttt{102018245:$-$851003014513514442} and geometry
                constraints. %
                \emph{Left}: %
                Same as \cref{fig:102014777-NEG780275758555920353-constraints}. %
                \emph{Right}: %
                Posterior constraints on the potential geometry. %
                The lower-left panel shows the full parameter set $(q, \phi, x, y)$; the upper-right shows the $(q, \phi)$ subspace with \gls{com} fixed to the galaxy's \gls{col}, used to colour the stream in the left panel. %
                The \gls{com} is well constrained. %
                When fixed, the flattening $q$ is constrained to be $q > 0.5$. %
                The dashed black lines show constraints from disconnected annotations (\cref{fig:zooniverse-annotatation-process}) whose differing end-point curvatures cause tighter limits at some orientations. %
            \label{fig:102018245-NEG851003014513514442-constraints}}
        \end{figure}

        The host in \cref{fig:102018245-NEG851003014513514442-constraints} is a bright spheroidal galaxy with a smooth, centrally concentrated light profile and pronounced bulge consistent with an elliptical or face-on S0. %
        Uniform broadband colour distinguishes the host from the unrelated, redder neighbouring galaxy to the lower right. %
        Faint diffuse material in the upper right, along with a plume-like feature in the lower-left body of the galaxy, point to tidal debris from a disrupted satellite. %
        Several stream segments encircle the host galaxy in \cref{fig:102018245-NEG851003014513514442-constraints} and together appear to form a single coherent stellar stream. %
        The stream is narrow and smoothly curved, with a prominent left arc, a loop above the galaxy, and another arc to the lower right. %
        It passes in front of the galaxy below the central bulge and appears to complete a full loop, although the segment behind the galaxy is not directly visible. %
        Its false-colour hue is similar to that of the host, making colour-based separation difficult, but its overall morphology is consistent with the tidal disruption of a satellite on a regular orbit. %

        The corner plot in \cref{fig:102018245-NEG851003014513514442-constraints} shows the posterior distribution over the host halo parameters inferred from the stream's morphology. %
        The \gls{com} is well constrained, with the highest-likelihood region enclosed within the top loop of the stream, just above the left arc. %
        This region includes the luminous centre of the galaxy at approximately $(x, y) \simeq (0, 0)$\,\unit{\pix} and excludes much of the visible asymmetry in the lower half of the galaxy, consistent with a centrally aligned baryonic and dark matter distribution. %
        The flattening $q$ is also constrained, with the posterior disfavouring values below $q \approx 0.5$ across all orientation angles. %
        However, the broadly looping stream nearly traces a great-circle path in projection, which limits the constraining power on $q$ and $\phi$. %
        At this viewing angle, many halo geometries produce similar projected acceleration fields and therefore remain consistent with the observed stream shape. %

        Fixing the \gls{com} to the galaxy's \gls{col} gives a posterior over halo flattening and orientation that closely matches the free-centre case, but with slightly tighter bounds (see upper right panel of \cref{fig:102018245-NEG851003014513514442-constraints}). %
        Constraints strengthen marginally for orientations in the range $\phi \in [\ang{10},\ang{80}]$, while remaining similar near $\phi \in [\ang{-90},\ang{0}]$. %
        The stream's coloured track shows that the left arc is the most constraining segment for $q$ and $\phi$, with the lower-right portion contributing second-most strongly, while the segments directly in front of the galaxy and near the top of the loop dominate the \gls{com} constraint. %


    \subsection[IV) 102018668:-595264901510604722]{IV) \texttt{102018668:$-$595264901510604722}}\label{app:all_streams:102018668_NEG595264901510604722}

        \begin{figure}[htbp]
            \centering
            \includegraphics[width=0.8\linewidth]{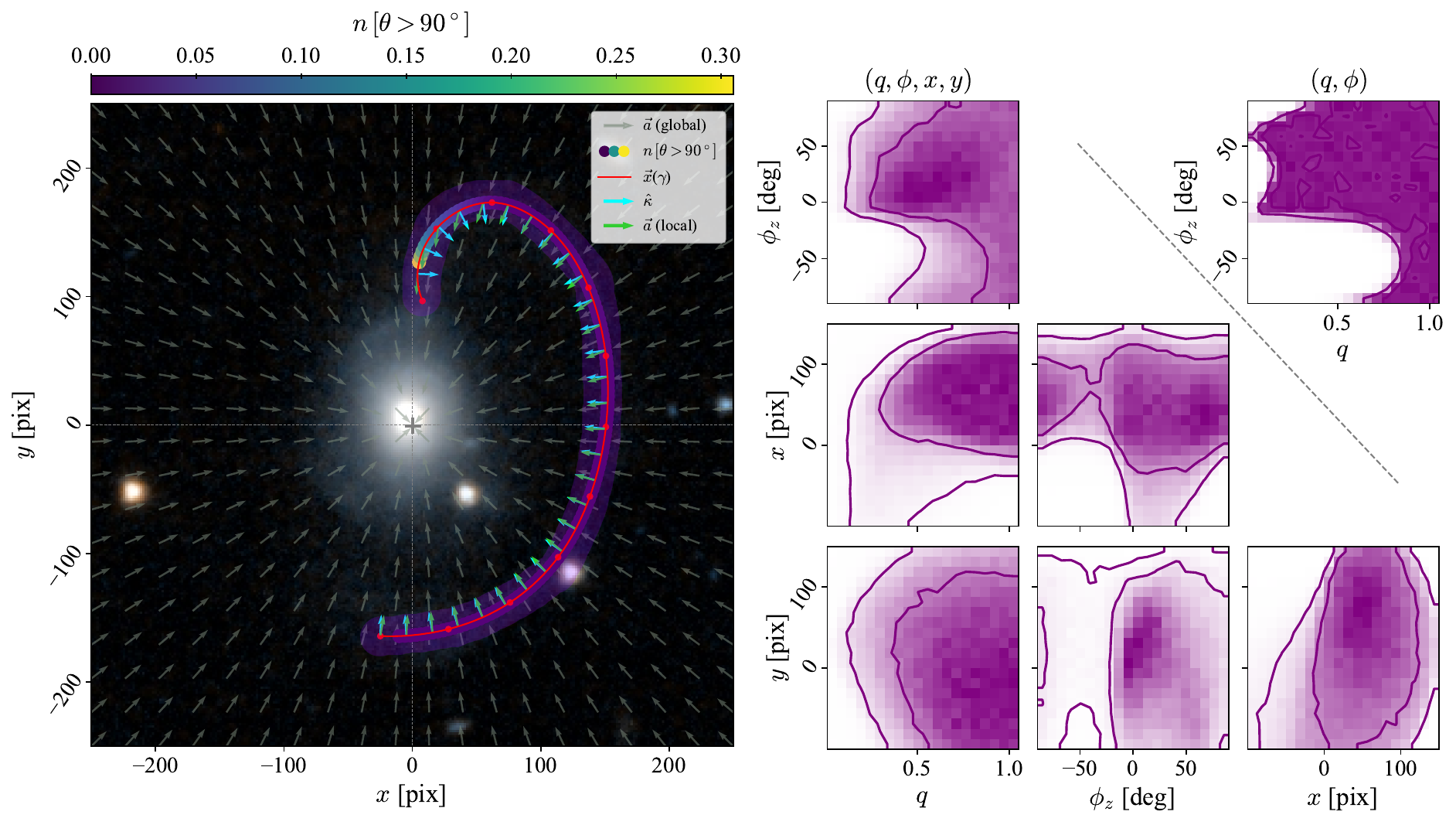}
            \caption{%
                Cutout~IV: \texttt{102018668:$-$595264901510604722} and geometry
                constraints. %
                \emph{Left}: %
                Same as \cref{fig:102014777-NEG780275758555920353-constraints}. %
                \emph{Right}: %
                Posterior constraints on the potential geometry. %
                The lower-left panel shows the full parameter set $(q, \phi, x, y)$; the upper-right shows the $(q, \phi)$ subspace with \gls{com} fixed to the galaxy's \gls{col}, used to colour the stream in the left panel. %
                The \gls{com} is lightly constrained. %
                When fixed, the flattening $q$ is strongly constrained for some $\phi$. %
            \label{fig:102018668-NEG595264901510604722-constraints}}
        \end{figure}

        The host in \cref{fig:102018668-NEG595264901510604722-constraints} is a bulge-dominated system with a bright compact core and faint, asymmetric outer structure. %
        The central bulge is slightly asymmetric and elongated toward the upper left, surrounded by a more extended component with an arc-like feature near the top that may be a spiral arm, a shell, or a remnant of a disrupted disc. %
        A luminous knot sits to the right at $(x, y) = (40, -10)$\,\unit{\pix}, and a background galaxy is visible in the lower right. %
        A very faint tidal stream extends from the top side of the galaxy in \cref{fig:102018668-NEG595264901510604722-constraints} and curves clockwise around the host. %
        Despite its low surface brightness, it remains morphologically coherent and narrow along most of its visible extent, even where it passes near a background galaxy. %
        Toward the lower left, however, it broadens into a fan and begins to lose coherence, consistent with debris that is starting to disperse at large separations. %
        Although partially embedded in the host light, the stream's shape remains informative for dynamical inference. %

        The corner plot in \cref{fig:102018668-NEG595264901510604722-constraints} shows the posterior distribution over the host halo parameters inferred from the stream's morphology. %
        In contrast to more tightly constrained systems, the \gls{com} is not well localised; the posterior spans a broad region within the large arc of the stream. %
        The flattening $q$ and orientation angle $\phi$ are likewise only weakly constrained, although highly flattened models are disfavoured, with $q \gtrsim 0.3$ overall and $q \gtrsim 0.5$ at orientations $\phi \in [\ang{-90},\ang{-30}], [\ang{50},\ang{90}]$. %
        This weak constraint arises because the stream closely follows a great circle in projection and is visible on only one side of the host. %
        In that geometry, the curvature aligns naturally with the radial acceleration expected for a wide range of potential shapes and orientations, so many halo models remain consistent with the observed morphology. %

        Fixing the \gls{com} to the galaxy's \gls{col} gives a posterior over flattening and orientation that differs noticeably from the free-\gls{com} case (see upper right panel of \cref{fig:102018668-NEG595264901510604722-constraints}). %
        This indicates that the full-parameter constraints were driven largely by the poorly constrained \gls{com}, which is strongly correlated with $q$ and $\phi$. %
        With a fixed \gls{com}, the distribution remains consistent with a spherical halo, but shows strong lower bounds at certain orientations, particularly $q \gtrsim 0.8$ for $\phi \in [\ang{-80},\ang{-10}]$, while remaining broad elsewhere. %


    \subsection[V) 102019123:-569554074506844623]{V) \texttt{102019123:$-$569554074506844623}}\label{app:all_streams:102019123_NEG569554074506844623}

        \begin{figure}[htbp]
            \centering
            \includegraphics[width=0.8\linewidth]{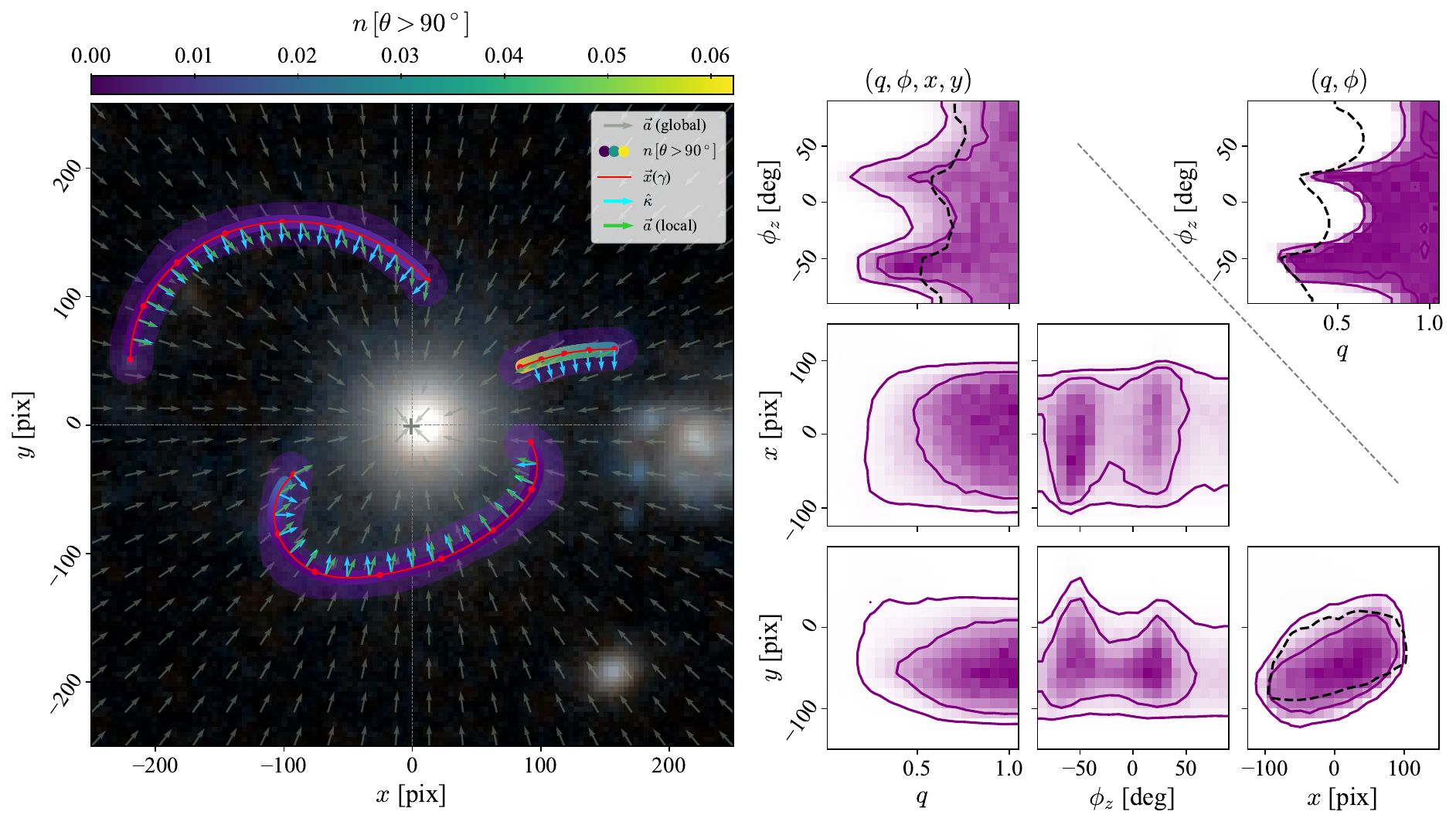}
            \caption{%
                Cutout~V: \texttt{102019123:$-$569554074506844623} and geometry
                constraints. %
                \emph{Left}: %
                Same as \cref{fig:102014777-NEG780275758555920353-constraints}.
                \emph{Right}: %
                Posterior constraints on the potential geometry. %
                The lower-left panel shows the full parameter set $(q, \phi, x, y)$; the upper-right shows the $(q, \phi)$ subspace with \gls{com} fixed to the galaxy's \gls{col}, used to colour the stream in the left panel. %
                The \gls{com} is well constrained. %
                When fixed, the flattening $q$ is strongly constrained except for $\phi \sim \ang{-60}, \ang{20}$. %
            \label{fig:102019123-NEG569554074506844623-constraints}}
        \end{figure}

        The host in \cref{fig:102019123-NEG569554074506844623-constraints} is a quiescent, bulge-dominated elliptical with a smooth, centrally concentrated light profile, uniform broadband colour, and no visible disc, bar, or shell features. %
        Faint outer material traces the tidal stream rather than the host morphology. %
        A second galaxy with strong substructure and significant colour variation lies just to the right, though photometric redshifts cannot confirm a physical association. %
        A stellar stream encircles the host galaxy in \cref{fig:102019123-NEG569554074506844623-constraints}, with multiple visible segments tracing a large loop. %
        The brightest clumps lie at approximately $(x, y) \simeq (150, -200)$\,\unit{\pix} and $(225, -20)$\,\unit{\pix}, and may be disrupted star clusters or the progenitor itself. %
        The stream arcs over the galaxy, continues from the right, curves beneath the host, and appears to reconnect on the left side. %
        Another stream-like feature extends from the diffuse material on the upper right toward the nearby galaxy, but does not appear to connect physically. %
        The stream is moderately thick but morphologically coherent, consistent with tidal debris from a disrupted dwarf galaxy. %

        The corner plot in \cref{fig:102019123-NEG569554074506844623-constraints} shows the posterior distribution over the host halo parameters inferred from the stream's morphology. %
        The \gls{com} is reasonably well constrained, with the highest-likelihood region near the galaxy's centre. %
        Where offsets are favoured, they point toward the nearby galaxy; if the two systems are associated this may be meaningful, and if not the result remains consistent with a centrally aligned potential. %
        
        Unlike many other systems in the sample, the marginalised posteriors of $\phi$ versus $x, y$ do not show the usual preference for solutions rotated by \ang{45} relative to one another. %
        Instead, both spatial dimensions prefer similar flattening angles. %
        The flattening $q$ and orientation angle $\phi$ are constrained only at specific orientations. %
        For $\phi \in [\ang{-90},\ang{-70}], [\ang{30},\ang{90}]$, the potential must be near-spherical; at other angles, less spherical configurations remain allowed. %
        This behaviour persists when the centre is fixed to the galaxy's centre: spherical halos are always permitted, and $q$ is tightly constrained only for specific orientations. %
        The joint $q$, $x$, and $y$ posteriors further show that centred potentials admit greater variation in shape, while off-centred geometries increasingly favour spherical solutions. %


    \subsection[VI) 102021493:-583998851481016349]{VI) \texttt{102021493:$-$583998851481016349}}\label{app:all_streams:102021493_NEG583998851481016349}

        \begin{figure}[htbp]
            \centering
            \includegraphics[width=0.8\linewidth]{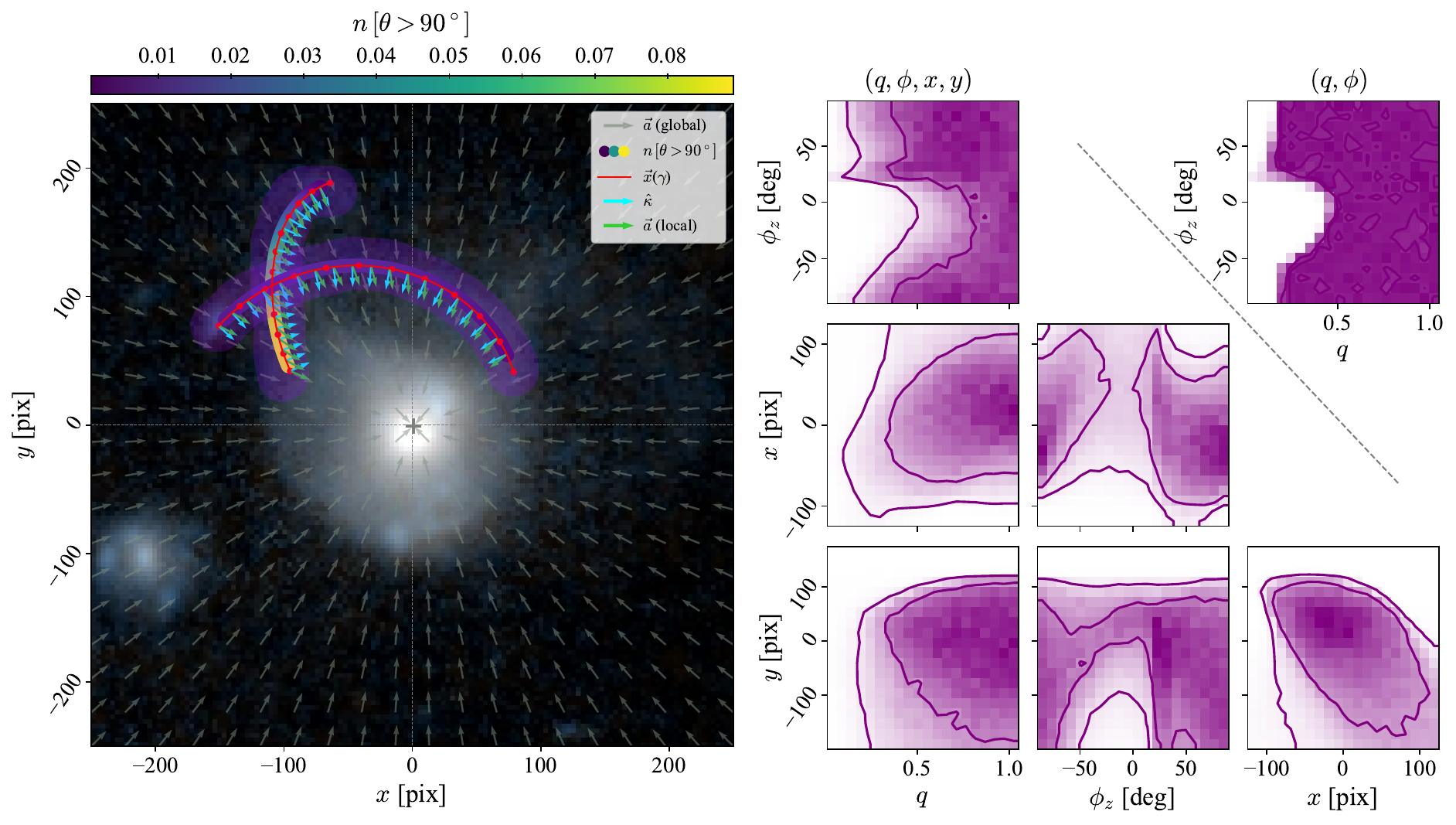}
            \caption{%
                Cutout~VI: \texttt{102021493:$-$583998851481016349} and geometry
                constraints. %
                \emph{Left}: %
                Same as \cref{fig:102014777-NEG780275758555920353-constraints}.
                \emph{Right}: %
                Posterior constraints on the potential geometry. %
                The lower-left panel shows the full parameter set $(q, \phi, x, y)$; the upper-right shows the $(q, \phi)$ subspace with centre of mass fixed to the galaxy's \gls{col}, used to colour the stream in the left panel. %
            \label{fig:102021493-NEG583998851481016349-constraints}}
        \end{figure}

        The host in \cref{fig:102021493-NEG583998851481016349-constraints} is a bulge-dominated system with a smooth core, no visible disc or spiral arms, and a disturbed asymmetric exterior with prominent luminosity and colour clumps near $(x, y) = (20, 10)$\,\unit{\pix} and $(-10, -100)$\,\unit{\pix}. %
        Faint arcs near the latter clump suggest a disrupting satellite or ongoing interaction. %
        A further clump near $(-150, 80)$\,\unit{\pix} may be a surviving satellite or stripped remnant. %
        A resolved galaxy in the lower left shows clumpy structure and varied colours, with unclear physical association with the primary. %
        
        The host galaxy in \cref{fig:102021493-NEG583998851481016349-constraints} is surrounded by broad arcing features that extend mainly from the top of the galaxy. %
        These structures are more diffuse and irregular than typical stellar streams and lack the thin, coherent morphology of dynamically cold tidal debris, although portions of the top arc may still follow ballistic trajectories. %
        Two features are annotated: the arc originating from the clump near $(-150, 80)$\,\unit{\pix} and extending to the top right of the galaxy, and another segment from the top left. %
        An additional unannotated loop extends from the top right to the bottom right of the image. %
        This system is plausibly the aftermath of a recent accretion or merger, in which some debris remains dynamically cold while other material has phase-mixed or been heated. %
        It may still yield useful constraints on the gravitational potential, but only if the reliably stream-like regions can be isolated. %


    \subsection[VII) 102022998:-787613526465934569]{VII) \texttt{102022998:$-$787613526465934569}}\label{app:all_streams:102022998_NEG787613526465934569}

        \begin{figure}[!h]
            \centering
            \includegraphics[width=0.8\linewidth]{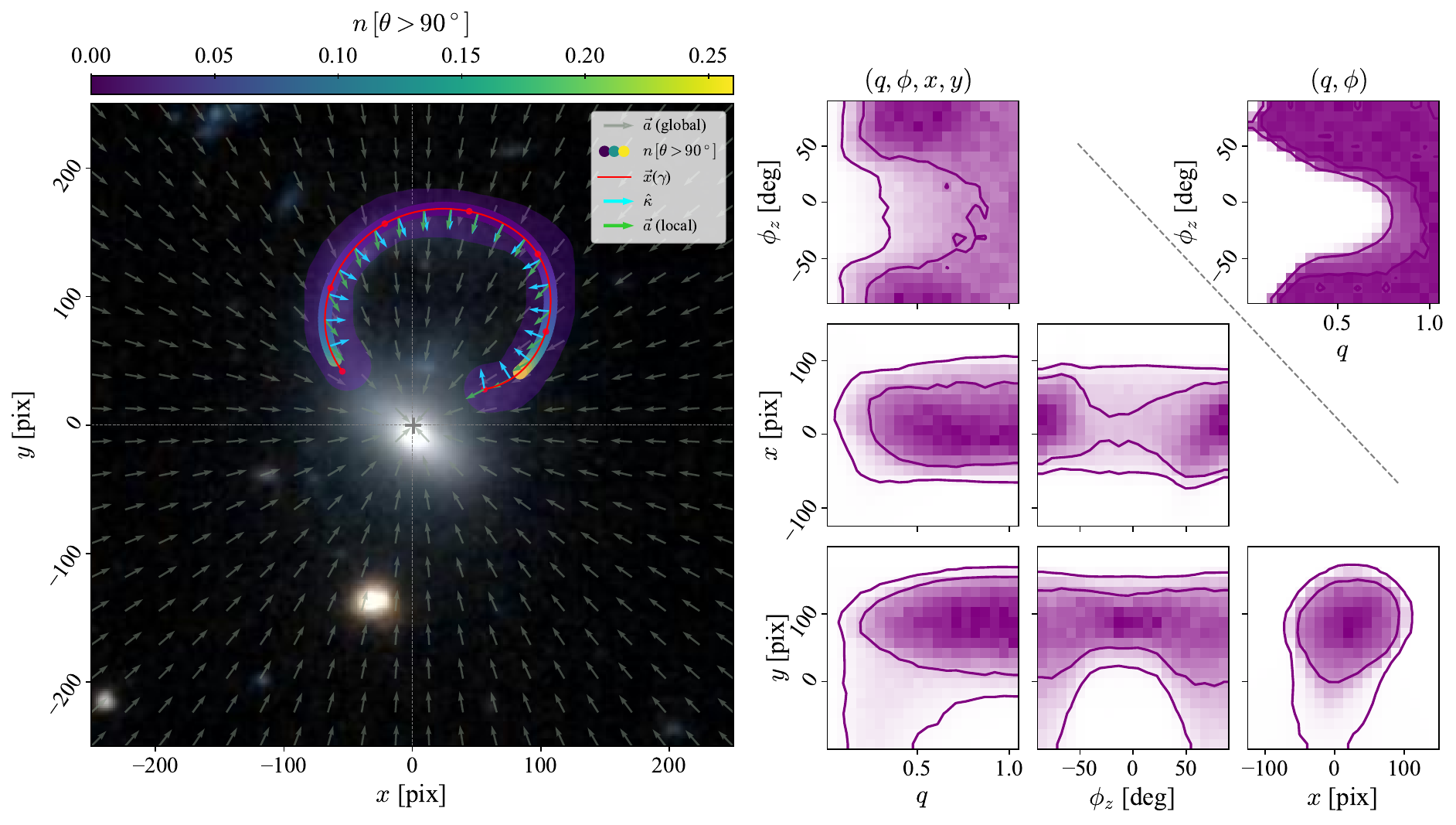}
            \caption{%
                Cutout~VII: \texttt{102022998:$-$787613526465934569} and geometry
                constraints. %
                \emph{Left}: %
                Same as \cref{fig:102014777-NEG780275758555920353-constraints}.
                \emph{Right}: %
                Posterior constraints on the potential geometry. %
                The lower-left panel shows the full parameter set $(q, \phi, x, y)$; the upper-right shows the $(q, \phi)$ subspace with centre of mass fixed to the galaxy's \gls{col}, used to colour the stream in the left panel. %
            \label{fig:102022998-NEG787613526465934569-constraints}}
        \end{figure}

        The host in \cref{fig:102022998-NEG787613526465934569-constraints} has a bright nucleus and elongated light profile consistent with a bar or inclined disc, and a smooth, symmetric inner structure with no spiral arms. %
        Small luminous features at $(x, y) = (0, 50)$\,\unit{\pix} and $(-100, -40)$\,\unit{\pix} may be host-associated or projected background sources. %
        A larger, partially resolved system near $(-20, -130)$\,\unit{\pix} is notably redder. %
        A diffuse, slightly elliptical outer halo may be phase-mixed tidal debris. %
        A moderately broad stellar stream emerges from the host galaxy in \cref{fig:102022998-NEG787613526465934569-constraints}, arcing over the top and reconnecting symmetrically on the opposite side to form a `kettle bell' shape in projection. %
        Although not especially narrow, it remains smooth and coherent, with nearly constant curvature consistent with dynamically cold tidal debris from a disrupted dwarf galaxy. %
        A small clump near $(-50, 120)$\,\unit{\pix} may be the remnant of the progenitor. %
        The stream may pass through the inner regions of the host, raising the possibility of a close-in merger rather than a distant wrapping stream, but its symmetry still makes it a promising tracer of the host potential. %

        The corner plot in \cref{fig:102022998-NEG787613526465934569-constraints} shows the posterior distribution over the host halo parameters inferred from the stream's morphology. %
        The \gls{com} is relatively well constrained, with the highest-likelihood region offset slightly from the galaxy's centre toward the stream arc, although the $1\;\sigma$ posterior remains consistent with the photometric centre. %
        The outermost part of the stream is most constraining: as it curves back toward the galaxy it becomes nearly radial in projection, strongly influencing the inferred acceleration direction. %

        The flattening $q$ and orientation angle $\phi$ are also constrained. %
        For some orientations the posterior favours near-spherical halos with $q \gtrsim 0.7$, while more flattened configurations are permitted when the halo orientation aligns closely with the curvature of the stream loop, $\phi \in [-90^\circ,-40^\circ] \cup [25^\circ,90^\circ]$ (i.e.\ $[25^\circ,140^\circ]$ modulo the domain convention). %
        This pattern is also visible in the fixed-centre posterior (upper-right panel of \cref{fig:102022998-NEG787613526465934569-constraints}), where the end of the annotation prefers flatter models but near-spherical configurations remain highly consistent with the data. %


    \subsection[VIII) 102030408:-732320008395115905]{VIII) \texttt{102030408:$-$732320008395115905}}\label{app:all_streams:102030408_NEG732320008395115905}

        \begin{figure}[ht]
            \centering
            \includegraphics[width=0.8\linewidth]{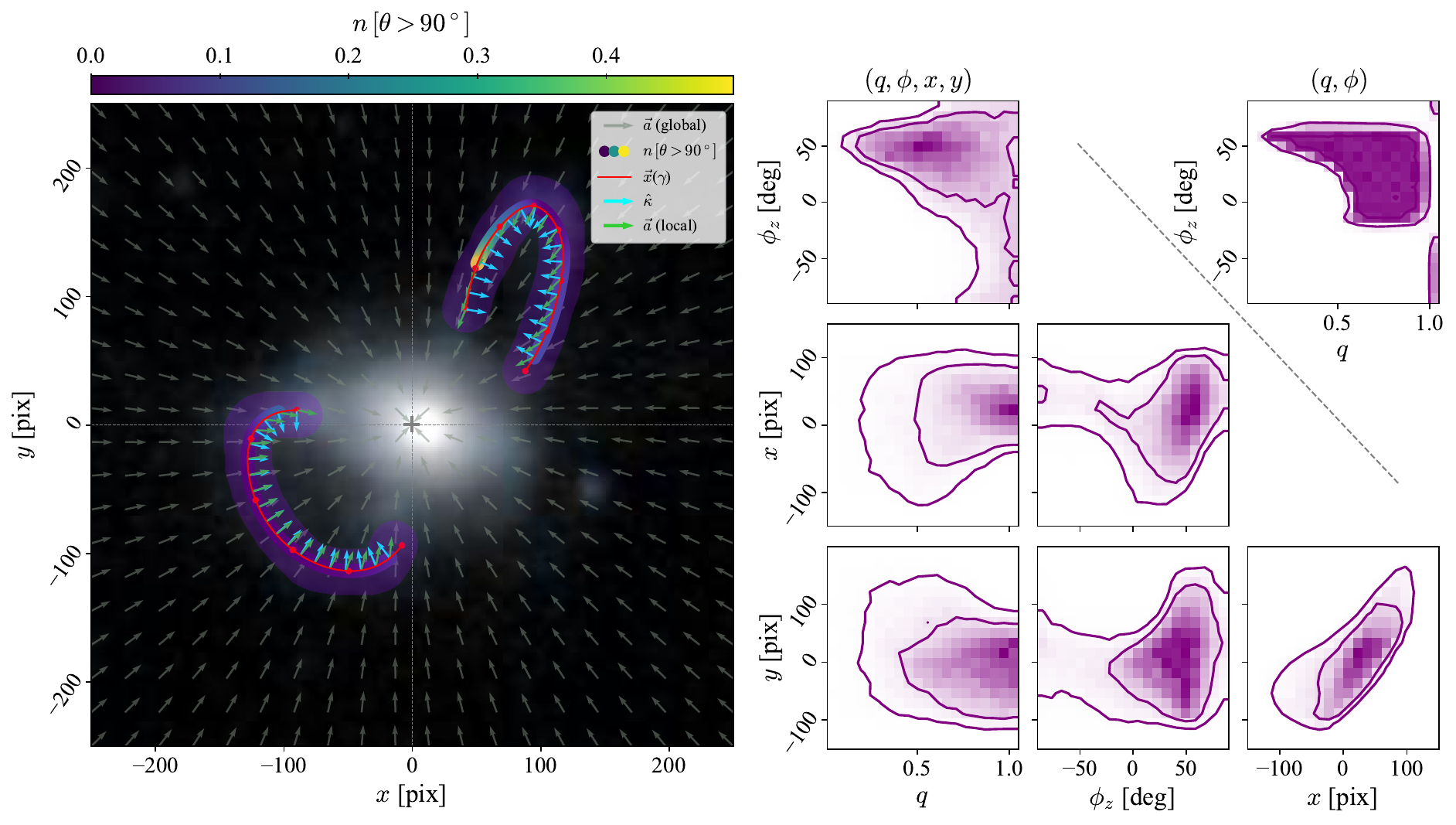}
            \caption{%
                Cutout~VIII: \texttt{102030408:$-$732320008395115905} and geometry
                constraints. %
                \emph{Left}: %
                Same as \cref{fig:102014777-NEG780275758555920353-constraints}.
                \emph{Right}: %
                Posterior constraints on the potential geometry. %
                The lower-left panel shows the full parameter set $(q, \phi, x, y)$; the upper-right shows the $(q, \phi)$ subspace with centre of mass fixed to the galaxy's \gls{col}, used to colour the stream in the left panel. %
            \label{fig:102030408-NEG732320008395115905-constraints}}
        \end{figure}

        The host in \cref{fig:102030408-NEG732320008395115905-constraints} is a spheroid-dominated early-type galaxy with a bright, centrally concentrated core and no visible spiral structure. %
        Mildly asymmetric outer isophotes and faint outer extensions suggest a disturbed or inclined disc component and past tidal interactions. %
        Two prominent tidal features extend from the host galaxy in \cref{fig:102030408-NEG732320008395115905-constraints}: a broad, fan-like structure to the lower left and a narrower, more coherent arm in the upper right. %
        The latter is the stronger candidate for curvature-based analysis, but based on their curvature and continuity we interpret both features as segments of a single stream-like feature. %
        In this picture, the feature arcs down the left side, passes behind the galaxy, re-emerges on the right, and loops upward before crossing in front of the host near the top, where it becomes harder to trace. %
        A third segment emerges from a compact clump near $(x, y) = (150, -50)$\,\unit{\pix}, which may be a star cluster. %
        This arm extends toward $(100, -100)$\,\unit{\pix} before fading into the diffuse background and is likely unrelated to the main stream-like feature. %

        The corner plot in \cref{fig:102030408-NEG732320008395115905-constraints} shows the posterior distribution over the host halo parameters inferred from the tidal features' morphology. %
        The \gls{com} is constrained to lie within the region enclosed by the arcs, including the tighter upper loop, and this region fully contains the galaxy's luminous centre. %
        Both annotated segments support this constraint, and the unmodelled short feature near $(150, -50)$\,\unit{\pix} is broadly consistent with the same centre. %

        The flattening $q$ is tightly constrained, with $q \approx 1$ for all angles except near $\phi \approx \ang{50}$, where significantly flatter models remain allowed. %
        The joint $\phi$-$x$ marginal shows that preferred orientations are those that align the potential with the arcs. %
        When the \gls{com} is fixed to the galaxy's \gls{col}, this trend becomes more pronounced. %
        The posterior then favours orientations tilted counter-clockwise from the $x$-axis, roughly matching the tighter loop, and spherical halos are no longer preferred. %
        However, this may partly reflect annotation uncertainty: the stream likely continues in front of the galaxy as a longer, straighter segment, which would reduce the apparent inward curvature at the end and make the track more consistent with a spherical halo. %
        This again highlights the value of ensemble annotations, which would allow us to marginalise over annotation noise and produce a statistically robust track. %


    \subsection[IX) 102032666:-674918847372942403]{IX) \texttt{102032666:$-$674918847372942403}}\label{app:all_streams:102032666_NEG674918847372942403}

        \begin{figure}[htbp]
            \centering
            \includegraphics[width=0.8\linewidth]{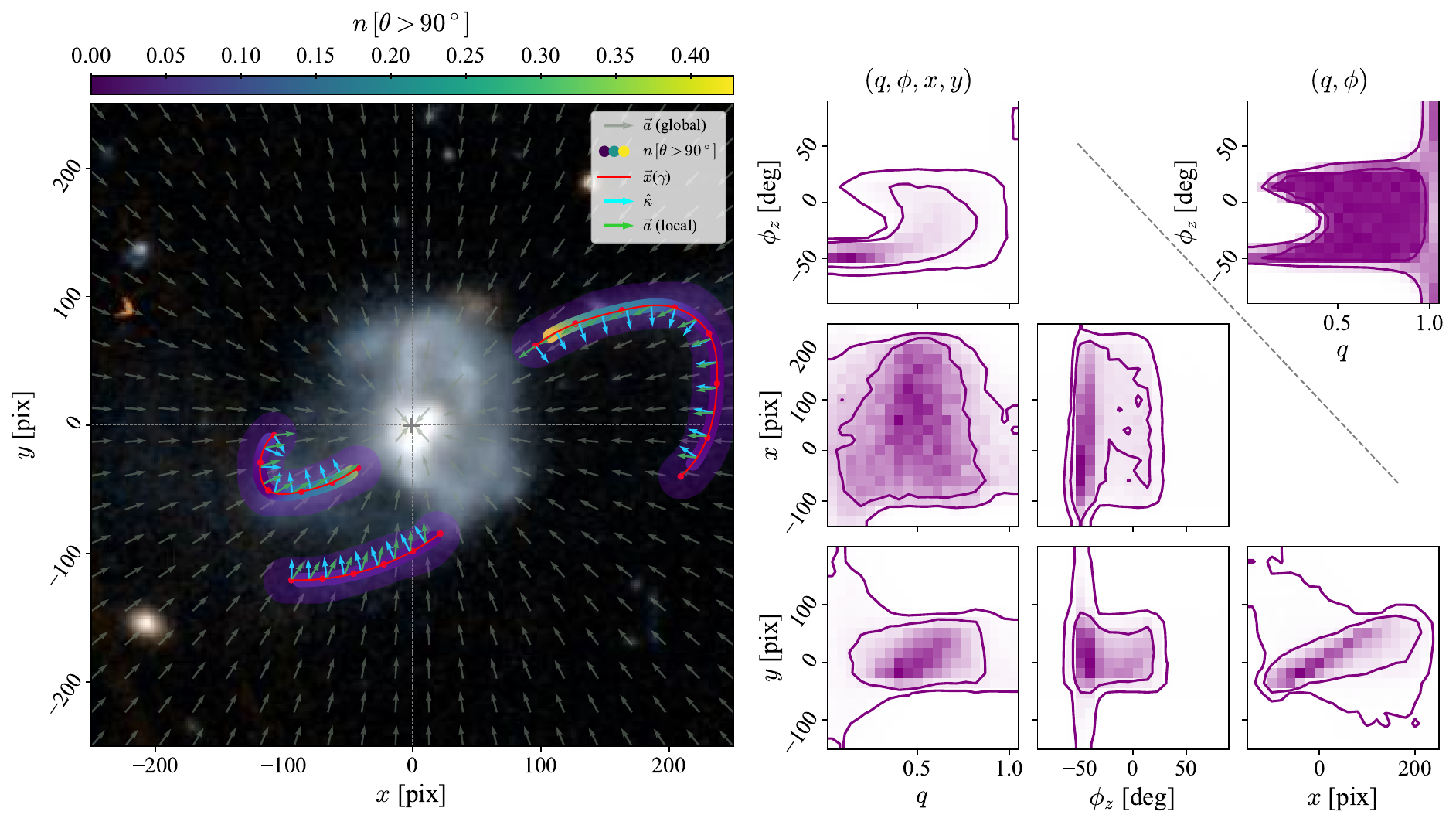}
            \caption{%
                Cutout~IX: \texttt{102032666:$-$674918847372942403} and geometry
                constraints. %
                \emph{Left}: %
                Same as \cref{fig:102014777-NEG780275758555920353-constraints}.
                \emph{Right}: %
                Posterior constraints on the potential geometry. %
                The lower-left panel shows the full parameter set $(q, \phi, x, y)$; the upper-right shows the $(q, \phi)$ subspace with centre of mass fixed to the galaxy's \gls{col}, used to colour the stream in the left panel. %
            \label{fig:102032666-NEG674918847372942403-constraints}}
        \end{figure} 

        The host in \cref{fig:102032666-NEG674918847372942403-constraints} is a dynamically disturbed, irregular galaxy with no visible disc, spiral arms, or bar, its core embedded in an extended asymmetric light distribution consistent with a recent or ongoing merger. %
        The false-colour image shows no strong colour gradients or blue clumps, indicating little recent star formation. %
        Multiple diffuse features extend from the host galaxy in \cref{fig:102032666-NEG674918847372942403-constraints}, including segments to the left, upper right, and bottom. %
        These arcs are morphologically consistent with tidal debris from a recent accretion or merger event. %
        The lower arc may form part of a larger unannotated loop that wraps around the galaxy and connects to the luminous clump near $(x, y) = (-25, 75)$\,\unit{\pix}, while the left and upper-right segments may also belong to the same tidal structure. %
        Because these connections cannot be traced confidently across the host in projection, we conservatively restrict the stream domain to $\gamma \in [-0.9, 0.9]$ for each segment when evaluating geometric constraints. %

        The corner plot in \cref{fig:102032666-NEG674918847372942403-constraints} shows the posterior distribution over the host halo parameters inferred from the annotated stream segments. %
        The \gls{com} is constrained to lie within or between the arcs spanning the left and upper-right regions of the host, forming a narrow bar-like structure in parameter space. %
        This region includes the photometric centre of the galaxy and is consistent with a centrally aligned dark matter distribution. %
        The orientation angle $\phi$ is relatively well constrained in both the free-centre and fixed-centre cases. %
        In the free-centre model, the posterior favours orientations near \ang{-50}, and the $q$--$\phi$ marginal prefers elongated halos. %
        In the fixed-centre case, the posterior retains a high-likelihood region near \ang{-50} but also develops a mode near \ang{15} and permits spherical configurations. %

        An important feature of this system is the agreement among all three stream segments. %
        Although the on-track colouring suggests that the inner tip of the upper-right segment is locally most constraining, each segment produces a similar likelihood surface and favours the same halo flattening and orientation. %
        The segments differ mainly in how they constrain the \gls{com}, since each anchors the potential in a different spatial location and only jointly determines the $\Delta x - \Delta y$ posterior. %

        This system departs most strongly from our modelling assumptions, with dynamical disturbance that a smooth logarithmic halo does not capture. %
        We note, however, that the arcsinh stretch amplifies the apparent severity of these perturbations, which are real but less dramatic than they appear. %
        The dynamical complexity is nonetheless reflected in the likelihood contours, which are less consistent with axisymmetry than those of more relaxed systems. %


    \subsection[X) 102046112:-526944083265011738]{X) \texttt{102046112:$-$526944083265011738}}\label{app:all_streams:102046112_NEG526944083265011738}

        \begin{figure}[htbp]
            \centering
            \includegraphics[width=0.8\linewidth]{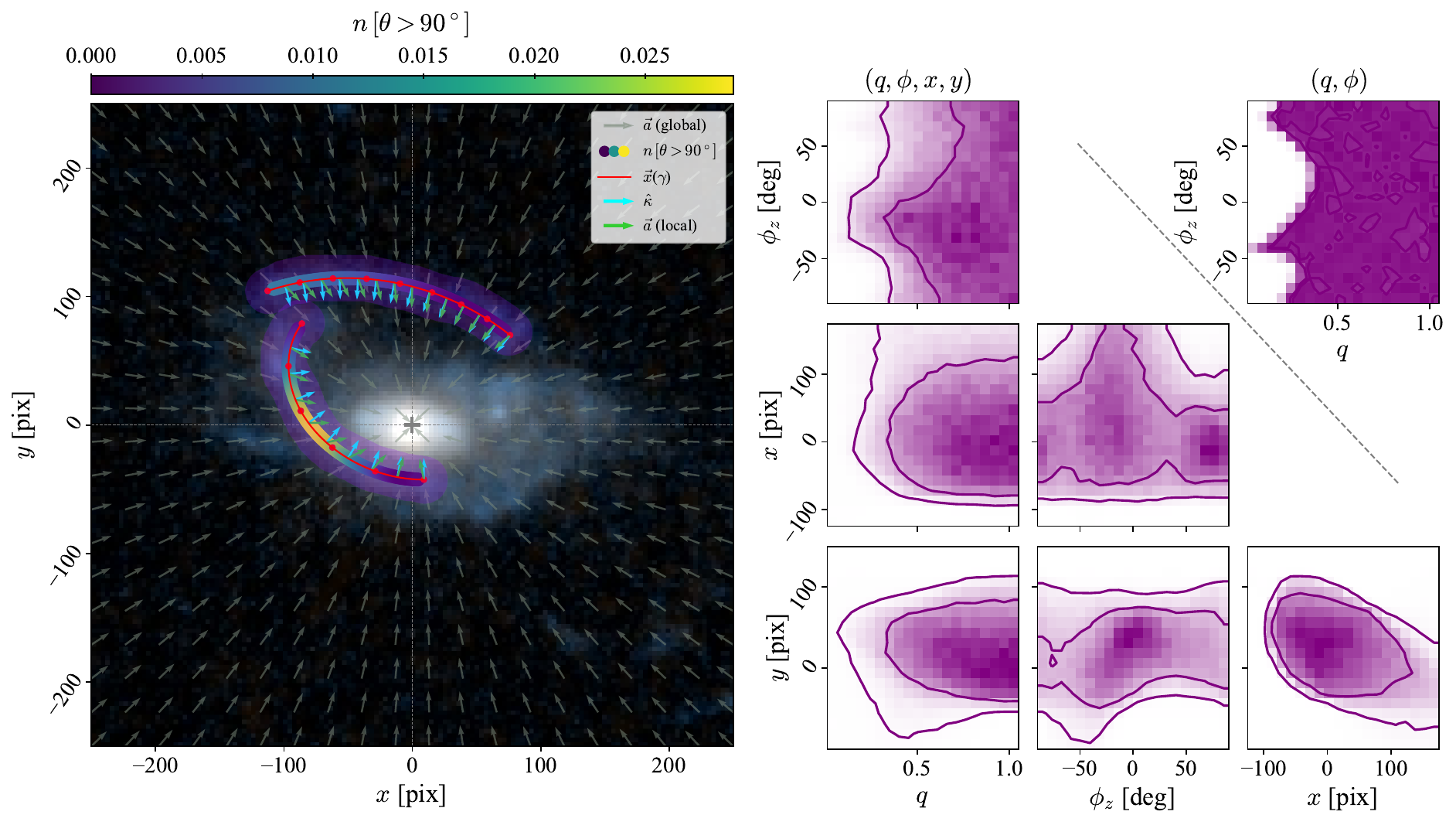}
            \caption{%
                Cutout~X: \texttt{102046112:$-$526944083265011738} and geometry
                constraints. %
                \emph{Left}: %
                Same as \cref{fig:102014777-NEG780275758555920353-constraints}.
                \emph{Right}: %
                Posterior constraints on the potential geometry. %
                The lower-left panel shows the full parameter set $(q, \phi, x, y)$; the upper-right shows the $(q, \phi)$ subspace with centre of mass fixed to the galaxy's \gls{col}, used to colour the stream in the left panel. %
            \label{fig:102046112-NEG526944083265011738-constraints}}
        \end{figure}

        The host in \cref{fig:102046112-NEG526944083265011738-constraints} is likely a disc galaxy with a luminous spheroidal core and a bright bar oriented at roughly \ang{-10} from the $x$-axis. %
        Asymmetric outer light includes a prominent clump near $(x, y) = (75, 10)$\,\unit{\pix} and extends farther to the right ($x \approx 150$) than to the left ($x \approx -100$). %
        Patchy blue and red colours suggest a mixed-age stellar population or localised dust obscuration. %
        This system features two annotated stream segments that are morphologically similar and nearly aligned, but separated by a small gap. %
        The lower segment curves around the bottom of the galaxy, arcing to the left and passing in front of the host in projection; it contains a bright stellar clump near $(-100, 50)$\,\unit{\pix} that may be the disrupting progenitor. %
        The upper segment begins just above the gap and extends over the top of the galaxy, curving to the right. %
        Although the two segments do not connect in projection, their similar thickness and curvature suggest that they may belong to the same stream. %
        If so, the break could be caused by an interaction with a structural feature such as the bar or with a massive subhalo. %

        The corner plot in \cref{fig:102046112-NEG526944083265011738-constraints} shows the posterior distribution over the host halo parameters inferred from the annotated stream segments. %
        The \gls{com} is constrained to remain consistent with the galaxy, with the allowed region bounded on the left by the stream tracks and overlapping the luminous centre. %
        The flattening is constrained to $q \gtrsim 0.5$ across most of parameter space, with more flattened halos permitted only for orientations in the narrow range $\phi \in [\ang{-40},\ang{0}]$ that align with the curvature of the upper segment. %
        These trends persist in the fixed-centre case, although the lower bound on $q$ becomes slightly less restrictive. %


    \subsection[XI) 102158269:2649560011643950805]{XI) \texttt{102158269:2649560011643950805}}\label{app:all_streams:102158269_2649560011643950805}

        \begin{figure}[!ht]
            \centering
            \includegraphics[width=0.8\linewidth]{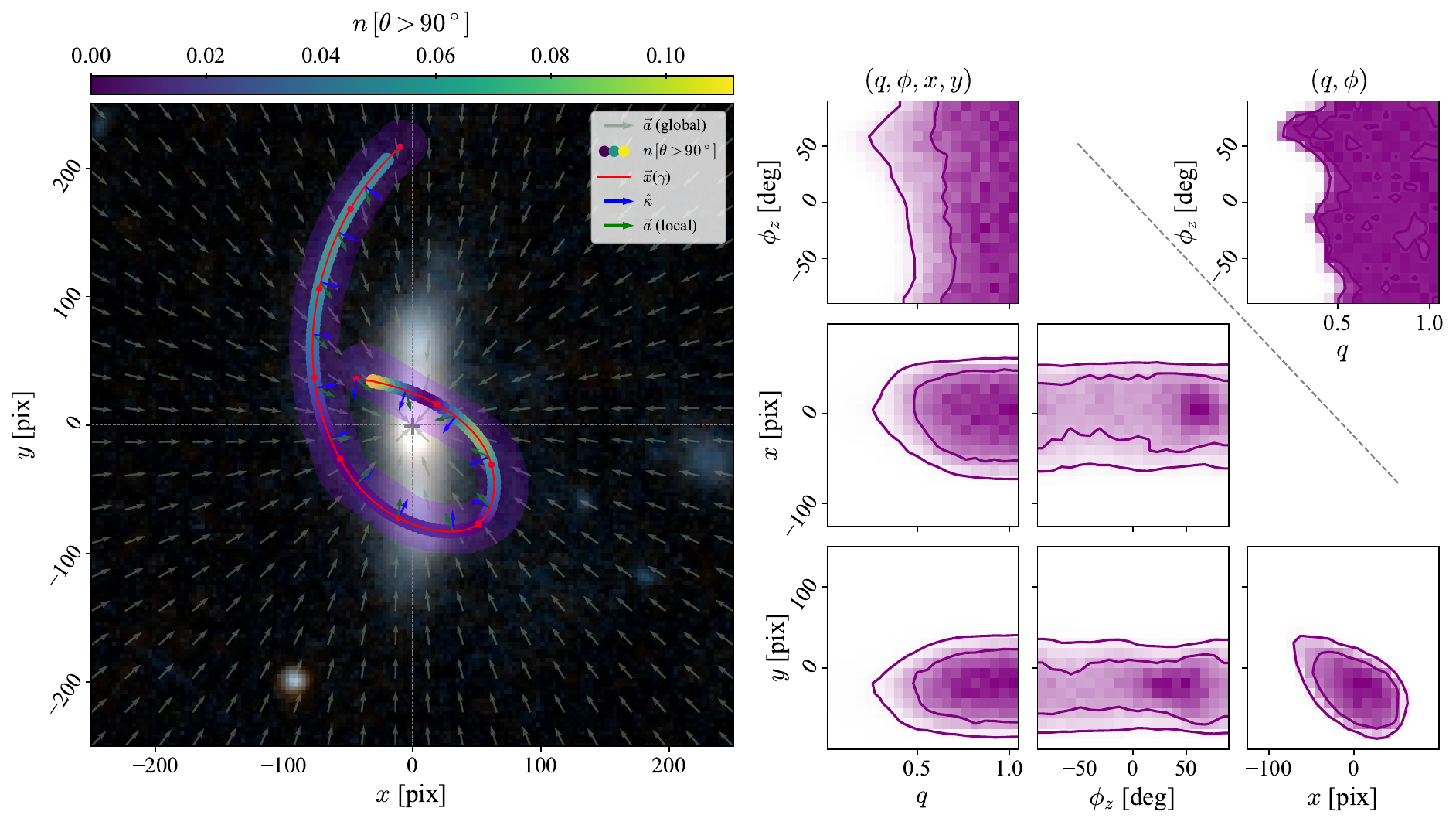}
            \caption{%
                Cutout~XI: \texttt{102158269:2649560011643950805} and geometry
                constraints. %
                \emph{Left}: %
                Same as \cref{fig:102014777-NEG780275758555920353-constraints}.
                \emph{Right}: %
                Posterior constraints on the potential geometry. %
                The lower-left panel shows the full parameter set $(q, \phi, x, y)$; the upper-right shows the $(q, \phi)$ subspace with centre of mass fixed to the galaxy's \gls{col}, used to colour the stream in the left panel. %
            \label{fig:102158269-2649560011643950805-constraints}}
        \end{figure}

        The host in \cref{fig:102158269-2649560011643950805-constraints} is an edge-on, disc-dominated galaxy with a bright central bulge and a thin vertical disc extending to $y \approx \pm 100$ pixels. %
        Slightly bluer luminosity enhancements at the disc edges may trace faint spiral arms or a ring, and a faint diffuse structure bending to the right from the galaxy's top may indicate a warped disc. %
        Several compact sources in the lower right, including a brighter object near $(x, y) = (100, -20)$\,\unit{\pix}, align with the disc and may be stellar clusters or background sources. %

        Most prominent is a long, coherent stellar stream arcing around the host galaxy in \cref{fig:102158269-2649560011643950805-constraints}. %
        The structure begins above the galaxy, curves along the left side, and intersects the disc in projection near $(x, y) = (-50, -80)$\,\unit{\pix}. %
        It likely continues through the disc and re-emerges on the opposite side, looping upward before crossing in front of the galaxy again just above the central bulge. %
        Its coherent morphology, consistent curvature, and projected close passage make it a valuable probe of the host's gravitational potential. %

        The corner plot in \cref{fig:102158269-2649560011643950805-constraints} shows the posterior distribution over the host halo parameters inferred from the stream morphology. %
        The \gls{com} is broadly constrained to lie within or near the arc traced by the stream, with the posterior bounded on the left by the stream's concavity and spanning a modest range in $x$ and $y$. %
        The stream's broad curvature places only weak constraints on the halo shape: most $(q, \phi)$ combinations remain allowed, and only a small subset of highly flattened models misaligned with the upper segment is disfavoured. %


    \subsection[XII) 102158893:2717847201657197076]{XII) \texttt{102158893:2717847201657197076}}\label{app:all_streams:102158893_2717847201657197076}

        \begin{figure}[!ht]
            \centering
            \includegraphics[width=0.85\linewidth]{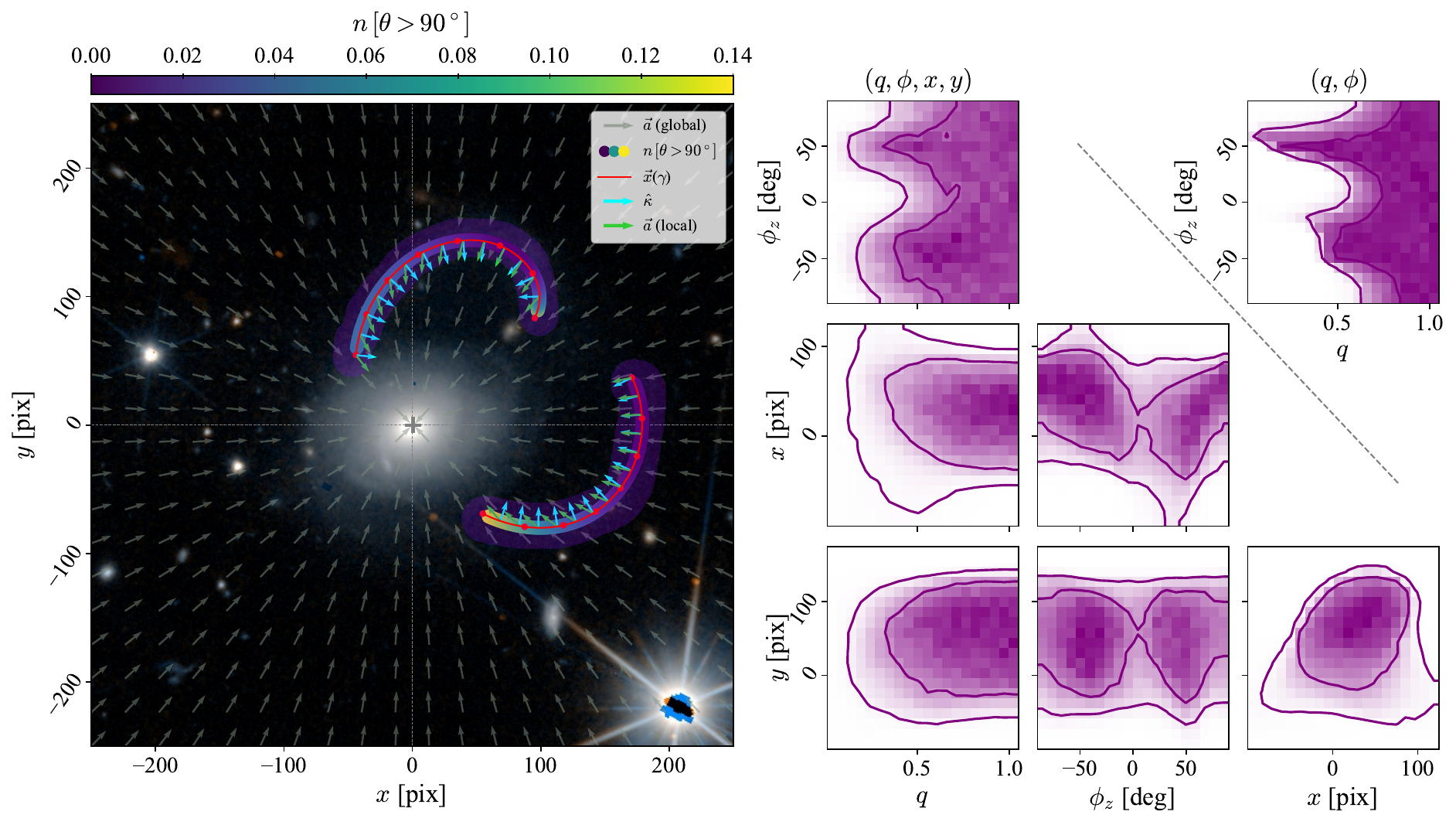}
            \caption{%
                Cutout~XII: \texttt{102158893:2717847201657197076} and geometry
                constraints. %
                \emph{Left}: %
                Same as \cref{fig:102014777-NEG780275758555920353-constraints}.
                \emph{Right}: %
                Posterior constraints on the potential geometry. %
                The lower-left panel shows the full parameter set $(q, \phi, x, y)$; the upper-right shows the $(q, \phi)$ subspace with centre of mass fixed to the galaxy's \gls{col}, used to colour the stream in the left panel. %
            \label{fig:102158893-2717847201657197076-constraints}}
        \end{figure}
        
        The host in \cref{fig:102158893-2717847201657197076-constraints} is an early-type galaxy with a bright central core and diffuse outer structure, slightly elongated at roughly \ang{20}. %
        The outer features may be a disc component or shell-like structures. %
        A saturated star is visible at the lower right, and redder patches may reflect extinction or projected background sources. %
        This system hosts two faint, narrow stellar stream segments that both intersect the galaxy and curve upward and to the right in projection. %
        They maintain a regular width along their visible extent, consistent with dynamically cold tidal debris from a disrupted dwarf-galaxy progenitor. %
        The upper segment is brighter than the lower one, with small stellar overdensity common in streams, though both share the same false-colour hue as the host. %
        A tentative segment below the galaxy may connect to the upper stream, suggesting a single stream that loops around the galaxy, whereas the second segment shows no obvious connection to that structure. %

        The corner plot in \cref{fig:102158893-2717847201657197076-constraints} shows the posterior distribution over the host halo parameters inferred from the two stream segments. %
        The highest-likelihood \glspl{com} lie within the concave regions traced by the streams and remain consistent with the galaxy, supporting a well-aligned dark matter halo. %
        Because both streams intersect the galaxy and curve in a similar direction, the posterior shows a strong degeneracy between the $(x, y)$ \gls{com} and the in-plane orientation angle $\phi$. %
        The flattening is constrained to satisfy $q \gtrsim 0.6$ except at orientations aligned with the luminous distribution, approximately $\phi \in [\ang{-50},\ang{50}]$, where more flattened halos remain allowed. %


    \subsection[XIII) 102159485:2680742453663022737]{XIII) \texttt{102159485:2680742453663022737}}\label{app:all_streams:102159485_2680742453663022737}

        \begin{figure}[!h]
            \centering
            \includegraphics[width=0.8\linewidth]{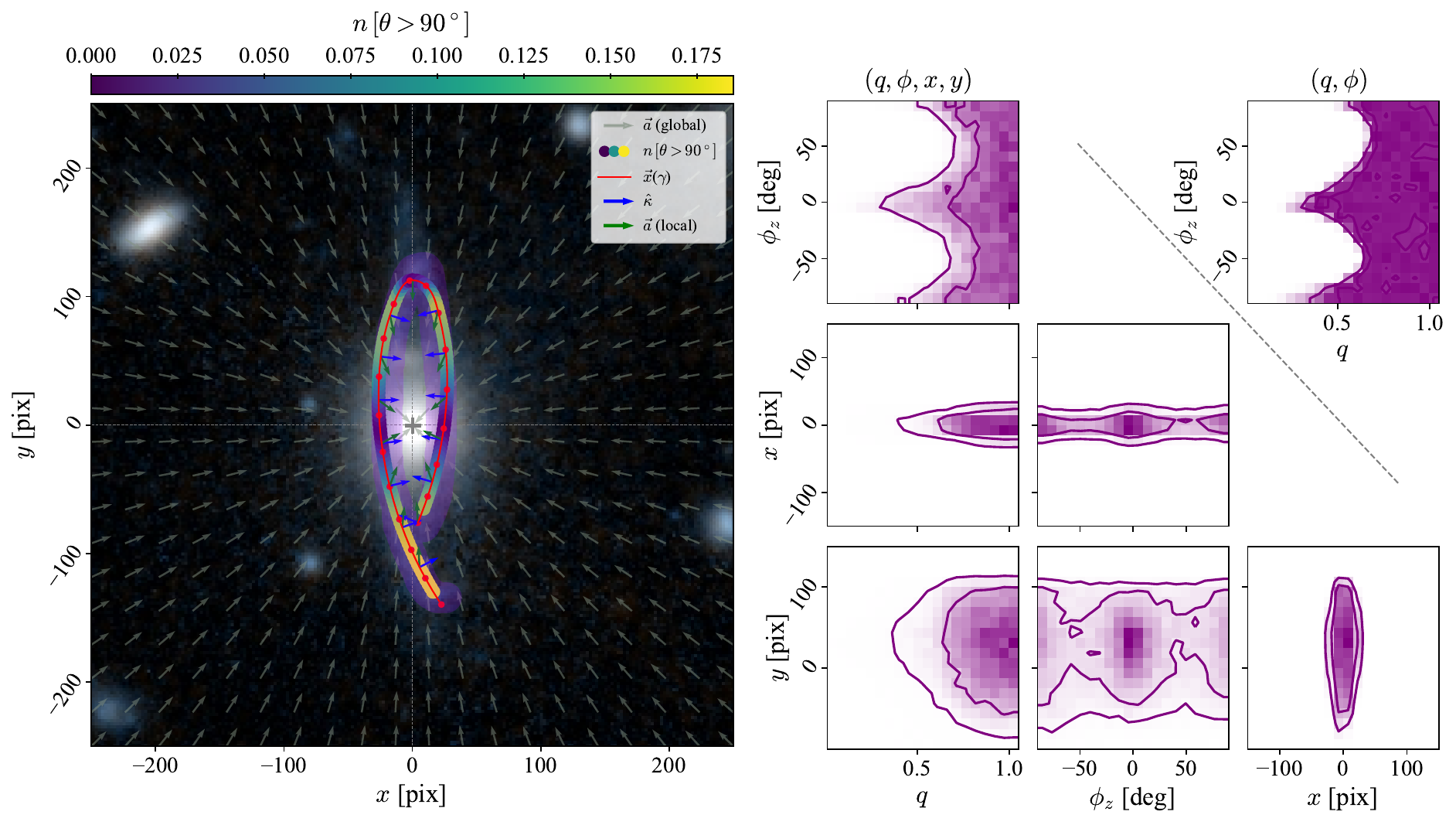}
            \caption{%
                Cutout~XIII: \texttt{102159485:2680742453663022737} and geometry
                constraints. %
                \emph{Left}: %
                Same as \cref{fig:102014777-NEG780275758555920353-constraints}.
                \emph{Right}: %
                Posterior constraints on the potential geometry. %
                The lower-left panel shows the full parameter set $(q, \phi, x, y)$; the upper-right shows the $(q, \phi)$ subspace with centre of mass fixed to the galaxy's \gls{col}, used to colour the stream in the left panel. %
            \label{fig:102159485-2680742453663022737-constraints}}
        \end{figure}

        The host in \cref{fig:102159485-2680742453663022737-constraints} is a featureless spheroidal galaxy with a bright core, no visible disc, spiral arms, or bar, and a uniform moderately red colour indicating a spatially homogeneous stellar population. %
        Two luminous star clusters to the left and lower left are likely host-associated, and a small feature at the galaxy's top may be a stellar overdensity or remnant substructure. %
        There is also a galaxy of comparable apparent mass, though not in the immediate vicinity of the host, which may be an interacting companion. %
        Encircling the host is a prominent stellar stream that loops tightly around the galaxy in a nearly vertical arc. %
        Its coherent path and narrow width are consistent with dynamically cold debris from the disruption of a dwarf satellite. %
        In the false-colour composite, the stream appears bluer than the host, suggesting a younger or more metal-poor stellar population. %
        If that neighbouring galaxy were projected substantially closer to the host, explicit modelling of its perturbation would become more important; at present, however, neither the galaxy morphology, the stream morphology, nor our posterior constraints indicate that such an additional component is required. %

        The corner plots in \cref{fig:102159485-2680742453663022737-constraints} show the posterior distribution over the host halo parameters inferred from the stream morphology. %
        The \gls{com} is tightly constrained to lie within the arc traced by the stream. %
        The flattening is broadly constrained to $q \gtrsim 0.5$, with highly flattened configurations allowed only for specific orientations. %
        In the free-centre case, modestly flattened models are permitted when the orientation angle $\phi$ matches the tilt of the projected stream. %
        With the centre fixed to the luminous core, a broader range of orientations is allowed, including configurations rotated by nearly \ang{90}, while spherical halos remain consistent with the data in both cases. %


\end{appendix}

\label{LastPage}
\end{document}